\documentclass[twocolumn]{aastex62}
\pdfoutput=1 %for arXiv submission
\usepackage{amsmath,amstext}
\usepackage[T1]{fontenc}
\usepackage{apjfonts}
\usepackage[figure,figure*]{hypcap}
\usepackage{rotating}
\usepackage{lineno}
\usepackage{multirow}
\usepackage{hyperref}
\usepackage{wrapfig}
\usepackage[shortlabels]{enumitem}
\graphicspath{{./}{figures/}}

\definecolor{chmagenta}{rgb}{0.54, 0.17, 0.88}

\shorttitle{$r$-process Enrichment of GCs from NSMs}
\shortauthors{Zevin et al.}

\begin{document}

\title{Can Neutron-Star Mergers Explain the $r$-process Enrichment in Globular Clusters?}

\author{Michael Zevin}\thanks{zevin@u.northwestern.edu}
\affiliation{Department of Physics and Astronomy, Northwestern University, 2145 Sheridan Road, Evanston, IL 60208, USA}
\affiliation{Center for Interdisciplinary Exploration and Research in Astrophysics (CIERA), 2145 Sheridan Road, Evanston, IL 60208, USA}

\author{Kyle Kremer}
\affiliation{Department of Physics and Astronomy, Northwestern University, 2145 Sheridan Road, Evanston, IL 60208, USA}
\affiliation{Center for Interdisciplinary Exploration and Research in Astrophysics (CIERA), 2145 Sheridan Road, Evanston, IL 60208, USA}

\author{Daniel M.\ Siegel}
\affiliation{Perimeter Institute for Theoretical Physics, Waterloo, Ontario, Canada, N2L 2Y5}
\affiliation{Department of Physics, University of Guelph, Guelph, Ontario, Canada, N1G 2W1}

\author{Scott Coughlin}
\affiliation{Center for Interdisciplinary Exploration and Research in Astrophysics (CIERA), 2145 Sheridan Road, Evanston, IL 60208, USA}

\author{Benny T.-H. Tsang}
\affiliation{Kavli Institute for Theoretical Physics, University of California, Santa Barbara, CA 93106, USA}

\author{Christopher P.\ L.\ Berry}
\affiliation{Center for Interdisciplinary Exploration and Research in Astrophysics (CIERA), 2145 Sheridan Road, Evanston, IL 60208, USA}

\author{Vicky Kalogera}
\affiliation{Department of Physics and Astronomy, Northwestern University, 2145 Sheridan Road, Evanston, IL 60208, USA}
\affiliation{Center for Interdisciplinary Exploration and Research in Astrophysics (CIERA), 2145 Sheridan Road, Evanston, IL 60208, USA}
\affiliation{CIFAR Fellow}

\begin{abstract}

Star-to-star dispersion of $r$-process elements has been observed in a significant number of old, metal-poor globular clusters (GCs). 
We investigate early-time neutron-star mergers as the mechanism for this enrichment. 
Through both numerical modeling and analytical arguments, we show that neutron-star mergers cannot be induced through dynamical interactions early in the history of the cluster, even when the most liberal assumptions about neutron-star segregation are assumed. 
Therefore, if neutron-star mergers are the primary mechanism for $r$-process dispersion in GCs, they likely result from the evolution of isolated, primordial binaries in the clusters. 
Through population modeling of double neutron-star progenitors, we find that most enrichment candidates are fast-merging systems that undergo a phase of mass transfer involving a naked He-star donor. 
Only models where a significant number of double neutron-star progenitors proceed through this evolutionary phase give rise to moderate fractions of GCs with enrichment; under various assumptions for the initial properties of GCs, a neutron-star merger with the potential for enrichment will occur in $\sim\,15-60\%$ ($\sim\,30-90\%$) of GCs if this phase of mass transfer proceeds stably (unstably). 
The strong anti-correlation between the pre-supernova orbital separation and post-supernova systemic velocity due to mass loss in the supernova leads to efficient ejection of most enrichment candidates from their host clusters. 
Thus, most enrichment events occur shortly after the double neutron stars are born. 
This requires star-forming gas that can absorb the $r$-process ejecta to be present in the globular cluster 30--50 Myr after the initial burst of star formation. 
If scenarios for redistributing gas in GCs cannot act on these timescales, the number of neutron-star merger enrichment candidates drops severely, and it is likely that another mechanism, such as $r$-process enrichment from collapsars, is at play.

\end{abstract}


\section{Introduction}

Astrophysical mechanisms for synthesizing the heaviest elements in the universe are poorly understood, yet essential in explaining the nucleosynthetic abundances observable today. 
Roughly half the elements heavier than iron are formed through the rapid capture of neutrons in a dense, neutron-rich environment, known as $r$-process nucleosynthesis \citep[e.g.,][]{Cowan2019,Kajino2019}. 
In these environments, the rate of neutron capture overcomes the rate of $\beta$-decay of radioactive nuclei, which converts the heavy nuclei into more stable isotopes with higher atomic numbers. 

%Core-collapse SNe (CCSNe) are a primary site for heavy-element nucleosynthesis, and act to synthesize many of the elements heavier than iron. 
%However, the conditions in these explosions are unlikely to favor $r$-process nucleosynthesis \citep[e.g.,][]{Arcones2007,Fischer2010,Hudepohl2010,Arcones2013}. 
Once the prevalent paradigm, regular core-collapse supernovae (CCSNe) are now strongly disfavored both theoretically \citep{Qian1996,Thompson2001,Roberts2012,Martinez-Pinedo2012} and observationally \citep{Wallner2015,Hotokezaka2015} as a production site for heavy $r$-process elements.
The neutron-rich ejecta from neutron-star mergers (NSMs) is theorized to fill these open gaps in the periodic table, polluting the universe with the heaviest naturally occurring elements \citep{Lattimer1974,Lattimer1976,Eichler1989,Meyer1989,Davies1994,Ruffert1997,Rosswog1999,Freiburghaus1999}. Over the last two decades, numerical simulations of the merger and post-merger phase have revealed several mechanisms by which neutron-rich material is ejected from these systems, including dynamical ejecta of tidal and shock-heated nature \citep{Ruffert1997,Rosswog1999,Oechslin2007,Hotokezaka2013}, neutrino-driven and magnetically driven winds from a (meta-)stable remnant \citep{Dessart2009,Siegel2014,Ciolfi2017}, and outflows from a post-merger accretion disk \citep{Fernandez2013,Just2015,Siegel2017}. The fact that NSMs can indeed synthesize $r$-process elements was corroborated in 2017 August; the discovery of GW170817 and observations of the subsequent kilonova explosion \citep{GW170817_MMA} provided unequivocal evidence that NSMs are a site for heavy $r$-process nucleosynthesis \citep[e.g.,][]{Chornock2017,Kasen2017,Tanvir2017}. 
If NSMs are the main channel for heavy $r$-process enrichment \citep[see][]{GW170817_kilonova,Cote2017,Hotokezaka2018}, $r$-process abundances in various astrophysical environments can thus probe the veiled physical mechanisms of binary stellar evolution, the rates of compact object mergers, and dynamical processes. 

Globular clusters (GCs) are one environment where $r$-process-enhanced, metal-poor stars have been observed, with a significant number of GCs exhibiting star-to-star dispersion of $r$-process species such as Eu and La \citep[e.g.,][]{Sneden1997, Roederer2011, Roederer2011a, Sobeck2011, Worley2013}. 
The homogeneity of iron-group abundances in these clusters, as well as the lack of correlation between the internal spread of $r$-process elements and light element dispersion, makes these observations difficult to explain in the standard CCSNe $r$-process scenario; such stellar explosions would likely introduce similarly high levels of Fe species into $r$-process-enhanced stars \citep[e.g.,][]{Ito2009,Placco2015}. 
Furthermore, measurements of [Pb/Eu] ratios indicate that these clusters have negligible dispersion of $s$-process elements \citep[e.g.,][]{Yong2006,Yong2008,Sobeck2011,Roederer2011}. 
Therefore, the astrophysical mechanism for introducing $r$-process elements must be inefficient at $s$-process production, or the dispersion of $s$-process elements is washed out from other, more frequent events. 

A number of observational campaigns have investigated the ubiquity of $r$-process enhancement in GCs. 
From a sample of $11$ low-metallicity GCs, \cite{Roederer2011a} found that four GCs showed clear signs of large $r$-process dispersion, five GCs showed no dispersion, and two GCs were more ambiguous with smaller levels of dispersion. 
Out of the five GCs that had no clear $r$-process dispersion, two showed bimodal chemical abundance ratios, possibly suggesting formation through the merger of two separate GCs that may have washed out initial star-to-star dispersion \citep{Yong2008,Marino2009,Carretta2010a}. 
Later work also found one of the enriched clusters from this sample to show no significant $r$-process dispersion \citep{Cohen2011}. 
More recent studies have expanded upon this sample and added more candidate GCs with and without evidence of $r$-process dispersion \citep[e.g.,][]{Roederer2015,Roederer2016}. 
Together, these samples show that heavy-element dispersion is a common feature of GCs, although further work is needed to robustly quantify the fraction of GCs that are enriched. 
Since heavy-element dispersion is apparent in a significant number of GCs but not ubiquitous across the full population, the mechanism for introducing $r$-process dispersion into these environments must occur in many, but not all, GCs. 

These observations have a few immediate consequences. 
First, they indicate some kind of extended or secondary star formation episode early in the history of the GC, leading to stellar populations with distinct chemical abundances due to astrophysical processes that transpired over this time.
Second, there must be an interstellar medium in the young GC with a high enough density to reduce the energy and momentum from the relativistic $r$-process ejecta and prevent it from leaving the GC \citep{Komiya2016}. 
Third, to provide this dense interstellar medium, there is likely a mechanism that replenishes the natal GC environment with gas for forming a second generation of stars \citep[such as asymptotic giant branch (AGB) ejecta;][]{Bekki2017}, as early-time SNe and feedback from of massive-star winds will efficiently remove the natal gas embedded in the GC. 
Finally, since GCs are believed to form the majority of their stars within the first $\sim$\,10 Myr after formation \citep[for reviews on star formation episodes and the formation of multiple stellar populations in GCs, see, e.g.,][]{Gratton2012,Bastian2018}, the enrichment mechanism had to proceed on a relatively rapid timescale. 

If NSMs are assumed to be responsible for the $r$-process enrichment in the second generation of stars formed in GCs, there are multiple stringent constraints on the properties of double neutron-star (DNS) systems at birth, most notably inspiral times and post-SN systemic velocities. 
Given the shallow gravitational potential expected of GC progenitors of $\sim$\,10 -- 100 km\,s$^{-1}$ and small physical sizes of only a few parsecs, if the newly formed DNSs attain appreciable post-SN systemic velocities, they are typically ejected from the cluster and must rapidly merge before fully evacuating the cluster environment. 
If newly formed DNSs receive small enough barycentric kicks to remain in the cluster environment, they still must merge before the cluster is void of the gas that will form the second generation of (enriched) stars. 
%This is also difficult, as the lower kicks will impart less eccentricity into the newly-formed DNS, resulting in a prolonged inspiral time. 
Dynamical interactions may also play a role in expediting NSMs in GCs \citep[e.g.,][]{Grindlay2006,Lee2010}, though these will need to have a significant impact early in the cluster lifetime. 
Though the NSM scenario for $r$-process enhancement has been explored in other environments such as ultra-faint dwarf galaxies \citep[UFDGs; e.g.,][]{Safarzadeh2018}, GCs provide a unique and complementary probe for investigating NSMs as the primary site for $r$-process enrichment due to their vastly different physical sizes and masses, shorter star formation timescales, and the possible role of dynamical encounters. 

In this paper, we examine multiple scenarios for rapid $r$-process enrichment in GCs using both of semi-analytic arguments and numerical modeling.
In Section~\ref{sec:dynamics}, we investigate NSMs produced by dynamical interactions, finding that this channel does not contribute to NSMs at early times. 
Section~\ref{sec:primordial} examines the impact that compact binaries formed from isolated primordial stellar pairs could have on enrichment. 
We show that primordial binaries can only lead to an appreciable number of $r$-process enhanced GCs if systems proceed through a phase of mass transfer (MT) involving a naked He-star donor. 
%Though the stability of MT and survival of DNS progenitors during this phase is highly uncertain \citep{Ivanova2003,Belczynski2008,Dominik2012,Tauris2015,Kruckow2016}, we find that even if MT proceeds stably an appreciable number of NSM enrichment candidates are produced. 
In Section~\ref{sec:discussion}, we discuss the effect that MT and SNe prescriptions have on the enrichment candidates in our models, review the implication of our results on the properties of GC progenitors and star formation timescales, compare our results to other environments that have observed $r$-process enhanced stars, and suggest alternative scenarios that may lead to $r$-process enhancement in GCs. 
Finally, we highlight our main conclusions in Section~\ref{sec:conclusions}. 
In all population modeling, we assume a Kroupa \citep{Kroupa2001} initial mass function (IMF) ranging from $0.08\,M_{\odot}\leq\,M\,\leq\,150\,M_{\odot}$, unless otherwise specified.

\section{Dynamical Assembly of DNS Systems}\label{sec:dynamics}

GCs host complex dynamics between stars and compact objects. 
Through dynamical friction, the most massive objects in a cluster sink to its core, where they readily take part in strong gravitational interactions. 
These interactions can induce rapid mergers of compact objects \citep[e.g., ][]{Samsing2014,Rodriguez2016a,Askar2017,Banerjee2017,Giesler2018,Hong2018,Fragione2018b,Rodriguez2018c,Zevin2019a,Kremer2019}. 
However, via energy equipartition, as the most massive objects migrate to the cluster core, lighter objects move further away from the cluster center. 
As a consequence, black holes (BHs) shape the dynamical evolution of the lower-mass ($M\,\lesssim 1\,M_{\odot}$) main-sequence stars and, because these lighter stars make up the bulk of the total cluster mass, the cluster as a whole \citep[e.g.,][]{Mackey2007,Mackey2008,Kremer2018,Askar2018}. 
At early times when a large BH population is present, BHs therefore inhibit the dynamical segregation of the NSs that could potentially merge and enrich the cluster with $r$-process elements. 
Nevertheless, it has been argued that dense stellar environments such as nuclear star clusters in the early universe can yield an appreciable rate of dynamically assembled NSMs and possibly explain the $r$-process enhancement of metal-poor stars \citep{Ramirez-Ruiz2015}. 

Over time, BHs will be ejected from the cluster through strong encounters with other BHs in the core \citep[e.g.,][]{Spitzer1987,Kulkarni1993,Sigurdsson1993,Morscher2015}, allowing lighter objects in the cluster to follow suit and migrate toward the core through a similar segregation process, ultimately leading to cluster core-collapse \citep[e.g.,][]{Kremer2018}. 
The depletion of BHs in the core also permits the segregation of NSs to the cluster core which leads to an increased formation rate of millisecond pulsars, DNSs, and possibly NSMs \citep{Ye2019}. 

Mass segregation timescales can be approximated by the time necessary for a GC to settle into equilibrium. 
The half-mass relaxation time is given by
\begin{equation}
    t_{\rm relax} = 0.138 \frac{M_{\rm c}^{1/2}R_{\rm h}^{3/2}}{\langle m \rangle G^{1/2} \ln \Lambda},
\end{equation}
where $M_{\rm c}$ is the total cluster mass, $\langle m \rangle$ is the average stellar mass, $R_{\rm h}$ is the half-mass radius, and $\ln \Lambda$ is the Coulomb logarithm, where $\Lambda \simeq 0.4\, N$ for GCs with $N$ being the total number of stars \citep{Spitzer1987}. 
For simple dynamical friction in a two-component model, the mass segregation timescale is
\begin{equation}\label{eq:mass_seq}
    t_{\rm ms}^i \sim \frac{\langle m \rangle}{m_i} t_{\rm relax},
\end{equation}
where $m_i$ is the mass of the segregating population \citep{Spitzer1987}. 
For a typical cluster with $M_{\rm c} \simeq 4 \times 10^5\,M_{\odot}$, $\langle m \rangle \simeq 1\,M_{\odot}$, $R_{\rm h} \simeq 1$ pc, and $N \simeq 8 \times 10^5$, and the half-mass relaxation time is $t_{\rm relax} \simeq 100$ Myr \citep{Meylan1997,Gurkan2004}. 
Therefore, $30\,M_{\odot}$ BHs segregate in $t_{\rm ms}^{\rm BH} \simeq 3 \times 10^6$ yr, whereas $1\,M_{\odot}$ NSs segregate in $t_{\rm ms}^{\rm NS} \simeq 10^8$ yr. 

To investigate this channel, we model GCs using the H\'enon-style Monte Carlo code \texttt{CMC} \citep[][]{Henon1971a,Henon1971b,Joshi2000,Joshi2001,Fregeau2003,Pattabiraman2013,Chatterjee2010,Chatterjee2013,Rodriguez2015a}. 
This code monitors and evolves the global properties of the GC via two-body relaxation, while accounting for binary stellar evolution \citep[using updated versions of the \texttt{SSE} and \texttt{BSE} codes;][]{Hurley2000,Hurley2002} and small-$N$ gravitational encounters using the \texttt{Fewbody} package \citep{Fregeau2004,Fregeau2007}, which now includes post-Newtonion effects in $N$-body integrations \citep{Antognini2014,Amaro-Seoane2016,Rodriguez2018c}. 
We record all strong encounters between objects, tracking the times at which new compact binary pairs are formed and calculating the gravitational-wave (GW) inspiral times of binaries synthesized due to these dynamical encounters. 

\begin{figure}[t!]\label{fig:cluster}
\includegraphics[width=0.5\textwidth]{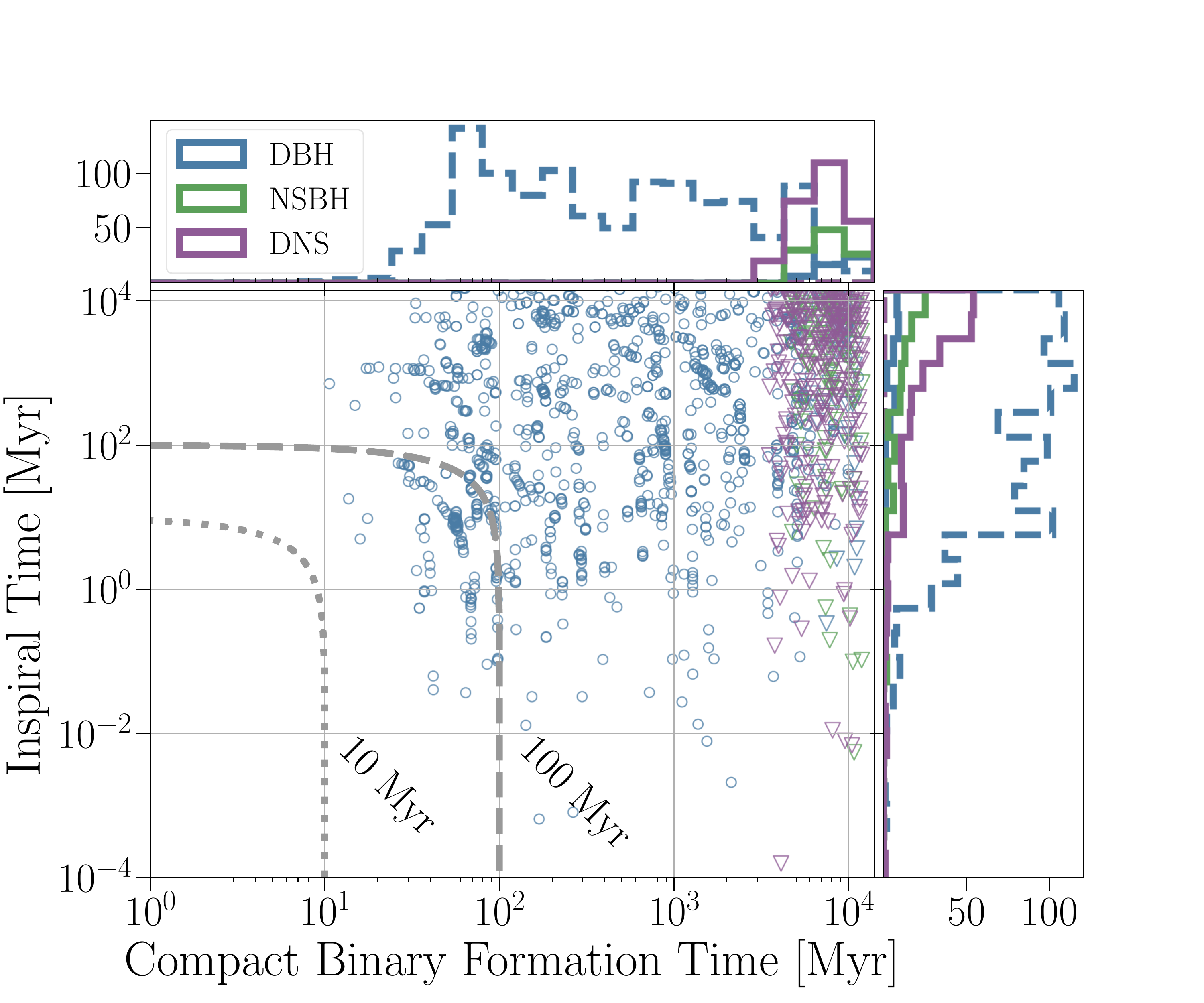}
\caption{Inspiral times of compact binaries synthesized from dynamical interactions, relative to the time when the interaction occurred in the cluster. 
Dotted (dashed) gray lines are lines of constant time at 10 (100) Myr, the time at which we assume enrichment events must occur by; points that fall below and to the left of these lines merge before star formation is assumed to cease. 
Colored points and histograms indicate different classes compact object mergers: double black hole (DBH), neutron star-black hole (NSBH), and double neutron star (DNS). 
Circle (triangle) markers and colored dashed (solid) histograms show the distribution of mergers from the standard (liberal) model. 
Even in the liberal model where BHs are artificially removed, no neutron-star mergers occur before a few Gyr after cluster formation.
}
\end{figure}

We simulate two clusters: one with typical GC properties and one with a number of liberal assumptions in an attempt to eliminate BHs from the environment so NSs will segregate to the cluster cores and interact as rapidly as possible. 
In this liberal model, we assume BHs have no mass fallback following the SN at their formation and therefore receive the full natal kicks typical of NSs, and truncate the IMF at 20 $M_{\odot}$ (which corresponds to a remnant mass of $\sim 5-8\,M_{\odot}$ at low metallicities; \citealt{Giacobbo2018a}). 
This acts to reduce the number of BHs that are created from the initial stellar population, and efficiently eject those that do happen to form. 
Furthermore, we reduce all NS natal kicks, drawing their magnitude from a Maxwellian with a dispersion of $20~\mathrm{km\,s}^{-1}$ instead of a dispersion of $265~\mathrm{km\,s}^{-1}$ \citep{Hobbs2005}. 
Figure \ref{fig:cluster} shows the time that compact binaries are dynamically formed in these models compared to their inspiral times. 
The sum of these two quantities approximates the time in the history of the cluster at which these compact binaries would merge, and in the case of systems with a NS component, possibly enrich the cluster with $r$-process material.

As expected, in the standard model binary BHs are the dominant dynamically induced systems and inhibit the formation of other types of compact binaries for most of the GC lifetime. 
Only a small number of BHs merge within 100 Myr of the formation of the GC, and none merge within 10 Myr. 
Primordial mass segregation in star clusters may help expedite dynamically induced BH mergers \citep[e.g.,][]{Parker2018,Alfaro2018}, though the interaction rate of other types of compact objects will be stifled until the higher-mass BH population is processed and ejected. 

In our liberal model that artificially removes most BHs and amplifies the number of retained NSs, we still do not find NSMs until a few Gyr after cluster formation --- far too late to enrich a second generation of stars with $r$-process material. 
Though BHs are efficiently removed from the cluster in this model, the 1--2\,$M_{\odot}$ NSs are still less massive than the surviving main-sequence stars. 
Therefore, mass segregation is still inefficient for $\sim$\,3 Gyr until these stars finish their main-sequence evolution. 
Though dynamically formed neutron star-black hole (NSBH) systems may also play a role in enriching our liberal GC model, the NSBH rate is lower than the DNS rate and they share similar inspiral time distributions; hence, NSBH enrichment should be subdominant to DNS enrichment. 
We therefore conclude that \emph{dynamical interactions in GCs do not contribute to early-time NSMs}, and it is unlikely that dynamics plays a significant role enhancing second-generation GC stars with $r$-process material.

\section{Mergers from Primordial Cluster Binaries}\label{sec:primordial}

We next investigate NSMs from primordial binaries in GCs following their isolated binary stellar evolution.\footnote{In this context, primordial binaries refer to stars that are born as binary pairs in the GC and have yet to segregate and undergo any dynamical interactions.}  
To pollute the cluster with $r$-process elements shortly after formation, DNSs must acquire short inspiral times through tight orbital separations and/or significant orbital eccentricities at birth. 
The standard picture by which compact binaries achieve hardened orbits during isolated binary evolution is via (at least) one common envelope (CE) phase, where one of the stars experiences Roche-lobe overflow (RLO) and unstable MT, enveloping both bodies in the outer layers of the donor star and exerting a drag on the orbiting binary system. 

The common envelope causes the binary to spiral inwards and harden its orbit \citep[e.g.,][]{Andrews2015,Tauris2017,Chruslinska2018}. 
For compact binaries that merge within a Hubble time, a CE phase will typically need to occur after one of the stars in the binary has already evolved into a compact object, so that the orbit can harden sufficiently without the bodies coming into contact and merging. 
Though the system circularizes following this evolutionary stage, eccentricity imparted into the system via the subsequent SN that forms the second NS will further reduce the GW inspiral time.

\pagebreak
\subsection{Enrichment Efficiency}

In addition to altering the orbital properties of the compact binary, the second SN can significantly kick the system through a combination of asymmetries in the explosion mechanism (the natal kick) and from symmetric mass loss of the exploding star \citep[the Blaauw kick;][]{Blaauw1961}. 
The post-SN center-of-mass velocity is referred to as the systemic velocity. 
Given a final mass $M_{\rm f} = m_1+m_2$ and initial mass $M_{\rm i} = m_1+m_2+\Delta M_{\rm SN}$ where $m_1$, $m_2$, and $\Delta M_{\rm SN}$ are the first-born NS mass, the second-born NS mass, and mass lost during the SN that creates the second NS, we can write the magnitude of the post-SN systemic velocity as
\begin{equation}\label{eq:Vsys}
    v_{\rm sys}^2 = \frac{1}{M_{\rm i}^2} 
    \left[
    (m_2 v_{\rm k})^2 
    + \frac{m_1 \Delta M_{\rm SN} v_{\rm r}}{M_{\rm f}^2}
    \left(m_1 \Delta M_{\rm SN} v_{\rm r} - 2 m_2 M_{\rm f} v_{\rm k \parallel}
    \right)
    \right],
\end{equation}
\noindent where $v_{\rm k}$ is the magnitude of the natal kick, $v_{\rm r}$ is the relative orbital velocity between the two objects prior to the SN, and $v_{\rm k \parallel}$ is the component of the natal kick aligned with the instantaneous orbital velocity of the exploding star \citep[e.g.,][]{Kalogera1996}. 

We consider DNS systems viable $r$-process polluters if the system merges within some enrichment radius of the cluster, $R_{\rm enrich}$, and within some span of time when new stars are still being born after the initial burst of star formation, $\Delta\tau_{\rm SF}$. 
This can happen in two ways: 
(i) the post-SN systemic velocity is greater than the cluster escape speed, but the DNS merges before leaving the cluster environment; (ii) the post-SN systemic velocity is less than the cluster escape speed, and the DNS delay time is less than $\Delta\tau_{\rm SF}$.
To derive an approximate escape velocity, we assume the mass distribution in our fiducial clusters follow a Plummer profile \citep{Plummer1911}: 
\begin{equation}
    \rho_\mathrm{p}(r) = \left(\frac{3 M_{\rm c}}{4 \pi R_\mathrm{p}^3} \right) \left( 1 + \frac{r^2}{R_\mathrm{p}^2} \right)^{-5/2},
\end{equation}
where $M_{\rm c}$ is the mass of the cluster and $R_\mathrm{p}$ is the Plummer radius. 
Given a Plummer radius, we assume DNS systems form at the half-mass-radius, which is $r_h \approx 1.3 R_\mathrm{p}$ for a Plummer sphere. 
The escape velocity is thus 
\begin{equation}
    v_{\rm esc}(r_h) = \sqrt{2 |\Phi_\mathrm{p} (r_h)|} 
    \approx 72 \left(\frac{M_{\rm c}}{10^6 M_{\odot}}\right)^{1/2} \left(\frac{R_{\rm p}}{1 {\rm pc}}\right)^{-1/2} {\rm km\,s}^{-1},
\end{equation}
where $\Phi_\mathrm{p}$ is the gravitational potential of the Plummer model. 
Though this is a simplistic description for the true potential of a young GC, we find our results to be robust to changes in the form of the potential since the post-SN systemic velocity of most enrichment candidates greatly exceeds the GC escape velocity. 

Assuming negligible deceleration from the gravitational potential, if $v_{\rm sys}$ > $v_{\rm esc}$, the time it takes for the DNS to go beyond the enrichment radius of the cluster is 
\begin{equation}
    \tau_{\rm eject} 
    \approx \frac{R_{\rm enrich}}{v_{\rm sys}}.
\end{equation} 
We typically assume that the enrichment radius is $R_{\rm enrich} = R_{\rm vir} \approx 1.7 R_{p}$ \citep[e.g.,][]{PortegiesZwart2010}, though later we will relax this assumption. 

Systems that remain bound to the cluster after being kicked and merge within $\Delta\tau_{\rm SF}$ are also viable $r$-process polluters. 
The delay time is defined as $t_{\rm delay} = t_{\rm DNS} + t_{\rm insp}$, where $t_{\rm DNS}$ is the time from zero-age main sequence (ZAMS) of the DNS progenitors to DNS formation and $t_{\rm insp}$ is the DNS inspiral time due to GW emission \citep{Peters1964}. 
We therefore categorize DNS systems that remain bound as viable polluters if $t_{\rm delay} < \Delta\tau_{\rm SF}$. 
Thus, the DNS enrichment efficiency is 
\begin{equation}\label{eq:enrich_efficiency}
    \epsilon_{\rm DNS} = \frac{1}{N_{\rm DNS}} \sum_{i\,=\,1}^{N_{\rm DNS}} \Theta(\Delta\tau_{\rm SF}^i - t_{\rm delay}^i) \times \mathcal{T},
\end{equation}
where $N_{\rm DNS}$ is the total number of DNS systems, $\Theta$ is the Heaviside step function, and $\mathcal{T} = \Theta(\tau_{\rm eject}^i - t_{\rm insp}^i)$ if $v_{\rm sys} > v_{\rm esc}$, and $\mathcal{T} = 1$ if $v_{\rm sys} \leq v_{\rm esc}$.

\subsection{NS Natal Kicks}

Proper motions of isolated pulsars in the Milky Way indicate that many NSs receive large natal kicks at birth, on the order of a few hundred kilometers per second (\citealt{Fryer1997,Hobbs2005,Bray2018}, though see also \citealt{Verbunt2017}). 
As described above, post-SN systemic velocities of DNS systems are affected by natal kicks, and can lead to systemic velocities that are $\lesssim$\,50\% larger than the pre-SN orbital velocity \citep{Kalogera1996}. 
%Large natal kicks will more often disrupt the binary and therefore decrease DNS formation rates \citep{Tauris1998}, though particular orientations and magnitudes of the natal kick, pre-SN orbital velocity, and mass loss in the SN can result in bound systems \citep{Wex2002}. 
Since the escape speeds of GCs are typically a few tens of kilometers per second, DNSs in relatively tight pre-SN orbits that are not disrupted from standard CCSNe natal kicks will usually lead to post-SN systemic velocities that unbind the systems from their host cluster. 

Evidence for some NSs receiving lower natal kicks has been determined by examining the proper motions of both the isolated NS population \citep{Brisken2002} and Galactic DNSs \citep{Wong2010,Schwab2010,Beniamini2016, Tauris2017}. 
These lower kick magnitudes of a few tens of kilometers per second are predicted for DNSs that explode due to electron capture in a strongly degenerate ONeMg core, known as electron-capture SNe \citep[ECSNe;][]{Miyagi1980,Nomoto1984,Nomoto1987}.
ECSNe are typically assumed to occur when a star has a He core mass of around 2 $M_{\odot}$ at the base of the AGB branch \citep[see][]{Ivanova2008}, though the extent and placement of this range is debated and can be strongly affected by binary interactions \citep{Podsiadlowski2004}. 
This pathway will make it more likely for a DNS to remain bound to the cluster (as well as survive the SN), since lower natal kicks and smaller amounts of mass loss lead to smaller post-SN systemic velocities. 
However, the eccentricities imparted into these systems from the SN will likewise be smaller, increasing the typical GW inspiral time. 

In our population modeling, we apply a bimodal distribution for natal kick magnitudes. 
Standard CCSNe have natal kicks drawn from a Maxwellian distribution with scale parameter $\sigma_{\rm high} = 265$\,km\,s$^{-1}$ \citep{Hobbs2005}, whereas for stars with He core masses at the base of the AGB branch in the range $1.4\,M_{\odot} \leq {\rm m}_{\rm core} \leq 2.5\,M_{\odot}$, which are predicted to undergo ECSNe \citep{Pfahl2002,Podsiadlowski2004}, we draw natal kicks from a Maxwellian distribution with $\sigma_{\rm low} = 20$ km\,s$^{-1}$. 
Another possible mechanism for stifling natal kicks of DNSs is by significantly stripping the atmosphere of the progenitor star via binary interactions prior to the SN, a scenario known as ultra-stripped SNe \citep[USSNe;][]{Tauris2013}, which we also take into account in one of our population models detailed below. 
We also apply natal kicks from the $\sigma_{\rm low}$ distribution for systems that form through accretion-induced collapse of an ONeMg white dwarf \citep{Nomoto1991,Saio2004} or a merger-induced collapse \citep{Saio1985}, though these channels (especially the latter) do not typically lead to DNS formation.

\subsection{Case BB MT}

One aspect of binary evolution that affects DNS populations is the onset, stability, and outcome of MT that results from a post-He main-sequence star overflowing its Roche lobe, known as Case BB MT \citep{Delgado1981,Dewi2002,Ivanova2003,Tauris2013}. 
Case BB MT is believed to occur during the He-burning analog of the Hertzprung gap (HG), when core He burning has ceased and shell burning causes the star to expand. 

Originally, it was assumed that the RLO of low-mass He stars that entered the HG was unstable and led to a successful CE phase \citep{Belczynski2002}, causing the spiral-in of the already-formed NS and the core of the He-star donor. 
However, the ability of binary systems to proceed through this phase is highly uncertain. 
For example, dynamical instability during RLO may lead to merger for donor stars without a core-envelope structure or a clear entropy jump at the core-envelope transition \citep{Ivanova2004,Belczynski2008}. 
\cite{Ivanova2003} demonstrated with detailed stellar evolution simulations that most unstable MT during this evolutionary phase leads to delayed dynamical instability and merger, and for DNS progenitors the MT typically proceeds stably during He-shell burning. 
More recent work by \cite{Tauris2015} has confirmed this, and found that the stability may be even more prevalent in DNS progenitors. 
This phase of binary evolution has important implications on fast-merging DNSs: Case BB MT is predicted to lead to hardened DNS systems with orbital periods of $\lesssim$\,10$^{-2}$ days at formation in the most extreme cases, corresponding to GW inspiral times of $\mathcal{O}(10^3)$ yr, \citep[e.g.,][]{Vigna-Gomez2018}.

\subsection{Population Models}\label{subsec:pop_models}

To determine $\epsilon_{\rm DNS}$, we simulate multiple populations of merging DNSs using the \texttt{COSMIC} population synthesis code \citep{COSMIC}.\footnote{\href{https://cosmic-popsynth.github.io/}{cosmic-popsynth.github.io}} 
\texttt{COSMIC} is a modified version of \texttt{BSE} \citep{Hurley2002}, which relies on polynomial fitting formulae for single-star evolution \citep{Hurley2000} and includes physical prescriptions for binary evolutionary processes such as tidal evolution, MT, CEs, and GW decay \citep{Hurley2002}. 
\texttt{COSMIC} is updated to include state-of-the-art prescriptions for mass loss in O and B stars \citep{Vink2001}, metallicity dependence in the evolution of Wolf-Rayet stars \citep{Vink2005}, new prescriptions for fallback and post-SN remnant masses \citep{Fryer2012}, variable prescriptions for the CE $\lambda$ parameter \citep{Claeys2014}, as well as prescriptions for ECSNe \citep{Podsiadlowski2004}, USSNe \citep[][]{Tauris2015}, and (pulsational) pair instability SNe \citep{Woosley2016}. 
A modification to \texttt{BSE} pertinent to this study is a correction to an inconsistency regarding the masses and orbital separations used when an SN occurs directly after a CE phase; more details can be found in Appendix \ref{Appendix}. 
In addition, \texttt{COSMIC} determines when particular populations of compact binaries have been adequately sampled by repeatedly checking for convergence in the distributions of various binary properties \citep{COSMIC}. 

The four models we explore in this study cover a range of uncertainties in binary evolution by varying certain aspects of Case BB MT, survival through CEs, and natal kick prescriptions. 
In all models, initial binary properties are determined according to the prescriptions in \cite{Moe2017}, and we assume that all stars form in a single burst of star formation with a metallicity of $Z_{\odot}/20$.\footnote{Though DNS merger rates are strongly impacted by metallicity, properties of merging DNSs vary only slightly with metallicity \citep{Dominik2012,Giacobbo2018a,Chruslinska2019,Neijssel2019}, and thus our choice of metallicity should have little effect on our results.}
In addition to providing astrophysically motivated distributions of post-SN orbital properties and systemic velocities, \texttt{COSMIC} tracks the total sampled mass necessary to generate our population of DNSs. 
The converged simulations resulted in $\mathcal{O}(10^4)$ DNS systems per population, corresponding to a total sampled mass of $\approx\,10^9\,M_{\odot}$. 
Details of each population model are as follows.

\begin{figure*}
\centering
\includegraphics[height=0.9\textheight]{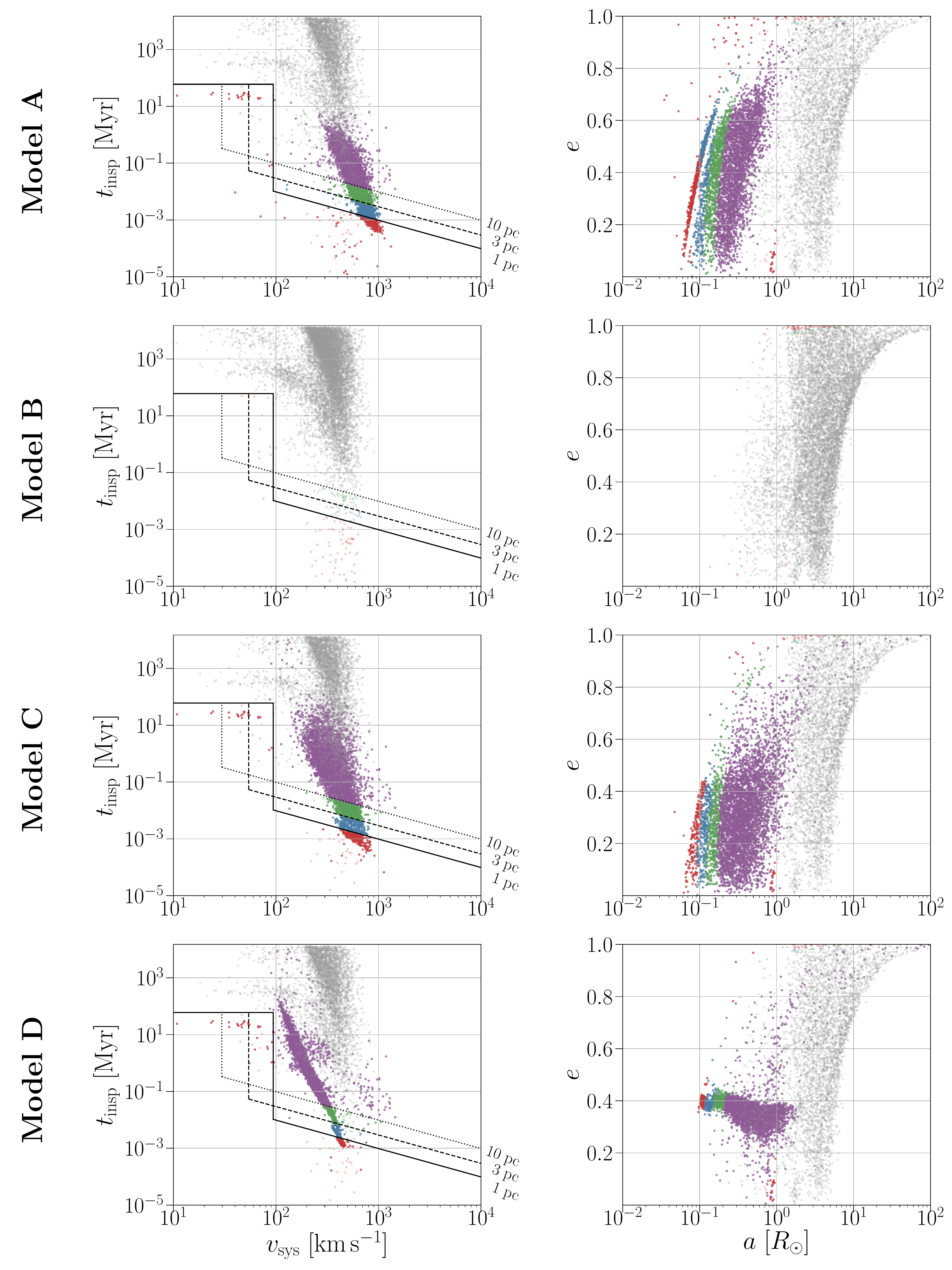}
\caption{Post-SN properties of DNS systems that merge within a Hubble time from binary population synthesis models with varying assumptions (see Section \ref{subsec:pop_models} for details). 
Black lines distinguish systems that are enrichment candidates for three assumed cluster virial radii: solid, dashed, and dotted lines correspond to 1 pc, 3 pc, and 10 pc, respectively. 
The diagonal component of the black lines marks a constant travel distance as a function of $v_{\rm sys}$, the vertical black lines mark the cluster escape velocity (assuming a GC progenitor mass of $10^6\,M_{\odot}$, and the horizontal black lines mark maximum inspiral time (assuming $\Delta\tau_{\rm SF} = 100$ Myr and $t_{\rm DNS} = 40$ Myr). 
Systems that fall below and to the left of the black lines are viable enrichment candidates assuming a virial radius of 1 pc (red), 3 pc (blue), and 10 pc (green), and are colored the same in both the left and right panels in a given row. 
For the remainder of the systems, purple points mark where the secondary star went through a stage of stable or unstable Case BB MT, and gray points mark systems where the secondary star went through only one CE. 
}
\label{fig:primordial}
\end{figure*}

\begin{enumerate}[A)]
%    \item Standard \texttt{BSE} model, which allows for evolved He stars to proceed through a successful CE.
%    NS masses are calculated using the Delayed prescription from \cite{Fryer2012}.\footnote{Models run using the Rapid prescription from \cite{Fryer2012} only led to percent-level differences in our results.}  
%    For naked He stars, we adopt maximum mass ratios for stable MT from \cite{Ivanova2003}: $q_{\rm crit} = 1.7$ for main sequence naked He stars (denoted as stellar type $k_{\star}=7$ in \texttt{BSE}) and $q_{\rm crit} = 3.5$ for HG or giant branch naked He stars ($k_{\star}=8,\ 9$). 
%    Other stellar types use the default \texttt{BSE} $q_{\rm crit}$ values. 
    
%($k_{\star}=7$), $q_{\rm crit} = 3.0$, and for HG or giant branch naked He stars ($k_{\star}=8,9$), $q_{\rm crit} = 0.784$. 
    
%    \item Same as Model A, except with $q_{\rm crit}$ values of naked He star donors updated according to \cite{Ivanova2003}: $q_{\rm crit} = 1.7$ for main sequence naked He stars and $q_{\rm crit} = 3.5$ for HG or giant branch naked He stars. 
%    This makes it more difficult for evolved naked He stars to enter a CE phase. 
    
    \item Standard model, which allows for evolved He stars to proceed through a successful CE.
    NS masses are calculated using the Delayed prescription from \cite{Fryer2012}.\footnote{Models run using the Rapid prescription from \cite{Fryer2012} only led to percent-level differences in our results.}  
    For naked He stars, we adopt maximum mass ratios for stable MT from \cite{Ivanova2003}: $q_{\rm crit} = 1.7$ for main-sequence naked He stars (denoted as stellar type $k_{\star}=7$ in \texttt{BSE}) and $q_{\rm crit} = 3.5$ for HG or giant branch naked He stars ($k_{\star}=8,\ 9$). 
    Other stellar types use the default \texttt{BSE} $q_{\rm crit}$ values. 
    %Same as Model A, except SNe that immediately follow a CE phase use the post-CE masses and orbital separations. 
    %By default, \texttt{BSE} uses the post-CE separation and pre-CE mass, which includes the mass of the ejected envelope. 
    %More details can be found in Appendix \ref{Appendix}. 

    \item Same as Model A, except that unstable MT from donor stars without a well-developed core-envelope structure always lead to a merger. 
    This is also assumed to be the case for donors on the HG, since they lack a clear entropy jump at the core-envelope boundary \citep{Ivanova2004}. 
    Analogous to the pessimistic CE model in \cite{Belczynski2008}, this drastically limits the number of short-period systems since it eliminates potential DNS progenitors that would have otherwise gone through a Case BB CE phase. 

    \item Same as Model A, except orbital properties and mass loss following MT with a He-star donor are calculated according to the prescriptions in \cite{Tauris2015}, who found that this phase of MT to typically proceeds stably and does not initiate a CE phase, and provided fitting formulae that map pre-RLO masses and separations to their values following stable Case BB RLO (i.e. the values immediately prior to the second SN).\footnote{Though further phases of stellar evolution occur before the SN, they proceed on rapid timescales and any stable or unstable MT will not strongly affect the orbital properties of the pre-SN system. }
    Though these fitting formulae are derived from systems at solar metallicity with pre-CE He-star masses between $2.5 \leq M_{\rm He}/M_{\odot} \leq 3.5$ and companion masses of $m_{\rm NS} = 1.35 M_{\odot}$, we use these fits for all He-star--NS systems that undergo Case BB MT. 
    Extrapolating to larger pre-RLO He-star masses may lead to larger uncertainties in post-RLO parameters, and exact results could potentially differ at the lower metallicities considered here. 

    \item Same as Model C, except all systems that undergo Case BB RLO are assumed to become ultra-stripped and receive the same natal kicks as ECSNe. 
\end{enumerate}

The time $t_{\rm DNS}$ between ZAMS and DNS formation for our population ranges from $\approx$\,11 to 68 Myr, with a median value of $t_{\rm DNS} = 41$ Myr. 
When calculating $t_{\rm delay}$ for an individual system, we use the $t_{\rm DNS}$ found from our modeling of that system and integrate its post-SN semi-major axis and eccentricity according to \cite{Peters1964} to determine $t_{\rm insp}$. 

\subsection{Enrichment probability from primordial DNSs}
\label{subsec:enrichment_prob}
%We assume a single NSM within the virial radius of the GC is sufficient to enrich its second generation of stars, an assumption corroborated by the expected Eu yield in NSM dynamical ejecta \citep{Macias2018}. 

Figure \ref{fig:primordial} shows the orbital properties immediately following the second SN for all DNSs that merge within a Hubble time in our population models, as well as the corresponding post-SN systemic velocities and inspiral times. 
We assume that the first generation of stars is born as a single burst at $t=0$ and that the second generation of (enriched) stars must be born within $\Delta \tau_{\rm SF} = 100\,\mathrm{Myr}$, though we investigate variations in the latter assumption in Section \ref{subsec:tauSF}. 
Black lines separate DNS systems that are viable enrichment candidates (lower left) from those that either merge outside the cluster or after the star formation has ceased (upper right) for three assumed cluster sizes. 
Therefore, the enrichment fraction $\epsilon_{\rm DNS}$ is given by the fraction of systems that lie to the lower left of these lines. 
For the remainder of the population, systems that undergo a Case BB MT phase are colored in purple whereas those that proceed through only a single CE after the first NS formed (typically during the hydrogen giant branch of the secondary) are in gray. 
Systems that undergo Case BB MT provide the largest contribution to $\epsilon_{\rm DNS}$, otherwise viable enrichment candidates can only result from systems that are kicked into an extremely eccentric orbit at birth (e.g., the small number of red points in Model B). 
However, as shown in Eq.~\eqref{eq:Vsys}, the post-SN systemic velocity scales with the orbital velocity of the binary prior to SN. 
Therefore, tighter pre-SN binaries will have larger Blaauw kicks; this anti-correlation between systemic velocity and inspiral time for the Case BB systems in evident in Figure \ref{fig:primordial}. 

To calculate the fraction of GCs we expect to exhibit $r$-process dispersion, we must first determine whether a single NSM can distribute enough $r$-process material to enrich the stars in a GC. 
The total Eu mass of a GC can be roughly estimated by $M_\mathrm{Eu} \approx X_{\mathrm{Eu},\odot} 10^{[\mathrm{Eu/H}]} M_\mathrm{gas}$, where $X_{\mathrm{Eu},\odot}$ is the solar Eu mass fraction and $M_\mathrm{gas}$ is the mass of intra-cluster gas in which the $r$-process was mixed into. For typical values, we have
\begin{equation}
    M_\mathrm{Eu} \approx (4.2 \times 10^{-5}) \times 10^{[\mathrm{Eu/H}]} \left(\frac{M_\mathrm{gas}}{10^{5}\,M_{\odot}}\right) M_{\odot},
\end{equation}
where we have used the solar abundances from \citet{Arnould2007}. 
This amounts to $M_\mathrm{Eu} \sim\,4 \times 10^{-6}\,M_{\odot}$ assuming typical GC abundances from \citet{Roederer2011a}. 
Assuming that an NSM can eject $\approx\,0.05\,M_{\odot}$ of $r$-process material, consistent with GW170817 \citep{Cowperthwaite2017,Chornock2017,Kasen2017,Villar2017}, and assuming a solar abundance pattern starting at mass number $A=69$ \citep{Arnould2007}, this translates into a Eu mass of $M_\mathrm{Eu}\approx 5.2\times 10^{-5}\,M_{\odot}$. 
We therefore conclude that a single NSM may be sufficient to enrich the second generation of stars in a typical GC.

Each model in Figure \ref{fig:primordial} is synthesized using a sample of binaries far larger than the stellar mass of a young GC. 
To quantify the probability that an enrichment event will occur in a GC with a stellar mass at birth of $M_{\rm c}$, we must determine the rate at which an NSM enrichment event would occur in a GC with a given stellar mass budget. 
%we must multiply $\epsilon_{\rm DNS}$ by the mass fraction of stars that become DNSs, and scale this number by the typical stellar mass of a young GC. 
The sampling procedure used in \texttt{COSMIC} from \cite{Moe2017} determines the probability that a system with a given primary mass and orbital period will have a companion.\footnote{The prescriptions from \cite{Moe2017} are based on observations of binaries in the Galactic field. 
Though there is significant uncertainty in the initial binary fraction of high-mass stars in GCs \citep{Ivanova2005}, observations of young stellar clusters still find that most massive stars are in binary pairs \citep{Sana2011}, and we expect slight variations in binary fractions to have a percent-level impact on our results. }
Though \texttt{COSMIC} only evolves binary systems, the total mass sampled includes the mass of both single and binary star systems. 
%Though over half of all high-mass stars in young stellar clusters are found in binary pairs \citep{Sana2011}, there is significant uncertainty in the initial binary fraction of high-mass stars in GCs \citep{Ivanova2005}. 
%Since low-mass stars dominate the IMF,\footnote{Assuming a Kroupa IMF \citep{Kroupa2001} and a stellar mass range of $0.05 M_{\odot} \leq M_{\star} \leq 300 M_{\odot}$, stars above $8 M_{\odot}$ only account for $\sim$13\% of the total stellar mass.} we use a binary fraction in GCs of $f_{\rm bin} = 0.1$, which is supported by observations of GCs \citep[e.g.,][]{Rubenstein1997,Bellazzini2002,Lucatello2015,Ji2015} and has been shown through numerical modeling to reproduce present-day binary fractions \citep{Hurley2007,Chatterjee2010}. 
Therefore, the expected number of enriching NSMs in a GC is 
\begin{equation}
    \Lambda_\mathrm{enrich} = \epsilon_{\rm DNS} N_{\rm DNS} \frac{M_{\rm c}}{M_{\rm samp}},
\end{equation}
where $M_{\rm samp}$ is the total mass sampled in our population models. 
Given the mean number $\Lambda_\mathrm{enrich}$, the probability that a GC has at least one enriching NSM follows a Poisson process as
\begin{equation}
    P_\mathrm{enrich} = 1 - \exp(-\Lambda_\mathrm{enrich}).
\end{equation}
The expected enrichment probabilities for our four population models using multiple assumptions for $R_{\rm enrich}$ and $M_{\rm c}$ are in Table \ref{table}.\footnote{We do not consider NSBH mergers in our enrichment probability calculations. 
The upper limit on the NSBH merger rate density measured by LIGO--Virgo is already lower than the median DNS merger rate density \citep{TheLIGOScientificCollaboration2019}. 
The mass fraction of binary stars that results in NSBH mergers should therefore be lower than that for DNSs, and thus their enrichment contribution in GCs should be subdominant. }

\begin{table}[b!]
\centering
 \begin{tabular}{c c || c c c}
 & & \multicolumn{3}{c}{$R_{\rm vir}$ [pc]} \\
 & $M_{\rm c}$ [$M_{\odot}$] & 1.0 & 3.0 & 10.0 \\ [0.2ex]
\hline\hline
 Model A   &  $5 \times 10^5$   & 0.31 & 0.55 & 0.72 \\
      &  $10^6$   & 0.52 & 0.80 & 0.92 \\
 \hline
  Model B   &  $5 \times 10^5$   & 0.02 & 0.02 & 0.03 \\
      &  $10^6$   & 0.04 & 0.05 & 0.06 \\
 \hline
  Model C   &  $5 \times 10^5$   & 0.14 & 0.27 & 0.46 \\
      &  $10^6$   & 0.27 & 0.48 & 0.70 \\
 \hline
  Model D   &  $5 \times 10^5$   & 0.12 & 0.22 & 0.38 \\
      &  $10^6$   & 0.23 & 0.40 & 0.62 \\
 \hline
 \end{tabular}
\caption{Probability that a GC will have at least one enrichment candidate $P_\mathrm{enrich}$ according to the various population models examined in Figure \ref{fig:primordial}. 
The numbers in the table are for three representative sizes and two representative masses for the natal GC. 
In all cases, $R_{\rm enrich}$ is assumed equal to $R_{\rm vir}$. 
The enrichment probabilities are typically $\approx\,2$ times larger for $M_{\rm c} = 10^6\,M_{\odot}$ compared to $M_{\rm c} = 5 \times 10^5\,M_{\odot}$; though the change in $M_{\rm c}$ affects the GC escape velocity, the primary contribution to $f_{\rm enrich}$ comes instead from the increased number of potential DNS progenitors, which scales linearly with $M_{\rm c}$.
}
\label{table}
\end{table}

%As anticipated, we find that enrichment fractions are sensitive to the assumptions made in population modeling. 
%Notably, Model B result in enrichment fractions that are far too low to be compatible with the $r$-process enhancement observed in GCs. 
%Assuming NSMs are the primary mechanism for this enhancement, this provides a unique constraint on the intricacies of binary evolution that lead to DNSs. 
%We discuss implications of each model's enrichment fraction in Section \ref{subsec:primordial_enrichment}. 

For a given population model, enrichment fractions generally increase with increasing cluster mass and cluster size. 
Increasing $R_{\rm vir}$ from $1$ to $10~\mathrm{pc}$ increases the enrichment probability in most models by a factor of $\approx\,2-3$. 
For low to moderate values of $P_{\rm enrich}$, enrichment fractions are typically $\approx$\,2 times larger for $M_{\rm c} = 10^6\ M_{\odot}$ compared to $M_{\rm c} = 5 \times 10^5\ M_{\odot}$: while the change in $M_{\rm c}$ affects the GC escape velocity (and thus $\epsilon_{\rm DNS}$), the primary contribution to $\Lambda_\mathrm{enrich}$ comes instead from the increased number of potential DNS progenitors, which scales linearly with $M_{\rm c}$.
However, as one pushes to extreme assumptions about the initial GC mass, this linear scaling breaks down since many more DNSs remain bound to the natal GC. 
To this end, we also examined the enrichment fractions recovered when assuming an extreme assumption about the initial mass: $M_{\rm c} = 10^7\ M_{\odot}$. 
In this case, regardless of the cluster size, Models A, C, and D all have expected numbers of enriching NSMs much greater than unity ($\Lambda_{\rm enrich} \gg 1$), and therefore enrichment probabilities of $P_{\rm enrich}\,\approx\,100\%$. 
However, for this assumed cluster mass, Model B still only reaches enrichment probabilities of $P_{\rm enrich}\,\approx\,50\%$.

\begin{figure}[t!]\label{fig:Renrich}
\includegraphics[width=0.48\textwidth]{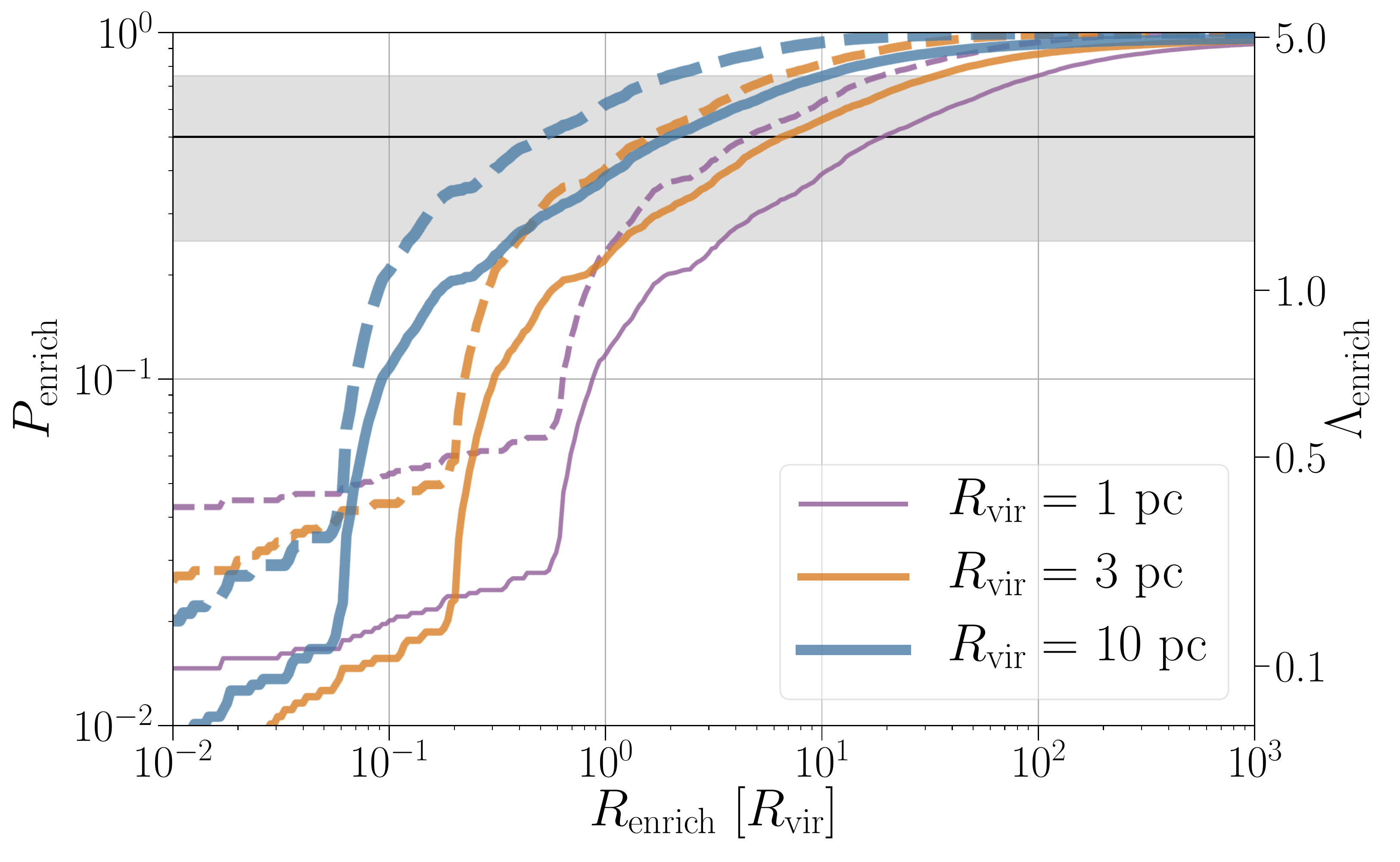}
\caption{Enrichment probability ($P_{\rm enrich}$) and expected number of enriching NSMs ($\Lambda_{\rm enrich}$) as a function of enrichment radius, in units of cluster virial radii, for DNS systems in Model D. 
The solid (dashed) line is for an assumed cluster mass of $5 \times 10^5$ ($10^6$) $M_{\odot}$, and different colors represent different virial radii for the fiducial cluster. 
The black line marks a $50\%$ enrichment probability and the gray shaded region marks enrichment probabilities between $25\%$ and $75\%$, broadly consistent with the enrichment fraction observed using the small sample of GCs that have been investigated for star-to-star $r$-process dispersion \citep[e.g.,][]{Roederer2011a}. }
\end{figure}

The $P_\mathrm{enrich}$ results in Table~\ref{table} assume that an NSM must occur within the virial radius of the young GC in order to enrich the still-forming second generation of stars. 
However, it may be possible to enrich the stellar populations from outside the cluster environment, though this will lead to a geometrical $R^{-2}$ reduction of $r$-process material available for pollution, assuming the ejecta is isotropically dispersed. 
We examine $P_\mathrm{enrich}$ as a function of enrichment radius in Figure \ref{fig:Renrich} for Model D. 
For a GC with an initial mass of $5 \times 10^5\,M_{\odot}$, enrichment probabilities hit $\approx\,50\%$ at $R_{\rm enrich} = 20~R_{\rm vir}$ ($2~R_{\rm vir}$) if $R_{\rm vir} = 1~\mathrm{pc}$ ($10~\mathrm{pc}$). 
For most assumptions about initial GC properties, Model B does not exceed enrichment probabilities of even $20\%$ unless $R_{\rm enrich}\,\gtrsim\,100\,R_{\rm vir}$. 
This indicates that the pessimistic CE model is incompatible with substantial enrichment fractions, regardless of the cluster initial conditions or assumed enrichment radius. 
%Though even these liberal values for the enrichment radii still struggle to match the observed enrichment fraction in GCs, certain changes to our initial assumptions can ameliorate this tension. 
%For example, since $\Lambda_\mathrm{enrich}$ scales linearly with the initial binary fraction, a higher initial binary fraction can act to increase these numbers further. 
%However, Model C is incompatible with observations of enriched GCs regardless of the assumed enrichment radius and binary fraction, indicating that Case BB MT is necessary to explain $r$-process enhanced clusters in the NSM scenario. 

\section{Discussion}\label{sec:discussion}

In the following sections, we highlight the implications of GC enrichment fractions on the physical assumptions of binary stellar evolution, discuss how initial properties and star formation timescales of GCs can be constrained by such observations, touch on other possible mechanisms for $r$-process enrichment in GCs, and comment on implications for other environments that have been observed to be $r$-process enhanced.

\subsection{$r$-process Enrichment from Primordial NSMs in GCs}\label{subsec:primordial_enrichment}

Model A, which allows for successful Case BB CEs, has enrichment probabilities ranging from $\sim\,30\%$ to $90\%$. 
Though the majority of DNS systems have post-SN systemic velocities greater than the escape velocity of the fiducial GCs, the tight orbital separations achieved from successful Case BB CEs still allow for an appreciable number of DNS mergers to occur within the virial radii of the GC. 
This model is consistent with most, if not all, GCs with virial radii $\gtrsim 3\,\mathrm{pc}$ and/or initial masses $\gtrsim 10^6\,M_{\odot}$ showing star-to-star $r$-process dispersion \textit{if} star formation does indeed persist for $\approx\,100$\,Myr. 
Since star-to-star $r$-process dispersion is not a ubiquitous phenomenon in GCs, this model may overpredict NSM enrichment candidates for most assumptions about initial GC properties. 

In Model B, which implements a pessimistic CE scenario not allowing for CEs involving an evolved He-star donor \citep[e.g.,][]{Belczynski2008}, we find few enrichment candidates. 
Regardless of initial GC size and compactness, $P_\mathrm{enrich}$ is consistently $\lesssim\,0.06$; the only enrichment candidates are those that proceed through a single CE and are kicked into highly eccentric orbits from the second SN. 
This shows that \emph{without some form of Case BB MT, NSMs are unable to explain $r$-process enhancement in more than a few percent of GCs}. 

Following \cite{Tauris2015}, in Models C and D we assume that Case BB MT typically proceeds stably (Case BB MT does not lead to a CE phase). 
However, as the mass of the donor star is always greater than that of the first-born NS when considering DNS progenitors, stable MT will still cause the binary orbit to shrink. 
Since the orbital hardening is less drastic than if a CE occurred, we find lower $\Lambda_{\rm enrich}$ compared to Model A by a factor of $\sim\,2$ for Model C and $\sim\,3$ for Model D. 
The reason for lower enrichment levels in Model D compared to Model C is the stifled natal kicks for systems that go through Case BB MT. 
This leads to less scatter in the post-SN orbital properties, a tighter trend in $v_{\rm sys} - t_{\rm insp}$ space, and a slightly lower enrichment probability. 
Both of these models are consistent with many, but not all, GCs having an NSM within its first 100 Myr. 
This demonstrates that in the NSM scenario with extended star formation, GC enrichment can be explained without invoking a Case BB CE phase.

\subsection{Sensitivity to CE Prescriptions for HG He stars}

As shown in Section \ref{sec:primordial}, the vast majority of viable enrichment candidates proceed through a phase of Case BB MT. 
The degree of orbital hardening from this stage of binary evolution strongly affects the post-SN systemic velocities and inspiral times, and thereby has an important impact on the number of NSM enrichment candidates in a GC. 

Since pre-SN orbits of systems that went through Case BB MT are slightly wider for Models C and D than for Model A, mass loss during the SNe in Model A typically leads to larger changes in barycentric velocity. 
In Model A the median post-SN systemic velocity for systems that undergo Case BB MT is $\sim$\,600 km\,s$^{-1}$, whereas Models C and D have median systemic velocities of $\sim$\,400 km\,s$^{-1}$ and $\sim$\,200 km\,s$^{-1}$, respectively. 
Though these systems are less likely to be ejected, as seen in Figure~\ref{fig:primordial}, the majority still exceed the escape velocity of a typical GC. 

Case BB systems in Models C and D have slightly longer inspiral times, which leads to slightly lower enrichment fractions than when Case BB MT proceeds unstably. 
Systems evolving through Case BB CE (Model A) generally lead to DNSs with smaller separations and hence shorter inspiral times than those evolving through stable Case BB MT (Models C, D; in these models $41\%$ and $27\%$ of systems have $t_{\rm insp} < 10^5\,\mathrm{yr}$, respectively, whereas this fraction reaches $73\%$ in Model A). 
This is particularly evident in Model D, where Case BB donors are assumed to be ultra-stripped and have stifled natal kicks. 

In the case of Model B, we impose the pessimistic CE scenario, causing all stars that undergo unstable MT as an evolved He-star to merge. 
This removes Case BB systems and thereby depletes the population of DNSs with short inspiral times, eliminating most viable enrichment candidates. 
With only large-separation DNS progenitors surviving, in this model only systems that get kicked into eccentricities close to unity merge quickly enough to potentially enrich the GC at early times; only $\approx\,0.2\%$ of DNS systems merging within a Hubble time in this model have inspiral times less than 100 Myr and systemic velocities less than 100 km\,s$^{-1}$. 

Overall, significant hardening of the pre-SN binary has a counteractive effect on the enrichment efficiency. 
Though hardened pre-SN orbits typically lead to shorter inspiral times, they also amplify post-SN systemic velocities; even if the natal kicks are small, high post-SN systemic velocities will be achieved through Blauuw kicks \citep{Kalogera1996}. 
This causes hardened systems to be more easily ejected from the GC. 
For example, DNS progenitors with inspiral times of $\lesssim\,100\,\mathrm{Myr}$ will typically require an envelope mass prior to the second SN of $\lesssim\,0.1\,M_{\odot}$ in order to remain bound to the GC. 
However, the hardest DNS progenitors, with pre-SN orbital separations of $\lesssim\,0.2\,R_{\odot}$, will typically be ejected from the GC regardless of the amount the secondary was stripped (see Appendix \ref{AppendixB}). 
Larger pre-SN orbital separations will reduce the effect of the Blaauw kick, but will also increase inspiral times. 
In the context of GCs, enrichment candidates are typically in the first of these regimes: ejected from the GC but residing in the tail of the inspiral time distribution such that they merge before leaving the GC entirely. 
We explore implications of Blauuw kicks on rapid-inspiral DNS progenitors in Appendix \ref{AppendixB}.

% REPEATED!
%First pointed out by \cite{Ivanova2003} and more recently investigated in detail by \cite{Tauris2015}, RLO between a naked He-star donor with a neutron-star companion typically proceeds stably if DNS systems are to form, thereby avoiding a second CE and subsequent merger. 
%In addition, \cite{Vigna-Gomez2018} found that populations where this phase of MT typically proceeds stably are in better agreement with the Milky Way DNS population than models that allow for Case BB CEs. 

\subsection{Initial Cluster Properties}

The GCs with observed $r$-process dispersion have varying properties; both core-collapsed and non-core-collapsed clusters show dispersion, and present-day masses of these clusters range from $\sim\,2 \times 10^5$ to $5 \times 10^5\,M_{\odot}$ \citep{Kimmig2015,Boyles2011,Leonard1992}. 
Since most of the Milky Way's GCs formed $\sim 10$\,Gyr ago, initial GC properties are highly uncertain. 
Observations of super star clusters, which are believed to be the progenitors of GCs, help to elucidate the initial conditions of these old stellar systems. 
These observations indicate that the stellar masses of GCs found in the Milky Way today are a factor of a few times less massive than they were at formation \citep[e.g.,][]{Leroy2018}. 
%in some extreme cases super star clusters can reach masses of $\sim 10^7\,M_{\odot}$ \citep{Herrera2017,Vanzella2017}. 

\begin{figure}[t!]\label{fig:cluster_props}
\includegraphics[width=0.48\textwidth]{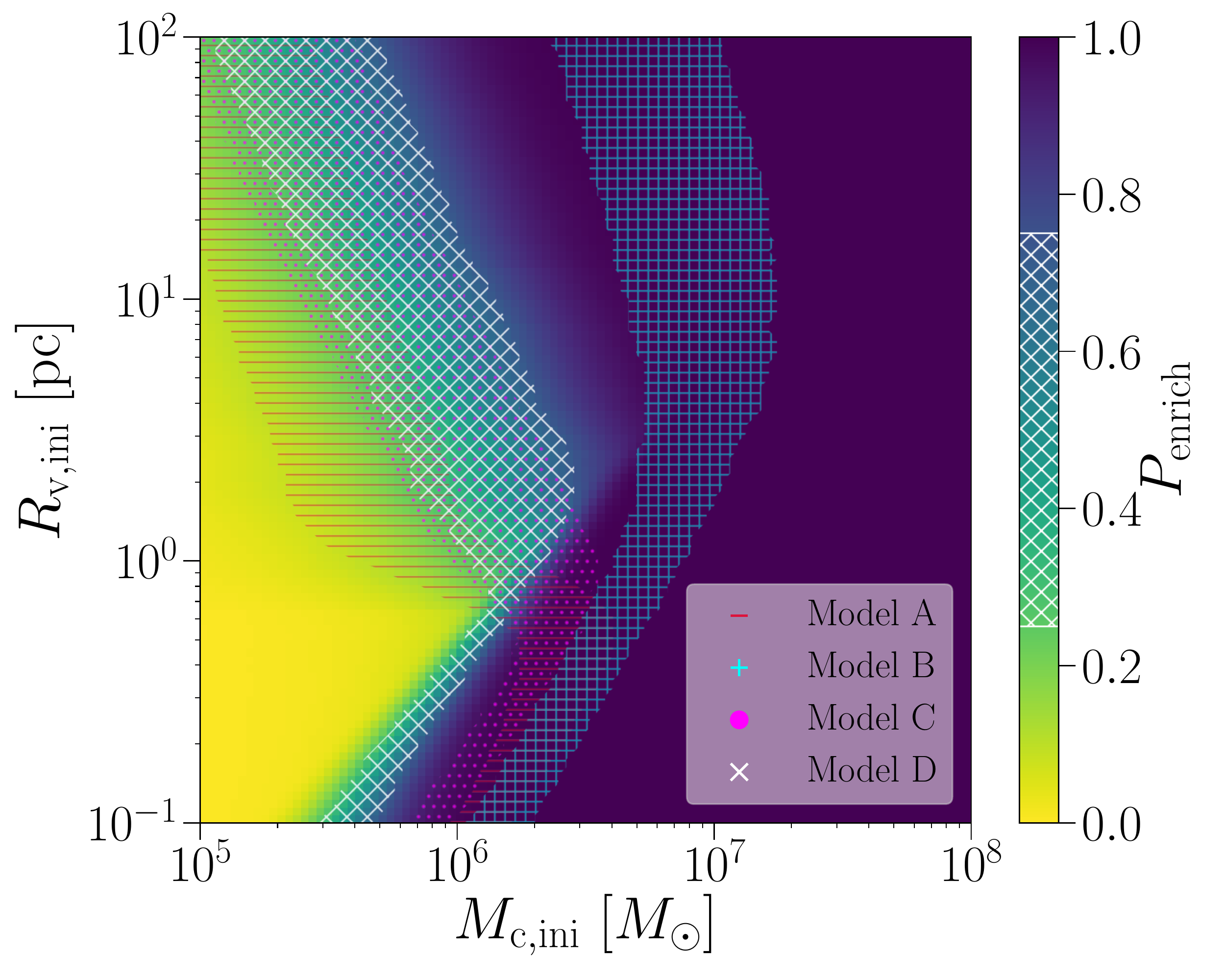}
\caption{Mapping of enrichment probabilities to initial GC properties, assuming a single NSM within the virial radius of the young GC is required for enrichment. 
Population Model D is used for determination of the $P_\mathrm{enrich}$ colormap. 
The various hatched regions indicate the constraints on initial cluster properties for different population models, assuming an enrichment fraction between $25\%$ and $75\%$; the hatched region on the colorbar marks values that correspond to this range in Model D. 
Due to its limited ability to enrich young GCs, Model B pushes to much more extreme initial cluster properties. 
}
\end{figure}

More massive GCs at formation lead to more enrichment candidate NSMs due to deeper gravitational potentials (higher escape velocities) and, more importantly, more stars sampled from the IMF that can potentially form DNS binaries. 
In our analysis, we investigated two representative natal GC masses of $5 \times 10^5$ and $10^6\,M_{\odot}$, working under the assumption that present-day GCs were a few times more massive at birth \citep[e.g.,][]{Webb2015}. 
However, with knowledge of the true population model representing DNSs in young GCs, 
%and that NSMs are the sole contributor to $r$-process enhancement in GCs
the inference can be reversed to place constraints on initial conditions of GCs necessary to match the number of enriched GCs.
Figure \ref{fig:cluster_props} demonstrates this by highlighting the regions of initial cluster mass--initial cluster size space that lead to the enrichment fractions between $25\%$ and $75\%$, broadly consistent with the fraction of GCs with star-to-star $r$-process dispersion in the limited observational samples to date \citep[e.g.,][]{Roederer2011a}. 

Assuming the enrichment event must occur within the virial radius of the cluster and Case BB MT proceeds stably (Models C and D), this maps to typical GCs with initial stellar masses of $\sim 8 \times 10^5$ ($5 \times 10^5$) $M_{\odot}$ given a physical size of $\sim$\,3 pc (10 pc) at formation. 
Though this is a few times more massive than the present-day masses of some of the GCs with observed $r$-process dispersion, simulations of massive star cluster formation do find stellar masses of up to $\approx 1$--$2 \times 10^6\ M_{\odot}$ (\citealt{Tsang2018}, see also \citealt{Skinner2015,Raskutti2016}). 
However, constraints on initial cluster properties should be taken with caution, since theoretical uncertainties in the evolution of DNS progenitors as well as the limited number of observational results concerning the $r$-process elements in GCs currently prohibit any definitive statements. 
If future observations of DNSs in both the Milky Way and via GWs can further constrain uncertain aspects of high-mass binary stellar evolution, the enrichment fraction observed in GCs could help probe these uncertain properties of GC progenitors.

\subsection{Extended Star Formation in Young GCs}\label{subsec:tauSF}

\begin{figure}[b!]\label{fig:tauSF}
\includegraphics[width=0.48\textwidth]{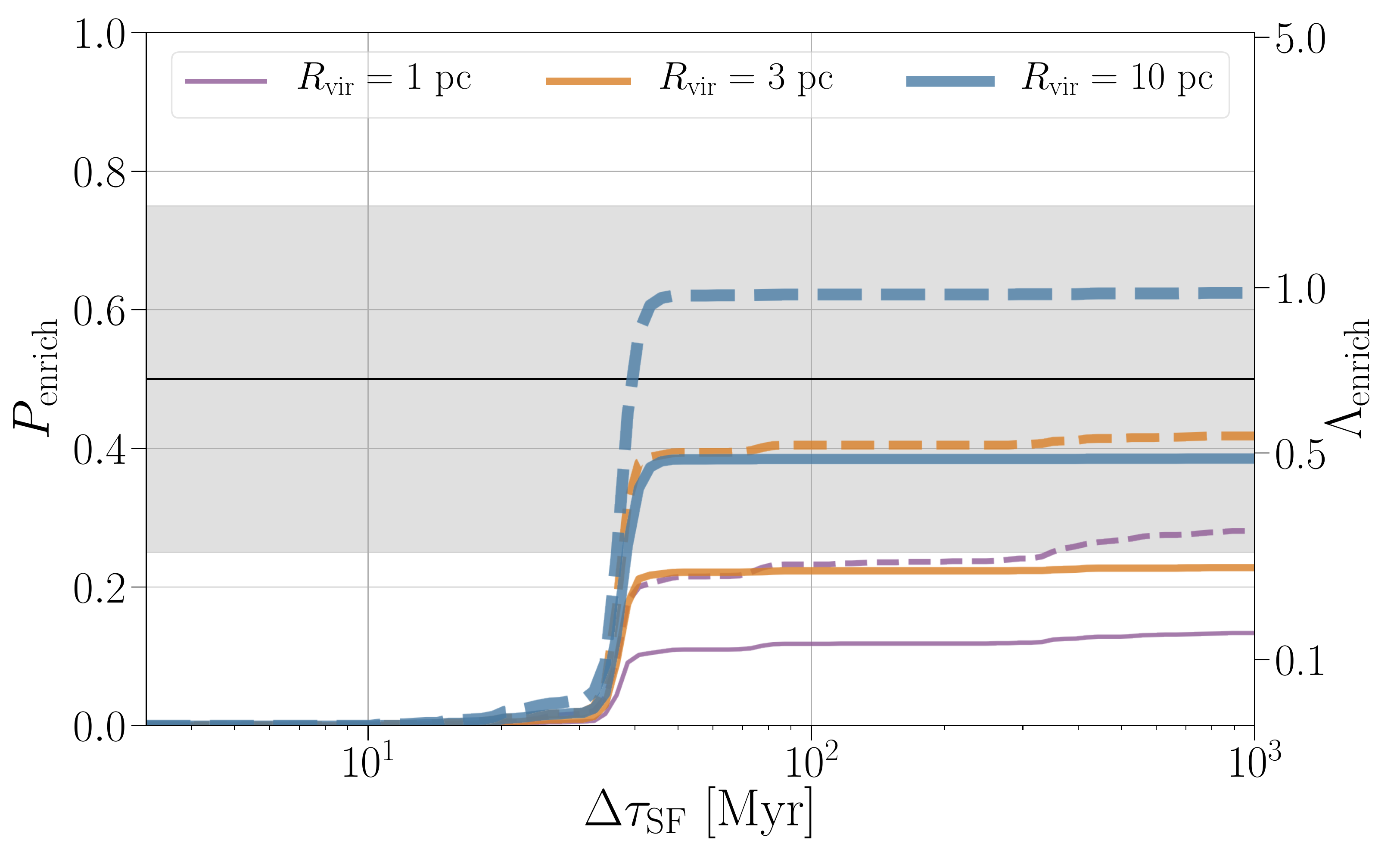}
\caption{Enrichment probability ($P_{\rm enrich}$) and expected number of enriching NSMs ($\Lambda_{\rm enrich}$) as a function of the timescale of star formation in a young GC. 
The solid (dashed) lines are for an assumed cluster mass of $5 \times 10^5$ ($10^6$) $M_{\odot}$, and different colors represent different virial radii for the fiducial cluster. 
For clarity, the black line marks a $50\%$ enrichment probability and the gray shaded region marks enrichment probabilities between $25\%$ and $75\%$, broadly consistent with the enrichment fraction observed using the small sample of GCs that have been investigated for star-to-star $r$-process dispersion \citep[e.g.,][]{Roederer2011a}. 
Model D is used in this figure; though different models lead to different peak values of $P_\mathrm{enrich}$, the steep build-up of $P_\mathrm{enrich}$ occurs at the same point in all models, as this is primarily determined by the stellar lifetime of DNS systems. 
The plateau of $P_\mathrm{enrich}$ at values of $\Delta \tau_{\rm SF} \gtrsim 50~\mathrm{Myr}$ is due to most enrichment candidates ($\gtrsim\,97\%$ for $R_{\rm vir} \geq 3$\,pc) being rapidly merging DNS systems that are unbound from their host cluster. }
\end{figure}

The timespan over which stars form from the natal gas in GCs is uncertain, though typically believed to be $\lesssim\,10$\,Myr \citep{Gratton2012,Bastian2018}. 
Observations of young GCs in the local universe find that they are already gas-free after $\sim\,2-4$\,Myr \citep{Bastian2014,Hollyhead2015}, and have no evidence for extended star-formation episodes \citep{Bastian2013,Cabrera-Ziri2014,Cabrera-Ziri2016,Martocchia2018}. 
The primary drivers for removing natal gas from young GCs are believed to be radiation pressure, photoionization, ram pressure, and feedback from the SNe of high-mass stars, which proceed through the entirety of their stellar evolution in only a few Myr. 
Despite this, mechanisms have been proposed for redistributing gas within young GCs, which needs to be sufficiently dense to halt the relativistic $r$-process ejecta from NSMs and to form new, $r$-processed-enhanced stars. 
For example, \cite{Bekki2017} suggest that this may be accomplished through the conglomeration of AGB ejecta from $6$ to $9\,M_{\odot}$ stars, which can form a high-density, compact gaseous region in the center of a young GC. 
This gas would begin to pollute the cluster $\sim\,100$ Myr after the initial burst of star formation, past the time that gas chemically polluted by SNe of high-mass stars has been removed. 

We take a conservative star formation timescale of $100~\mathrm{Myr}$ for most of the values presented in this paper, which is corroborated by certain observations of young massive star clusters \citep{Vanzella2018}. 
Though the freefall timescale of a $10^6\,M_{\odot}$ molecular cloud about $10~\mathrm{pc}$ in radius is only $\approx\,1~\mathrm{Myr}$, star clusters in general do not form from a single isolated cloud but rather through the hierarchical merging of smaller clusters. 
If clusters form hierarchically in turbulent molecular clouds with freefall times of tens of millions of years, it may be possible for new star formation to occur $\approx\,100~\mathrm{Myr}$ into the cluster evolution as gas clouds continue to stream down the potentials of the newly formed clusters. 

Our results are largely insensitive to increases in $\Delta \tau_\mathrm{SF}$; as seen in Figure \ref{fig:primordial}, few systems fall directly above the horizontal black line at $t_{\rm delay} = 100~\mathrm{Myr}$, so increasing $\Delta\tau_{\rm SF}$ has little impact on $\epsilon_{\rm DNS}$. 
For example, assuming a progenitor GC mass of $10^6\,M_{\odot}$ and virial radius of $3~\mathrm{pc}$, increasing $\Delta \tau_{\rm SF}$ from $100~\mathrm{Myr}$ to $1~\mathrm{Gyr}$ increases the enrichment probabilities by only $0.5\%$, $2.3\%$, and $3.2\%$ in Models A, C, and D, respectively. 
Therefore, cluster formation scenarios that invoke more extended star formation episodes will not lead a to significant amplification of enrichment candidates. 

In contrast, results can change drastically if we decrease $\Delta \tau_{\rm SF}$. 
In Figure \ref{fig:tauSF}, we show the enrichment fraction as a function of $\Delta\tau_{\rm SF}$ for various assumptions about the initial conditions of natal GCs. 
Regardless of GC properties, $P_\mathrm{enrich}$ plateaus at $\approx\, 40~\mathrm{Myr}$, which is the median formation time of DNS systems in our models. 
For example, a value of $\Delta\tau_{\rm SF} = 30$ Myr produces only $\approx\,4$--$7\%$ of the enrichment candidates as $\Delta\tau_{\rm SF} = 50$ Myr across our population model. 
This is because the evolutionary timescales of DNS progenitors typically exceed $\Delta\tau_{\rm SF}$ for $\Delta\tau_{\rm SF} \lesssim$\,30 Myr. 

The steep buildup of $P_\mathrm{enrich}$ at $30-50$ Myr has important implications for \emph{when} the redistribution of gas in the GC must have occurred by. 
Since most DNS mergers that contribute to enrichment are those that are efficiently ejected from the GC, NSMs that are viable enrichment candidates must merge shortly after DNS formation. 
Therefore, for NSMs to be a viable scenario for $r$-process enhancement, star formation in GCs \emph{must} be ongoing or rejuvenated at the time that DNSs form, $\approx$\,40 Myr after the initial burst of star formation. 
This criterion may be hard to accomplish using, for example, lower-mass AGB stars as the mechanism for re-polluting the GC with gas $\approx$\,100 Myr after formation, and may indicate that another mechanism for $r$-process enhancement in young GCs is at play. 
%However, if GCs truly form hierarchically in turbulent %molecular clouds, it is not unreasonable 
%
%dynamical time for the self-gravitating, turbulent molecular clouds can be tens of Myr. So we do expect gas to get collected and stream down the potentials of newly formed clusters as the global gas distribution collapses. 

\subsection{Alternative Scenarios for Enrichment}

As demonstrated in the above sections, the NSM scenario for $r$-process enrichment can explain moderate to high fractions of GCs with enrichment \emph{if} star formation for forming the second generation of stars is ongoing at the time of DNS formation, $\approx$\,30--50\,Myr after the initial bout of star formation. 
Though this may be possible, either through extended star formation or
a redistribution of gas from stellar processes after the first wave
of massive stars heat and eject the natal gas, other scenarios that can contaminate the cluster with $r$-process material at early times may also be viable. 

Various other astrophysical scenarios have been proposed for contributing to the observed abundances of $r$-process elements \citep[see][for a review]{Kajino2019}.
One such scenario is via the formation of a dense accretion disk during the collapse of a massive, rapidly rotating star --- the \emph{collapsar} scenario \citep{Woosley1993,Siegel2019,Siegel2019a}. This scenario has also been suggested to be the cause of long $\gamma$-ray bursts (GRBs; e.g., \citealt{MacFadyen1999}). 
% Daniel: I have commented the following statement, as I doubt the argument by Macias 2019 is correct (as discussed earlier this week). Will look into this more.
%Though this mechanism is expected to be prevalent in low-metallicity environments, recent work by \cite{Macias2019} argues that $r$-process enhanced metal-poor stars require early $r$-process production that is spatially uncorrelated with Fe production, which is difficult to explain using the collapsar scenario but fits the NSM scenario well. 
%On the other hand,
Collapsars arguably better satisfy existing constraints on $r$-process enrichment and may overcome problems or alleviate tensions that enrichment scenarios solely based on NSMs face (see \citealt{Siegel2019a} for a discussion of some of the issues). 
This relates to issues of prompt enrichment at low metallicities (e.g., \citealt{VandeVoort2015,Shen2015,Wehmeyer2015,Cescutti2015,Safarzadeh2019b}), and also to the high-metallicity stars of the Milky Way disk \citep[e.g.,][]{Hotokezaka2018,Siegel2019,Cote2019}. 
Though potential concerns have been raised regarding the co-production of iron in the context of extremely metal-poor stars \citep{Macias2019}, this may not be applicable to $r$-process enrichment in GCs; given the total mass and metallicity of GCs that show internal $r$-process dispersion, many SNe likely occurred in young GCs.
%\cite{Ji2019} found that the lanthanide production from GW170817 is inconsistent with the most highly $r$-process enhanced stars, and if future kilonovae observations are similar to GW170817 then an alternative astrophysical mechanism, such as collapsars, may significantly contribute to $r$-process enrichment in metal-poor stars. 
%\cite{Siegel2019} found that the typical nucleosynthetic yield of $r$-process elements from a single collapsar event is $\sim$10x the yield from a BNS merger, and thus may provide the young GC with more $r$-process material to pollute a second generation of stars. 
%Furthermore, \cite{Cote2017} found that the Milky Way's $r$-process abundance is best explained as a mixture of NSMs and another site, which enriches early in the Universe's history, potentially a rare SN type such as collapsars. 

Unlike enrichment from NSMs, the collapsar scenario is not susceptible to ejection from the GC due to large post-SN systemic velocities. 
Furthermore, as collapsars result from massive star evolution, they could occur within a few million years from the initial burst of star formation, when natal gas is still present in the cluster environment and the first wave of stars are still forming. 
However, for this scenario to be viable, the rate density of collapsars per unit of stellar mass must be in line with the total stellar mass available to the young GC. 

By assuming collapsars are the cause of long GRBs, the number of collapsars per unit stellar mass formed can be empirically estimated from the observed local long GRB rate. 
Assuming the rate of collapsars (events per unit comoving volume per unit time) tracks the star formation rate $\rho_\mathrm{SF}$ (stellar mass per unit comoving volume per unit time) with negligible delay ($\sim$\,few Myr), $R_\mathrm{coll}(t) \propto \rho_\mathrm{SF}(t)$, where $t$ denotes cosmic time. Thus, one can write the total number of collapsars per unit stellar mass formed as
\begin{equation}
	n_\mathrm{coll} = \frac{\int R_\mathrm{coll}(t)\,\mathrm{d}t}{\int \rho_\mathrm{SF}(t)\,\mathrm{d}t} = \frac{R_\mathrm{coll}(z=0)}{\rho_\mathrm{SF}(z=0)}. \label{eq:n_coll}
\end{equation}
The collapsar rate is given by $R_\mathrm{coll}(z=0)= R_\mathrm{LGRB}(z=0)/f_\mathrm{b}$, where $R_{\rm LGRB}(z = 0) \approx 1.3_{-0.7}^{+0.6}$ Gpc$^{-3}$ yr$^{-1}$ is the rate of local ($z=0$) long GRBs \citep{Wanderman2010} and $f_\mathrm{b} \approx 5\times 10^{-3}$ is the GRB beaming fraction \citep{Goldstein2015}. 
This GRB rate does not include the separate class of low-luminosity GRBs \citep{Liang2007}, which likely produce little to no $r$-process material at all \citep{Siegel2019}. 
Employing the cosmic star formation history as reported by \cite{Madau2017}, we obtain from Eq.~\eqref{eq:n_coll}:
\begin{equation}
	n_\mathrm{coll} = 2.6^{+1.2}_{-1.4} \times 10^{-5} \left(\frac{f_\mathrm{b}}{5\times 10^{-3}}\right)^{-1} M_{\odot}^{-1}. \label{eq:n_coll_num}
\end{equation}
This estimate also roughly applies to collapsars in our Milky Way. 

The estimate in Eq.~\eqref{eq:n_coll_num} is unchanged when considering that collapsars and GRBs may only occur up to a certain metallicity threshold. 
Host galaxy studies of long GRBs show that they preferentially occur below a certain stellar metallicity and thus may have shut off in recent Galactic history \citep{Stanek2006,Perley2016}. 
The metallicity threshold for collapsars is slightly sub-solar \citep{Stanek2006}, and is thus not relevant at the metallicities of old, metal-poor GCs considered here. 
Assuming that the estimate in Eq.~\eqref{eq:n_coll_num} also roughly applies to star formation in GCs, this would indicate that, given typical initial GC masses of $5\times10^{5}\,M_{\odot}$--$10^{6}\,M_{\odot}$, most young GCs could have of order unity collapsar events early in their cosmic histories. 
As collapsars tend to produce more $r$-process material per event than NSMs \citep{Siegel2019}, a single collapsar event per GC may be enough to explain the observed internal $r$-process dispersion of GCs (see Sec.~\ref{subsec:enrichment_prob}). 
This simple estimate thus indicates that collapsars may indeed be a viable $r$-process enrichment scenario for GCs. 
Future work, however, is required to provide a better quantitative estimate. 
One open question is whether a collapsar event early in the history of the GC can lead to sufficient inhomogeneity of its $r$-process ejecta into a new generation of stars to explain the observed dispersion in $r$-process elements. While this seems conceivable at first sight, future work is needed to address this question quantitatively.

Finally, magnetohydrodynamic (MHD) SNe \citep{Winteler2012,Thompson2004,Metzger2008} might provide another source of $r$-process enrichment. 
If MHD SNe indeed occur at a rate of $0.1 - 0.3\%$ of CCSNe as assumed by \citet{Wehmeyer2015}, we obtain in analogy to Eq.~\eqref{eq:n_coll} $n_\mathrm{MHDSN}\approx (7\times 10^{-6} - 2\times 10^{-5})\,M_{\odot}^{-1}$, again order unity enrichment events per GC. 
Here, we have used the observed local CCSN rate of $7.05^{+1.43}_{-1.25}\times 10^{-5}\,\mathrm{Mpc}^{-3}\mathrm{yr}^{-1}$ \citep{Li2011}. 
However, when considering the three-dimensional stability of the magnetized jets that give rise to the fast-expanding neutron-rich ejecta in these systems \citep{Moesta2014}, such events are challenged to eject significant amounts of heavy $r$-process nuclei \citep{Moesta2018,Halevi2018}. 
Additionally, if MHD SNe did produce significant amounts of heavy $r$-process elements, the high opacity of the lanthanide material would be mixed with the $^{56}\mathrm{Ni}$ of the supernova ejecta in a way that would likely be incompatible with present observations of CCSNe \citep{Siegel2019}.

%\fixme{Nick Stone also suggested $r$-process from neutron star AIC disks? Spin up NS from donor star...}

\subsection{$r$-process Enrichment in Other Environments}

Enhancement of $r$-process material in metal-poor stars has been observed in various other environments, including UFDGs \citep{Ji2016,Hansen2017} and the Galactic halo \citep{Hansen2018,Sakari2018}. 
Similar to GCs, the enhancement in UFDGs is difficult to explain using the NSM scenario due to their shallow gravitational potential, and depends strongly on SN kicks and distance traveled prior to merger \citep[e.g.,][]{Safarzadeh2017,Andrews2019}. 
Even with stifled natal kick velocities for DNS systems, the extremely shallow gravitational potentials of UFDGs lead to NSMs that are typically at large offsets from their hosts \citep{Bonetti2019}. 

Using the population synthesis models from \cite{Dominik2012}, which use the \texttt{StarTrack} code \citep{Belczynski2002}, \cite{Safarzadeh2018} found that the only way NSMs can explain the enrichment in the UFDG Tucana III is if their progenitors proceed through unstable Case BB MT and form extremely hardened DNSs that merge before their post-SN systemic velocity causes them to evacuate the UFDG. 
However, the modeling in \cite{Safarzadeh2018} still found it difficult to explain the enrichment in Reticulum II from this scenario. 
From our population models, we find less tension in the NSM enrichment scenario for UFDGs; for Tucana III-like and Reticulum II-like initial stellar masses and $r_{\rm vir} \approx$\,4.5 kpc, we find that $\approx$\,32\% and $\approx$\,10\% of Tucana III-like and Reticulum II-like UFDGs would have an NSM enrichment event, respectively. 
This is broadly consistent with the observed number of $r$-processed enhanced UFDGs, which constitutes $\approx$\,20\% of the population \citep{Ji2019a}. 
These numbers are nearly identical whether or not unstable (Model A) or stable (Models C and D) Case BB MT is assumed; the reason that unstable Case BB models produce about double the enrichment candidates than stable Case BB models in GCs is due to the small physical sizes of GCs (and therefore the necessity for extremely short inspiral times to merge before ejection) compared to UFDGs.
This indicates that if the NSM scenario is the main contributor to $r$-process enhancement in UFDGs the onset of Case BB CEs is not necessary, which is in better agreement with studies that find this phase of MT to typically proceed stably \citep{Tauris2015, Vigna-Gomez2018}.

\section{Conclusions}\label{sec:conclusions}

In this study, we examine the efficiency of NSMs at enriching young GCs with $r$-process material. 
The enrichment mechanism must occur relatively early in the history of the GC in order to pollute the second generation of stars. 
We focus on NSMs that are induced from dynamical encounters, and those that are born and evolve as isolated binary pairs in the GC. 
Our main conclusions are as follows.

\begin{enumerate}
    \item Dynamically hardened DNSs merge far too late to contribute to $r$-process enrichment, even when the most liberal assumptions about NS segregation are assumed. 
    
    \item For the primordial binary population to contribute enough NSM enrichment candidates, DNSs \emph{must} be allowed to proceed through a phase of Case BB MT. 
    
    \item The stability of Case BB MT affects the number of enrichment candidates in GCs by about a factor of 2; if Case BB MT typically proceeds stably we find NSM enrichment events in $\sim\,15-60\%$ of GCs, and if it typically proceeds unstably, $\sim\,30-90\%$ of GCs. 
    This is comparable to the fraction of GCs with star-to-star $r$-process dispersion in current observational samples \citep[$\sim$\,50\%,  e.g.,][]{Roederer2011a}.  
    %However, the current observational sample is not large enough to make definitive statements about the exact enrichment fraction in old, metal-poor GCs, and variations in the assumed initial cluster properties, binary fraction, and enrichment radius can further act to reconcile these numbers. 
    
    \item Significant stripping (such that the envelope mass prior to the second SN is $\lesssim\,0.1\,M_{\odot}$) of the donor star during Case BB MT is required for DNS enrichment candidates to remain bound to the GC.
    However, the tightest DNS progenitors, with pre-SN orbital separations of $\lesssim\,0.2\,R_{\odot}$, will typically be ejected from the GC regardless of the amount the donor was stripped. 
    
    \item Since most enrichment candidates are efficiently ejected from the GC and quickly evacuate the cluster environment, for NSMs to be viable enrichment candidates there \emph{must} be dense, star-forming gas in the GC at the formation time of DNSs, $\sim\,30-50$\,Myr after the initial burst of star formation. 

\end{enumerate}

Though the current sample of $r$-process enriched GCs cannot provide a definitive enrichment fraction, future observations may be able to provide a representative sample with well understood biases. 
As we show in this paper, this can act to constrain the late mass-transfer phase of high-mass binaries --- which is essential in the classical evolutionary scenario for the formation of merging DNSs --- as well as the uncertain initial properties of GC progenitors. 
However, if mechanisms for quickly replenishing young GCs with star-forming gas cannot be established, another mechanism for $r$-process enhancement in GCs is needs to be at play. 
The predicted rate density of collapsars is in line with the stellar mass available to a young GC, though a more detailed investigation of this scenario's ability to produce the observed chemical abundances observed in $r$-process enriched, metal-poor GCs is required to asses this channel's viability.

\acknowledgments

M.Z. thanks Jeff Andrews, Pablo Marchant, Mario Spera, Katie Breivik, and Chris Belczynski for useful discussions regarding binary evolution and DNS populations, Steinn Sigurdsson and Fred Rasio for insights into GC initial conditions and multiple populations, Enrico Ramirez-Ruiz for many helpful talks about $r$-process enrichment, Nate Bastian for useful references regarding star formation in young GCs, Claire Ye for discussions on NS populations in GCs, Nick Stone for conversations about other potential $r$-process scenarios, Seth Jacobson for constructive interpretations of $r$-process dispersion, and Chris Pankow for general all-round helpfulness. 
We also thank the anonymous referee for thoughtful and constructive comments on the content and structure of the manuscript. 
M.Z. greatly appreciates financial support from the IDEAS Fellowship, a research traineeship program supported by the National Science Foundation under grant DGE-1450006. 
C.P.L.B. is supported by the CIERA Board of Visitors Research Professorship.
V.K. is supported by a CIFAR G+EU Fellowship and Northwestern University. 
Research at Perimeter Institute is supported in part by the Government of Canada through the Department of Innovation, Science and Economic Development Canada and by the Province of Ontario through the Ministry of Economic Development, Job Creation and Trade.
The majority of our analysis was performed using the computational resources of the Quest high performance computing facility at Northwestern University, which is jointly supported by the Office of the Provost, the Office for Research, and Northwestern University Information Technology. 
This work was performed in part at Aspen Center for Physics, which is supported by National Science Foundation grant PHY-1607611.

%\facility{facility ID}
%\facilities{facility ID, facility ID, facility ID} 
\software{\texttt{Astropy}~\citep{TheAstropyCollaboration2013,TheAstropyCollaboration2018}, \texttt{IPython}~\citep{ipython}, \texttt{matplotlib}~\citep{matplotlib}, \texttt{numpy}~\citep{numpy}, \texttt{pandas}~\citep{pandas}, \texttt{scipy}~\citep{scipy}.}

\bibliographystyle{aasjournal}

\bibliography{library}

\begin{thebibliography}{}
\expandafter\ifx\csname natexlab\endcsname\relax\def\natexlab#1{#1}\fi
\providecommand{\url}[1]{\href{#1}{#1}}

\bibitem[{Abbott {et~al.}(2017{\natexlab{a}})Abbott, Abbott, Abbott, Acernese,
  Ackley, Adams, Adams, Addesso, Adhikari, Adya, Affeldt, Afrough, Agarwal,
  Agathos, Agatsuma, Aggarwal, Aguiar, Aiello, Ain, Ajith, Allen, Allen,
  Allocca, Altin, Amato, Ananyeva, Anderson, Anderson, Angelova, Antier,
  Appert, Arai, Araya, Areeda, Arnaud, Arun, Ascenzi, Ashton, Ast, Aston,
  Astone, Atallah, Aufmuth, Aulbert, AultONeal, Austin, Avila-Alvarez, Babak,
  Bacon, Bader, Bae, Baker, Baldaccini, Ballardin, Ballmer, Banagiri, Barayoga,
  Barclay, Barish, Barker, Barkett, Barone, Barr, Barsotti, Barsuglia, Barta,
  Barthelmy, Bartlett, Bartos, Bassiri, Basti, Batch, Bawaj, Bayley, Bazzan,
  B{\'{e}}csy, Beer, Bejger, Belahcene, Bell, Berger, Bergmann, Bero, Berry,
  Bersanetti, Bertolini, Betzwieser, Bhagwat, Bhandare, Bilenko, Billingsley,
  Billman, Birch, Birney, Birnholtz, Biscans, Biscoveanu, Bisht, Bitossi,
  Biwer, Bizouard, Blackburn, Blackman, Blair, Blair, Blair, Bloemen, Bock,
  Bode, Boer, Bogaert, Bohe, Bondu, Bonilla, Bonnand, Boom, Bork, Boschi, Bose,
  Bossie, Bouffanais, Bozzi, Bradaschia, Brady, Branchesi, Brau, Briant,
  Brillet, Brinkmann, Brisson, Brockill, Broida, Brooks, Brown, Brown, Brunett,
  Buchanan, Buikema, Bulik, Bulten, Buonanno, Buskulic, Buy, Byer, Cabero,
  Cadonati, Cagnoli, Cahillane, Bustillo, Callister, Calloni, Camp, Canepa,
  Canizares, Cannon, Cao, Cao, Capano, Capocasa, Carbognani, Caride, Carney,
  Diaz, Casentini, Caudill, Cavagli{\`{a}}, Cavalier, Cavalieri, Cella, Cepeda,
  Cerd{\'{a}}-Dur{\'{a}}n, Cerretani, Cesarini, Chamberlin, Chan, Chao,
  Charlton, Chase, Chassande-Mottin, Chatterjee, Chatziioannou, Cheeseboro,
  Chen, Chen, Chen, Cheng, Chia, Chincarini, Chiummo, Chmiel, Cho, Cho, Chow,
  Christensen, Chu, Chua, Chua, Chung, Chung, Ciani, Ciolfi, Cirelli, Cirone,
  Clara, Clark, Clearwater, Cleva, Cocchieri, Coccia, Cohadon, Cohen, Colla,
  Collette, Cominsky, Jr., Conti, Cooper, Corban, Corbitt,
  Cordero-Carri{\'{o}}n, Corley, Cornish, Corsi, Cortese, Costa, Coughlin,
  Coughlin, Coulon, Countryman, Couvares, Covas, Cowan, Coward, Cowart, Coyne,
  Coyne, Creighton, Creighton, Cripe, Crowder, Cullen, Cumming, Cunningham,
  Cuoco, Canton, D{\'{a}}lya, Danilishin, D'Antonio, Danzmann, Dasgupta, Costa,
  Dattilo, Dave, Davier, Davis, Daw, Day, De, DeBra, Degallaix, Laurentis,
  Del{\'{e}}glise, Pozzo, Demos, Denker, Dent, Pietri, Dergachev, Rosa, DeRosa,
  Rossi, DeSalvo, de~Varona, Devenson, Dhurandhar, D{\'{i}}az, Fiore, Giovanni,
  Girolamo, Lieto, Pace, Palma, Renzo, Doctor, Dolique, Donovan, Dooley,
  Doravari, Dorrington, Douglas, {\'{A}}lvarez, Downes, Drago, Dreissigacker,
  Driggers, Du, Ducrot, Dupej, Dwyer, Edo, Edwards, Effler, Eggenstein, Ehrens,
  Eichholz, Eikenberry, Eisenstein, Essick, Estevez, Etienne, Etzel, Evans,
  Evans, Factourovich, Fafone, Fair, Fairhurst, Fan, Farinon, Farr, Farr,
  Fauchon-Jones, Favata, Fays, Fee, Fehrmann, Feicht, Fejer, Fernandez-Galiana,
  Ferrante, Ferreira, Ferrini, Fidecaro, Finstad, Fiori, Fiorucci, Fishbach,
  Fisher, Fitz-Axen, Flaminio, Fletcher, Fong, Font, Forsyth, Forsyth,
  Fournier, Frasca, Frasconi, Frei, Freise, Frey, Frey, Fries, Fritschel,
  Frolov, Fulda, Fyffe, Gabbard, Gadre, Gaebel, Gair, Gammaitoni, Ganija,
  Gaonkar, Garcia-Quiros, Garufi, Gateley, Gaudio, Gaur, Gayathri, Gehrels,
  Gemme, Genin, Gennai, George, George, Gergely, Germain, Ghonge, Ghosh, Ghosh,
  Ghosh, Giaime, Giardina, Giazotto, Gill, Glover, Goetz, Goetz, Gomes,
  Goncharov, Gonz{\'{a}}lez, Castro, Gopakumar, Gorodetsky, Gossan, Gosselin,
  Gouaty, Grado, Graef, Granata, Grant, Gras, Gray, Greco, Green, Gretarsson,
  Griswold, Groot, Grote, Grunewald, Gruning, Guidi, Guo, Gupta, Gupta, Gushwa,
  Gustafson, Gustafson, Halim, Hall, Hall, Hamilton, Hammond, Haney, Hanke,
  Hanks, Hanna, Hannam, Hannuksela, Hanson, Hardwick, Harms, Harry, Harry,
  Hart, Haster, Haughian, Healy, Heidmann, Heintze, Heitmann, Hello, Hemming,
  Hendry, Heng, Hennig, Heptonstall, Heurs, Hild, Hinderer, Hoak, Hofman, Holt,
  Holz, Hopkins, Horst, Hough, Houston, Howell, Hreibi, Hu, Huerta, Huet,
  Hughey, Husa, Huttner, Huynh-Dinh, Indik, Inta, Intini, Isa, Isac, Isi, Iyer,
  Izumi, Jacqmin, Jani, Jaranowski, Jawahar, Jim{\'{e}}nez-Forteza, Johnson,
  Jones, Jones, Jonker, Ju, Junker, Kalaghatgi, Kalogera, Kamai, Kandhasamy,
  Kang, Kanner, Kapadia, Karki, Karvinen, Kasprzack, Katolik, Katsavounidis,
  Katzman, Kaufer, Kawabe, K{\'{e}}f{\'{e}}lian, Keitel, Kemball, Kennedy,
  Kent, Key, Khalili, Khan, Khan, Khan, Khazanov, Kijbunchoo, Kim, Kim, Kim,
  Kim, Kim, Kim, Kimbrell, King, King, Kinley-Hanlon, Kirchhoff, Kissel,
  Kleybolte, Klimenko, Knowles, Koch, Koehlenbeck, Koley, Kondrashov, Kontos,
  Korobko, Korth, Kowalska, Kozak, Kr{\"{a}}mer, Kringel, Krishnan,
  Kr{\'{o}}lak, Kuehn, Kumar, Kumar, Kumar, Kuo, Kutynia, Kwang, Lackey, Lai,
  Landry, Lang, Lange, Lantz, Lanza, Larson, Lartaux-Vollard, Lasky, Laxen,
  Lazzarini, Lazzaro, Leaci, Leavey, Lee, Lee, Lee, Lee, Lee, Lehmann, Lenon,
  Leonardi, Leroy, Letendre, Levin, Li, Linker, Littenberg, Liu, Lo, Lockerbie,
  London, Lord, Lorenzini, Loriette, Lormand, Losurdo, Lough, Lousto, Lovelace,
  L{\"{u}}ck, Lumaca, Lundgren, Lynch, Ma, Macas, Macfoy, Machenschalk,
  MacInnis, Macleod, Hernandez, Maga{\~{n}}a-Sandoval, Zertuche, Magee,
  Majorana, Maksimovic, Man, Mandic, Mangano, Mansell, Manske, Mantovani,
  Marchesoni, Marion, M{\'{a}}rka, M{\'{a}}rka, Markakis, Markosyan, Markowitz,
  Maros, Marquina, Marsh, Martelli, Martellini, Martin, Martin, Martynov,
  Mason, Massera, Masserot, Massinger, Masso-Reid, Mastrogiovanni, Matas,
  Matichard, Matone, Mavalvala, Mazumder, McCarthy, McClelland, McCormick,
  McCuller, McGuire, McIntyre, McIver, McManus, McNeill, McRae, McWilliams,
  Meacher, Meadors, Mehmet, Meidam, Mejuto-Villa, Melatos, Mendell, Mercer,
  Merilh, Merzougui, Meshkov, Messenger, Messick, Metzdorff, Meyers, Miao,
  Michel, Middleton, Mikhailov, Milano, Miller, Miller, Miller, Millhouse,
  Milovich-Goff, Minazzoli, Minenkov, Ming, Mishra, Mitra, Mitrofanov,
  Mitselmakher, Mittleman, Moffa, Moggi, Mogushi, Mohan, Mohapatra, Montani,
  Moore, Moraru, Moreno, Morriss, Mours, Mow-Lowry, Mueller, Muir, Mukherjee,
  Mukherjee, Mukherjee, Mukund, Mullavey, Munch, Mu{\~{n}}iz, Muratore, Murray,
  Napier, Nardecchia, Naticchioni, Nayak, Neilson, Nelemans, Nelson, Nery,
  Neunzert, Nevin, Newport, Newton, Ng, Nguyen, Nguyen, Nichols, Nielsen,
  Nissanke, Nitz, Noack, Nocera, Nolting, North, Nuttall, Oberling, O'Dea,
  Ogin, Oh, Oh, Ohme, Okada, Oliver, Oppermann, Oram, O'Reilly, Ormiston,
  Ortega, O'Shaughnessy, Ossokine, Ottaway, Overmier, Owen, Pace, Page, Page,
  Pai, Pai, Palamos, Palashov, Palomba, Pal-Singh, Pan, Pan, Pang, Pang,
  Pankow, Pannarale, Pant, Paoletti, Paoli, Papa, Parida, Parker, Pascucci,
  Pasqualetti, Passaquieti, Passuello, Patil, Patricelli, Pearlstone, Pedraza,
  Pedurand, Pekowsky, Pele, Penn, Perez, Perreca, Perri, Pfeiffer, Phelps,
  Piccinni, Pichot, Piergiovanni, Pierro, Pillant, Pinard, Pinto, Pirello,
  Pitkin, Poe, Poggiani, Popolizio, Porter, Post, Powell, Prasad, Pratt,
  Pratten, Predoi, Prestegard, Price, Prijatelj, Principe, Privitera, Prodi,
  Prokhorov, Puncken, Punturo, Puppo, P{\"{u}}rrer, Qi, Quetschke, Quintero,
  Quitzow-James, Raab, Rabeling, Radkins, Raffai, Raja, Rajan, Rajbhandari,
  Rakhmanov, Ramirez, Ramos-Buades, Rapagnani, Raymond, Razzano, Read,
  Regimbau, Rei, Reid, Reitze, Ren, Reyes, Ricci, Ricker, Rieger, Riles, Rizzo,
  Robertson, Robie, Robinet, Rocchi, Rolland, Rollins, Roma, Romano, Romel,
  Romie, Rosi{\'{n}}ska, Ross, Rowan, R{\"{u}}diger, Ruggi, Rutins, Ryan,
  Sachdev, Sadecki, Sadeghian, Sakellariadou, Salconi, Saleem, Salemi,
  Samajdar, Sammut, Sampson, Sanchez, Sanchez, Sanchis-Gual, Sandberg, Sanders,
  Sassolas, Sathyaprakash, Saulson, Sauter, Savage, Sawadsky, Schale, Scheel,
  Scheuer, Schmidt, Schmidt, Schnabel, Schofield, Sch{\"{o}}nbeck, Schreiber,
  Schuette, Schulte, Schutz, Schwalbe, Scott, Scott, Seidel, Sellers, Sengupta,
  Sentenac, Sequino, Sergeev, Shaddock, Shaffer, Shah, Shahriar, Shaner, Shao,
  Shapiro, Shawhan, Sheperd, Shoemaker, Shoemaker, Siellez, Siemens,
  Sieniawska, Sigg, Silva, Singer, Singh, Singhal, Sintes, Slagmolen, Smith,
  Smith, Smith, Somala, Son, Sonnenberg, Sorazu, Sorrentino, Souradeep,
  Spencer, Srivastava, Staats, Staley, Steinke, Steinlechner, Steinlechner,
  Steinmeyer, Stevenson, Stone, Stops, Strain, Stratta, Strigin, Strunk,
  Sturani, Stuver, Summerscales, Sun, Sunil, Suresh, Sutton, Swinkels,
  Szczepa{\'{n}}czyk, Tacca, Tait, Talbot, Talukder, Tanner, T{\'{a}}pai,
  Taracchini, Tasson, Taylor, Taylor, Tewari, Theeg, Thies, Thomas, Thomas,
  Thomas, Thorne, Thorne, Thrane, Tiwari, Tiwari, Tokmakov, Toland, Tonelli,
  Tornasi, Torres-Forn{\'{e}}, Torrie, T{\"{o}}yr{\"{a}}, Travasso, Traylor,
  Trinastic, Tringali, Trozzo, Tsang, Tse, Tso, Tsukada, Tsuna, Tuyenbayev,
  Ueno, Ugolini, Unnikrishnan, Urban, Usman, Vahlbruch, Vajente, Valdes, van
  Bakel, van Beuzekom, van~den Brand, Broeck, Vander-Hyde, van~der Schaaf, van
  Heijningen, van Veggel, Vardaro, Varma, Vass, Vas{\'{u}}th, Vecchio,
  Vedovato, Veitch, Veitch, Venkateswara, Venugopalan, Verkindt, Vetrano,
  Vicer{\'{e}}, Viets, Vinciguerra, Vine, Vinet, Vitale, Vo, Vocca, Vorvick,
  Vyatchanin, Wade, Wade, Wade, Walet, Walker, Wallace, Walsh, Wang, Wang,
  Wang, Wang, Wang, Ward, Warner, Was, Watchi, Weaver, Wei, Weinert, Weinstein,
  Weiss, Wen, Wessel, Wessels, Westerweck, Westphal, Wette, Whelan, Whitcomb,
  Whiting, Whittle, Wilken, Williams, Williams, Williamson, Willis, Willke,
  Wimmer, Winkler, Wipf, Wittel, Woan, Woehler, Wofford, Wong, Worden, Wright,
  Wu, Wysocki, Xiao, Yamamoto, Yancey, Yang, Yap, Yazback, Yu, Yu, Yvert,
  Zadro{\.{z}}ny, Zanolin, Zelenova, Zendri, Zevin, Zhang, Zhang, Zhang, Zhang,
  Zhao, Zhou, Zhou, Zhu, Zhu, Zimmerman, Zucker, Zweizig, Collaboration,
  Collaboration, Wilson-Hodge, Bissaldi, Blackburn, Briggs, Burns, Cleveland,
  Connaughton, Gibby, Giles, Goldstein, Hamburg, Jenke, Hui, Kippen, Kocevski,
  McBreen, Meegan, Paciesas, Poolakkil, Preece, Racusin, Roberts, Stanbro,
  Veres, von Kienlin, GBM, Savchenko, Ferrigno, Kuulkers, Bazzano, Bozzo,
  Brandt, Chenevez, Courvoisier, Diehl, Domingo, Hanlon, Jourdain, Laurent,
  Lebrun, Lutovinov, Martin-Carrillo, Mereghetti, Rodi, Roques, Sunyaev,
  Ubertini, INTEGRAL, Aartsen, Ackermann, Adams, Aguilar, Ahlers, Ahrens,
  Samarai, Altmann, Andeen, Anderson, Ansseau, Anton, Arg{\"{u}}elles,
  Auffenberg, Axani, Bagherpour, Bai, Barron, Barwick, Baum, Bay, Beatty, Tjus,
  Bernardini, Besson, Binder, Bindig, Blaufuss, Blot, Bohm, B{\"{o}}rner, Bos,
  Bose, B{\"{o}}ser, Botner, Bourbeau, Bourbeau, Bradascio, Braun, Brayeur,
  Brenzke, Bretz, Bron, Brostean-Kaiser, Burgman, Carver, Casey, Casier,
  Cheung, Chirkin, Christov, Clark, Classen, Coenders, Collin, Conrad, Cowen,
  Cross, Day, de~Andr{\'{e}}, Clercq, DeLaunay, Dembinski, Ridder, Desiati,
  de~Vries, de~Wasseige, de~With, DeYoung, D{\'{i}}az-V{\'{e}}lez, di~Lorenzo,
  Dujmovic, Dumm, Dunkman, Dvorak, Eberhardt, Ehrhardt, Eichmann, Eller,
  Evenson, Fahey, Fazely, Felde, Filimonov, Finley, Flis, Franckowiak,
  Friedman, Fuchs, Gaisser, Gallagher, Gerhardt, Ghorbani, Giang, Glauch,
  Gl{\"{u}}senkamp, Goldschmidt, Gonzalez, Grant, Griffith, Haack, Hallgren,
  Halzen, Hanson, Hebecker, Heereman, Helbing, Hellauer, Hickford, Hignight,
  Hill, Hoffman, Hoffmann, Hokanson-Fasig, Hoshina, Huang, Huber, Hultqvist,
  H{\"{u}}nnefeld, In, Ishihara, Jacobi, Japaridze, Jeong, Jero, Jones,
  Kalaczynski, Kang, Kappes, Karg, Karle, Keivani, Kelley, Kheirandish, Kim,
  Kim, Kintscher, Kiryluk, Kittler, Klein, Kohnen, Koirala, Kolanoski,
  K{\"{o}}pke, Kopper, Kopper, Koschinsky, Koskinen, Kowalski, Krings, Kroll,
  Kr{\"{u}}ckl, Kunnen, Kunwar, Kurahashi, Kuwabara, Kyriacou, Labare,
  Lanfranchi, Larson, Lauber, Lesiak-Bzdak, Leuermann, Liu, Lu, L{\"{u}}nemann,
  Luszczak, Madsen, Maggi, Mahn, Mancina, Maruyama, Mase, Maunu, McNally,
  Meagher, Medici, Meier, Menne, Merino, Meures, Miarecki, Micallef,
  Moment{\'{e}}, Montaruli, Moore, Moulai, Nahnhauer, Nakarmi, Naumann, Neer,
  Niederhausen, Nowicki, Nygren, Pollmann, Olivas, O'Murchadha, Palczewski,
  Pandya, Pankova, Peiffer, Pepper, de~los Heros, Pieloth, Pinat, Price,
  Przybylski, Raab, R{\"{a}}del, Rameez, Rawlins, Rea, Reimann, Relethford,
  Relich, Resconi, Rhode, Richman, Robertson, Rongen, Rott, Ruhe, Ryckbosch,
  Rysewyk, S{\"{a}}lzer, Herrera, Sandrock, Sandroos, Santander, Sarkar,
  Sarkar, Satalecka, Schlunder, Schmidt, Schneider, Schoenen, Sch{\"{o}}neberg,
  Schumacher, Seckel, Seunarine, Soedingrekso, Soldin, Song, Spiczak, Spiering,
  Stachurska, Stamatikos, Stanev, Stasik, Stettner, Steuer, Stezelberger,
  Stokstad, St{\"{o}}ssl, Strotjohann, Stuttard, Sullivan, Sutherland, Taboada,
  Tatar, Tenholt, Ter-Antonyan, Terliuk, Tilav, Toale, Tobin, Toscano, Tosi,
  Tselengidou, Tung, Turcati, Turley, Ty, Unger, Usner, Vandenbroucke,
  Driessche, van Eijndhoven, Vanheule, van Santen, Vehring, Vogel, Vraeghe,
  Walck, Wallace, Wallraff, Wandler, Wandkowsky, Waza, Weaver, Weiss, Wendt,
  Werthebach, Whelan, Wiebe, Wiebusch, Wille, Williams, Wills, Wolf, Wood,
  Woolsey, Woschnagg, Xu, Xu, Xu, Yanez, Yodh, Yoshida, Yuan, Zoll,
  Collaboration, Balasubramanian, Mate, Bhalerao, Bhattacharya, Vibhute,
  Dewangan, Rao, Vadawale, Team, Svinkin, Hurley, Aptekar, Frederiks,
  Golenetskii, Kozlova, Lysenko, Oleynik, Tsvetkova, Ulanov, Cline,
  Collaboration, Li, Xiong, Zhang, Lu, Song, Cao, Chang, Chen, Chen, Chen,
  Chen, Chen, Chen, Cui, Cui, Deng, Dong, Du, Fu, Gao, Gao, Gao, Ge, Gu, Guan,
  Guo, Han, Hu, Huang, Huo, Jia, Jiang, Jiang, Jin, Jin, Li, Li, Li, Li, Li,
  Li, Li, Li, Li, Li, Li, Liang, Liao, Liu, Liu, Liu, Liu, Liu, Liu, Liu, Lu,
  Lu, Luo, Ma, Meng, Nang, Nie, Ou, Qu, Sai, Sun, Tan, Tao, Tao, Tuo, Wang,
  Wang, Wang, Wang, Wang, Wen, Wu, Wu, Xiao, Xu, Xu, Yan, Yang, Yang, Yang,
  Zhang, Zhang, Zhang, Zhang, Zhang, Zhang, Zhang, Zhang, Zhang, Zhang, Zhang,
  Zhang, Zhang, Zhang, Zhang, Zhang, Zhang, Zhang, Zhao, Zhao, Zhao, Zheng,
  Zhu, Zhu, Zou, Collaboration, Albert, Andr{\'{e}}, Anghinolfi, Ardid, Aubert,
  Aublin, Avgitas, Baret, Barrios-Mart{\'{i}}, Basa, Belhorma, Bertin, Biagi,
  Bormuth, Bourret, Bouwhuis, Br{\^{a}}nzaş, Bruijn, Brunner, Busto, Capone,
  Caramete, Carr, Celli, Chiarusi, Circella, Coelho, Coleiro, Coniglione,
  Costantini, Coyle, Creusot, D{\'{i}}az, Deschamps, Bonis, Distefano, Palma,
  Domi, Donzaud, Dornic, Drouhin, Eberl, Els{\"{a}}sser, Enzenh{\"{o}}fer,
  Ettahiri, Fassi, Felis, Fusco, Gay, Giordano, Glotin, Gr{\'{e}}goire, Ruiz,
  Graf, Hallmann, van Haren, Heijboer, Hello, Hern{\'{a}}ndez-Rey, H{\"{o}}ssl,
  Hofest{\"{a}}dt, Hugon, Illuminati, James, de~Jong, Jongen, Kadler, Kalekin,
  Katz, Kiessling, Kouchner, Kreter, Kreykenbohm, Kulikovskiy, Lachaud,
  Lahmann, Lef{\`{e}}vre, Leonora, Lotze, Loucatos, Marcelin, Margiotta,
  Marinelli, Mart{\'{i}}nez-Mora, Mele, Melis, Michael, Migliozzi, Moussa,
  Navas, Nezri, Organokov, Pellegrino, Perrina, Piattelli, Popa, Pradier,
  Quinn, Racca, Riccobene, S{\'{a}}nchez-Losa, Salda{\~{n}}a, Salvadori,
  Samtleben, Sanguineti, Sapienza, Sieger, Spurio, Stolarczyk, Taiuti,
  Tayalati, Trovato, Turpin, T{\"{o}}nnis, Vallage, Elewyck, Versari, Vivolo,
  Vizzoca, Wilms, Zornoza, Z{\'{u}}{\~{n}}iga, Collaboration, Beardmore,
  Breeveld, Burrows, Cenko, Cusumano, D'A{\`{i}}, de~Pasquale, Emery, Evans,
  Giommi, Gronwall, Kennea, Krimm, Kuin, Lien, Marshall, Melandri, Nousek,
  Oates, Osborne, Pagani, Page, Palmer, Perri, Siegel, Sbarufatti, Tagliaferri,
  Tohuvavohu, Collaboration, Tavani, Verrecchia, Bulgarelli, Evangelista,
  Pacciani, Feroci, Pittori, Giuliani, Monte, Donnarumma, Argan, Trois, Ursi,
  Cardillo, Piano, Longo, Lucarelli, Fuschino, Labanti, Marisaldi, Minervini,
  Fioretti, Parmiggiani, Gianotti, Trifoglio, Persio, Antonelli, Barbiellini,
  Caraveo, Cattaneo, Costa, Colafrancesco, D'Amico, Ferrari, Morselli,
  Paoletti, Picozza, Pilia, Rappoldi, Soffitta, Vercellone, Team, Foley,
  Coulter, Kilpatrick, Drout, Piro, Shappee, Siebert, Simon, Ulloa, Kasen,
  Madore, Murguia-Berthier, Pan, Prochaska, Ramirez-Ruiz, Rest, Rojas-Bravo,
  Team, Berger, Soares-Santos, Annis, Alexander, Allam, Balbinot, Blanchard,
  Brout, Butler, Chornock, Cook, Cowperthwaite, Diehl, Drlica-Wagner, Drout,
  Durret, Eftekhari, Finley, Fong, Frieman, Fryer, Garc{\'{i}}a-Bellido,
  Gruendl, Hartley, Herner, Kessler, Lin, Lopes, Louren{\c{c}}o, Margutti,
  Marshall, Matheson, Metzger, Muir, Nicholl, Nugent, Palmese,
  Paz-Chinch{\'{o}}n, Quataert, Sako, Sauseda, Schlegel, Scolnic, Smith,
  Sobreira, Villar, Vivas, Wester, Williams, Yanny, Zenteno, Zhang, Abbott,
  Banerji, Bechtol, Benoit-L{\'{e}}vy, Bertin, Brooks, Buckley-Geer, Burke,
  Capozzi, Rosell, Kind, Castander, Crocce, Cunha, D'Andrea, Davis, DePoy,
  Desai, Dietrich, Eifler, Fernandez, Flaugher, Fosalba, Gaztanaga, Gerdes,
  Giannantonio, Goldstein, Gruen, Gschwend, Gutierrez, Honscheid, James,
  Jeltema, Johnson, Johnson, Kent, Krause, Kron, Kuehn, Lahav, Lima, Maia,
  March, Martini, McMahon, Menanteau, Miller, Miquel, Mohr, Nichol, Ogando,
  Plazas, Romer, Roodman, Rykoff, Sanchez, Scarpine, Schindler, Schubnell,
  Sevilla-Noarbe, Sheldon, Smith, Smith, Stebbins, Suchyta, Swanson, Tarle,
  Thomas, Troxel, Tucker, Vikram, Walker, Wechsler, Weller, Carlin, Gill, Li,
  Marriner, Neilsen, Collaboration, Collaboration, Haislip, Kouprianov,
  Reichart, Sand, Tartaglia, Valenti, Yang, Collaboration, Benetti, Brocato,
  Campana, Cappellaro, Covino, D'Avanzo, D'Elia, Getman, Ghirlanda, Ghisellini,
  Limatola, Nicastro, Palazzi, Pian, Piranomonte, Possenti, Rossi, Salafia,
  Tomasella, Amati, Antonelli, Bernardini, Bufano, Capaccioli, Casella, Dadina,
  Cesare, Paola, Giuffrida, Giunta, Israel, Lisi, Maiorano, Mapelli, Masetti,
  Pescalli, Pulone, Salvaterra, Schipani, Spera, Stamerra, Stella, Testa,
  Turatto, Vergani, Aresu, Bachetti, Buffa, Burgay, Buttu, Caria, Carretti,
  Casasola, Castangia, Carboni, Casu, Concu, Corongiu, Deiana, Egron, Fara,
  Gaudiomonte, Gusai, Ladu, Loru, Leurini, Marongiu, Melis, Melis, Orlati,
  Ortu, Palmas, Pellizzoni, Perrodin, Pisanu, Poppi, Righini, Saba, Serra,
  Serrau, Stagni, Surcis, Vacca, Vargiu, Hunt, Jin, Klose, Kouveliotou,
  Mazzali, M{\`{e}}ller, Nava, Piran, Selsing, Vergani, Wiersema, Toma,
  Higgins, Mundell, Alighieri, G{\'{o}}tz, Gao, Gomboc, Kaper, Kobayashi,
  Kopac, Mao, Starling, Steele, van~der Horst, TeAm, Acero, Atwood, Baldini,
  Barbiellini, Bastieri, Berenji, Bellazzini, Bissaldi, Blandford, Bloom,
  Bonino, Bottacini, Bregeon, Buehler, Buson, Cameron, Caputo, Caraveo,
  Cavazzuti, Chekhtman, Cheung, Chiang, Ciprini, Cohen-Tanugi, Cominsky,
  Costantin, Cuoco, D'Ammando, de~Palma, Digel, Lalla, Mauro, Venere, Dubois,
  Fegan, Focke, Franckowiak, Fukazawa, Funk, Fusco, Gargano, Gasparrini,
  Giglietto, Giordano, Giroletti, Glanzman, Green, Grondin, Guillemot, Guiriec,
  Harding, Horan, J{\'{o}}hannesson, Kamae, Kensei, Kuss, Mura, Latronico,
  Lemoine-Goumard, Longo, Loparco, Lovellette, Lubrano, Magill, Maldera,
  Manfreda, Mazziotta, McEnery, Meyer, Michelson, Mirabal, Monzani, Morselli,
  Moskalenko, Negro, Nuss, Ojha, Omodei, Orienti, Orlando, Palatiello, Paliya,
  Paneque, Pesce-Rollins, Piron, Porter, Principe, Rain{\`{o}}, Rando, Razzano,
  Razzaque, Reimer, Reimer, Reposeur, Rochester, Parkinson, Sgr{\`{o}},
  Siskind, Spada, Spandre, Suson, Takahashi, Tanaka, Thayer, Thayer, Thompson,
  Tibaldo, Torres, Torresi, Troja, Venters, Vianello, Zaharijas, Collaboration,
  Allison, Bannister, Dobie, Kaplan, Lenc, Lynch, Murphy, Sadler, Array, Hotan,
  James, Oslowski, Raja, Shannon, Whiting, Pathfinder, Arcavi, Howell, McCully,
  Hosseinzadeh, Hiramatsu, Poznanski, Barnes, Zaltzman, Vasylyev, Maoz, Group,
  Cooke, Bailes, Wolf, Deller, Lidman, Wang, Gendre, Andreoni, Ackley,
  Pritchard, Bessell, Chang, M{\"{o}}ller, Onken, Scalzo, Ridden-Harper, Sharp,
  Tucker, Farrell, Elmer, Johnston, Krishnan, Keane, Green, Jameson, Hu, Ma,
  Sun, Wu, Wang, Shang, Hu, Ashley, Yuan, Li, Tao, Zhu, Zhang, Suntzeff, Zhou,
  Yang, Orange, Morris, Cucchiara, Giblin, Klotz, Staff, Thierry, Schmidt,
  OzGrav, {DWF (Deeper, Wider}, AST3, CAASTRO, Tanvir, Levan, Cano, Evans,
  Gonz{\'{a}}lez-Fern{\'{a}}ndez, Greiner, Hjorth, Irwin, Kr{\"{u}}hler,
  Mandel, Milvang-Jensen, O'Brien, Rol, Rosetti, Rosswog, Rowlinson, Steeghs,
  Th{\"{o}}ne, Ulaczyk, Watson, Bruun, Cutter, Fujii, Fruchter, Gompertz,
  Jakobsson, Hodosan, J{\`{e}}rgensen, Kangas, Kann, Rabus, Schr{\`{e}}der,
  Stanway, Wijers, Collaboration, Lipunov, Gorbovskoy, Kornilov, Tyurina,
  Balanutsa, Kuznetsov, Vlasenko, Podesta, Lopez, Podesta, Levato, Saffe,
  Mallamaci, Budnev, Gress, Kuvshinov, Gorbunov, Vladimirov, Zimnukhov,
  Gabovich, Yurkov, Sergienko, Rebolo, Serra-Ricart, Tlatov, Ishmuhametova,
  Collaboration, Abe, Aoki, Aoki, Asakura, Baar, Barway, Bond, Doi, Finet,
  Fujiyoshi, Furusawa, Honda, Itoh, Kanda, Kawabata, Kawabata, Kim, Koshida,
  Kuroda, Lee, Liu, Matsubayashi, Miyazaki, Morihana, Morokuma, Motohara,
  Murata, Nagai, Nagashima, Nagayama, Nakaoka, Nakata, Ohsawa, Ohshima, Ohta,
  Okita, Saito, Saito, Sako, Sekiguchi, Sumi, Tajitsu, Takahashi, Takayama,
  Tamura, Tanaka, Tanaka, Tanaka, Terai, Tominaga, Tristram, Uemura, Utsumi,
  Yamaguchi, Yasuda, Yoshida, Zenko, J-GEM, Adams, Allison, Anupama, Bally,
  Barway, Bellm, Blagorodnova, Cannella, Chandra, Chatterjee, Clarke, Cobb,
  Cook, Copperwheat, De, Emery, Evans, Feindt, Foster, Fox, Frail, Fremling,
  Frohmaier, Garcia, Ghosh, Giacintucci, Goobar, Gottlieb, Grefenstette,
  Hallinan, Harrison, Heida, Helou, Ho, Horesh, Hotokezaka, Ip, Itoh, Jacobs,
  Jencson, Kasen, Kasliwal, Kassim, Kim, Kiran, Kuin, Kulkarni, Kupfer, Lau,
  Madsen, Mazzali, Miller, Miyasaka, Mooley, Myers, Nakar, Ngeow, Nugent, Ofek,
  Palliyaguru, Pavana, Perley, Peters, Pike, Piran, Qi, Quimby, Rana, Rosswog,
  Rusu, Sadler, Sistine, Sollerman, Xu, Yan, Yatsu, Yu, Zhang, Zhao, GROWTH,
  JAGWAR, Caltech-NRAO, TTU-NRAO, NuSTAR, Chambers, Huber, Schultz, Bulger,
  Flewelling, Magnier, Lowe, Wainscoat, Waters, Willman, Pan-STARRS, Ebisawa,
  Hanyu, Harita, Hashimoto, Hidaka, Hori, Ishikawa, Isobe, Iwakiri, Kawai,
  Kawai, Kawamuro, Kawase, Kitaoka, Makishima, Matsuoka, Mihara, Morita,
  Morita, Nakahira, Nakajima, Nakamura, Negoro, Oda, Sakamaki, Sasaki, Serino,
  Shidatsu, Shimomukai, Sugawara, Sugita, Sugizaki, Tachibana, Takao, Tanimoto,
  Tomida, Tsuboi, Tsunemi, Ueda, Ueno, Yamada, Yamaoka, Yamauchi, Yatabe,
  Yoneyama, Yoshii, Team, Coward, Crisp, Macpherson, Andreoni, Laugier,
  Noysena, Klotz, Gendre, Thierry, Turpin, Consortium, Im, Choi, Kim, Yoon,
  Lim, Lee, Lee, Kim, Ko, Joe, Kwon, Kim, Lim, Choi, Collaboration, Fynbo,
  Malesani, Xu, Telescope, Smartt, Jerkstrand, Kankare, Sim, Fraser, Inserra,
  Maguire, Leloudas, Magee, Shingles, Smith, Young, Kotak, Gal-Yam, Lyman,
  Homan, Agliozzo, Anderson, Angus, Ashall, Barbarino, Bauer, Berton,
  Botticella, Bulla, Cannizzaro, Cartier, Cikota, Clark, Dennefeld, Dessart,
  Dimitriadis, Elias-Rosa, Firth, Fl{\"{o}}rs, Frohmaier, Galbany,
  Gonz{\'{a}}lez-Gait{\'{a}}n, Gromadzki, Guti{\'{e}}rrez, Hamanowicz,
  Harmanen, Heintz, Hernandez, Hodgkin, Hook, Izzo, James, Jonker, Kerzendorf,
  Kostrzewa-Rutkowska, Kromer, Kuncarayakti, Lawrence, Manulis, Mattila,
  McBrien, M{\"{u}}ller, Nordin, O'Neill, Onori, Palmerio, Pastorello, Patat,
  Pignata, Podsiadlowski, Razza, Reynolds, Roy, Ruiter, Rybicki, Salmon, Pumo,
  Prentice, Seitenzahl, Smith, Sollerman, Sullivan, Szegedi, Taddia,
  Taubenberger, Terreran, Vos, Walton, Wright, Wyrzykowski, Yaron, EPESSTO,
  Chen, Kr{\"{u}}hler, Schady, Wiseman, Greiner, Rau, Schweyer, Klose, {Nicuesa
  Guelbenzu}, GROND, Palliyaguru, University, Shara, Williams, Vaisanen,
  Potter, Colmenero, Crawford, Buckley, Mao, Group, D{\'{i}}az, Macri, Ribeiro,
  S{\'{a}}nchez, Schoenell, Abramo, Akras, Alcaniz, Artola, Beroiz, Bonoli,
  Cabral, Camuccio, Chavushyan, Coelho, Colazo, Costa-Duarte, Dultzin,
  Fern{\'{a}}ndez, Garc{\'{i}}a, Girardini, Gon{\c{c}}alves, Gon{\c{c}}alves,
  Gurovich, Jim{\'{e}}nez-Teja, Kanaan, Lares, L{\'{o}}pez-Cruz, Melia, Molino,
  Padilla, Pe{\~{n}}uela, Placco, Qui{\~{n}}ones, Renzi, Riguccini,
  R{\'{i}}os-L{\'{o}}pez, Rodriguez, Sampedro, Schneiter, Sodr{\'{e}}, Starck,
  Torres-Flores, Tornatore, Zadro{\.{z}}ny, Collaboration, Castro-Tirado,
  Tello, Hu, Zhang, Cunniffe, Castell{\'{o}}n, Hiriart, Caballero-Garc{\'{i}}a,
  Jel{\'{i}}nek, Kub{\'{a}}nek, Park, Jeong, Pandey, Yock, Querel, Fan, Wang,
  Collaboration, Beardsley, Brown, Crosse, Emrich, Franzen, Gaensler, Horsley,
  Johnston-Hollitt, Kenney, Morales, Pallot, Sokolowski, Steele, Tingay, Trott,
  Walker, Wayth, Williams, Wu, Array, Yoshida, Sakamoto, Kawakubo, Yamaoka,
  Takahashi, Asaoka, Ozawa, Torii, Shimizu, Tamura, Ishizaki, Cherry,
  Ricciarini, Penacchioni, Marrocchesi, Collaboration, Pozanenko, Volnova,
  Mazaeva, Minaev, Krugov, Kusakin, Reva, Moskvitin, Rumyantsev, Inasaridze,
  Klunko, Tungalag, Schmalz, Burhonov, Collaboration, Abdalla, Abramowski,
  Aharonian, Benkhali, Ang{\"{u}}ner, Arakawa, Arrieta, Aubert, Backes, Balzer,
  Barnard, Becherini, Tjus, Berge, Bernhard, Bernl{\"{o}}hr, Blackwell,
  B{\"{o}}ttcher, Boisson, Bolmont, Bonnefoy, Bordas, Bregeon, Brun, Brun,
  Bryan, B{\"{u}}chele, Bulik, Capasso, Caroff, Carosi, Casanova, Cerruti,
  Chakraborty, Chaves, Chen, Chevalier, Colafrancesco, Condon, Conrad, Davids,
  Decock, Deil, Devin, DeWilt, Dirson, Djannati-Ata{\"{i}}, Donath, Drury,
  Dutson, Dyks, Edwards, Egberts, Emery, Ernenwein, Eschbach, Farnier, Fegan,
  Fernandes, Fiasson, Fontaine, Funk, F{\"{u}}ssling, Gabici, Gallant,
  Garrigoux, Gat{\'{e}}, Giavitto, Giebels, Glawion, Glicenstein, Gottschall,
  Grondin, Hahn, Haupt, Hawkes, Heinzelmann, Henri, Hermann, Hinton, Hofmann,
  Hoischen, Holch, Holler, Horns, Ivascenko, Iwasaki, Jacholkowska, Jamrozy,
  Jankowsky, Jankowsky, Jingo, Jouvin, Jung-Richardt, Kastendieck,
  Katarzy{\'{n}}ski, Katsuragawa, Khangulyan, Kh{\'{e}}lifi, King, Klepser,
  Klochkov, Klu{\'{z}}niak, Komin, Kosack, Krakau, Kraus, Kr{\"{u}}ger, Laffon,
  Lamanna, Lau, Lees, Lefaucheur, Lemi{\`{e}}re, Lemoine-Goumard, Lenain,
  Leser, Lohse, Lorentz, Liu, Lypova, Malyshev, Marandon, Marcowith, Mariaud,
  Marx, Maurin, Maxted, Mayer, Meintjes, Meyer, Mitchell, Moderski, Mohamed,
  Mohrmann, Mor{\aa}, Moulin, Murach, Nakashima, de~Naurois, Ndiyavala,
  Niederwanger, Niemiec, Oakes, O'Brien, Odaka, Ohm, Ostrowski, Oya, Padovani,
  Panter, Parsons, Pekeur, Pelletier, Perennes, Petrucci, Peyaud, Piel, Pita,
  Poireau, Poon, Prokhorov, Prokoph, P{\"{u}}hlhofer, Punch, Quirrenbach, Raab,
  Rauth, Reimer, Reimer, Renaud, de~los Reyes, Rieger, Rinchiuso, Romoli,
  Rowell, Rudak, Rulten, Sahakian, Saito, Sanchez, Santangelo, Sasaki,
  Schlickeiser, Sch{\"{u}}ssler, Schulz, Schwanke, Schwemmer, Seglar-Arroyo,
  Settimo, Seyffert, Shafi, Shilon, Shiningayamwe, Simoni, Sol, Spanier,
  Spir-Jacob, Stawarz, Steenkamp, Stegmann, Steppa, Sushch, Takahashi,
  Tavernet, Tavernier, Taylor, Terrier, Tibaldo, Tiziani, Tluczykont, Trichard,
  Tsirou, Tsuji, Tuffs, Uchiyama, van~der Walt, van Eldik, van Rensburg, van
  Soelen, Vasileiadis, Veh, Venter, Viana, Vincent, Vink, Voisin, V{\"{o}}lk,
  Vuillaume, Wadiasingh, Wagner, Wagner, Wagner, White, Wierzcholska, Willmann,
  W{\"{o}}rnlein, Wouters, Yang, Zaborov, Zacharias, Zanin, Zdziarski, Zech,
  Zefi, Ziegler, Zorn, {\.{Z}}ywucka, Collaboration, Fender, Broderick,
  Rowlinson, Wijers, Stewart, Shulevski, Collaboration, Kavic, Simonetti,
  League, Tsai, Obenberger, Nathaniel, Taylor, Dowell, Liebling, Estes,
  Lippert, Sharma, Vincent, Farella, Array, Abeysekara, Albert, Alfaro,
  Alvarez, Arceo, Arteaga-Vel{\'{a}}zquez, Barber, Becerril, Belmont-Moreno,
  BenZvi, Berley, Bernal, Braun, Brisbois, Caballero-Mora, Capistr{\'{a}}n,
  Carrami{\~{n}}ana, Casanova, Castillo, Cotti, Cotzomi, Dichiara, Dingus,
  DuVernois, D{\'{i}}az-V{\'{e}}lez, Ellsworth, Engel, Enr{\'{i}}quez-Rivera,
  Fiorino, Fleischhack, Fraija, Garc{\'{i}}a-Gonz{\'{a}}lez, Garfias, Gerhardt,
  Gonz{\'{a}}lez, Goodman, Hampel-Arias, Harding, Hernandez, Hernandez-Almada,
  Hona, H{\"{u}}ntemeyer, Iriarte, Jardin-Blicq, Joshi, Kaufmann, Kieda, Lara,
  Lauer, Lennarz, Linnemann, Longinotti, Luna-Garc{\'{i}}a, L{\'{o}}pez-Coto,
  Malone, Marinelli, Martinez, Martinez-Castellanos, Mart{\'{i}}nez-Castro,
  Mart{\'{i}}nez-Huerta, Matthews, Miranda-Romagnoli, Moreno, Mostaf{\'{a}},
  Nellen, Newbold, Nisa, Noriega-Papaqui, Pelayo, Pretz,
  P{\'{e}}rez-P{\'{e}}rez, Ren, Rho, Rivi{\`{e}}re, Rosa-Gonz{\'{a}}lez,
  Rosenberg, Ruiz-Velasco, Salazar, Sandoval, Schneider, Schoorlemmer, Sinnis,
  Smith, Springer, Surajbali, Tibolla, Tollefson, Torres, Ukwatta, Weisgarber,
  Westerhoff, Wisher, Wood, Yapici, Yodh, Younk, Zhou, {\'{A}}lvarez,
  Collaboration, Aab, Abreu, Aglietta, Albuquerque, Albury, Allekotte, Almela,
  Alvarez-Mu{\~{n}}iz, Anastasi, Anchordoqui, Andrada, Andringa, Aramo, Arsene,
  Asorey, Assis, Avila, Badescu, Balaceanu, Barbato, Becker, Bellido, Berat,
  Bertaina, Bertou, Biermann, Biteau, Blaess, Blanco, Blazek, Bleve,
  Boh{\'{a}}{\v{c}}ov{\'{a}}, Bonifazi, Borodai, Botti, Brack, Brancus, Bretz,
  Bridgeman, Briechle, Buchholz, Bueno, Buitink, Buscemi, Caballero-Mora,
  Caccianiga, Cancio, Canfora, Caruso, Castellina, Catalani, Cataldi, Cazon,
  Chavez, Chinellato, Chudoba, Clay, Colalillo, Coleman, Collica, Coluccia,
  Concei{\c{c}}{\~{a}}o, Consolati, Contreras, Cooper, Coutu, Covault, Cronin,
  D'Amico, Daniel, Dasso, Daumiller, Dawson, Day, de~Almeida, de~Jong, Mauro,
  Neto, Mitri, de~Oliveira, de~Souza, Debatin, Deligny, Diogo, Dobrigkeit,
  D'Olivo, Dorosti, Dova, Dundovic, Ebr, Engel, Erdmann, Erfani, Escobar,
  Espadanal, Etchegoyen, Falcke, Farmer, Farrar, Fauth, Fazzini, Feldbusch,
  Fenu, Fick, Figueira, Freire, Fujii, Fuster, Ga{\"{i}}or, Garc{\'{i}}a,
  Gat{\'{e}}, Gemmeke, Gherghel-Lascu, Ghia, Giaccari, Giammarchi, Giller,
  G{\l}as, Glaser, Golup, Gonz{\'{a}}lez, Gorgi, Gottowik, Grillo, Grubb,
  Guarino, Guedes, Halliday, Hampel, Hansen, Harari, Harrison, Harvey, Haungs,
  Hebbeker, Heck, Heimann, Herve, Hill, Hojvat, Holt, Homola, H{\"{o}}randel,
  Horvath, Huege, Hulsman, Insolia, Isar, Jandt, Johnsen, Josebachuili,
  Jurysek, K{\"{a}}{\"{a}}p{\"{a}}, Kampert, Keilhauer, Kemmerich, Kemp,
  Kieckhafer, Klages, Kleifges, Kleinfeller, Krause, Krohm, Kuempel, Kunka,
  Lago, LaHurd, Lang, Lauscher, Legumina, Letessier-Selvon, Lhenry-Yvon, Link,
  Lopes, L{\'{o}}pez, Lorek, Luce, Lucero, Malacari, Mallamaci, Mandat,
  Mantsch, Mariazzi, Marsella, Martello, Martinez, Mathes, Mathys, Matthews,
  Matthiae, Mayotte, Mazur, Medina, Medina-Tanco, Melo, Menshikov, Merenda,
  Michal, Micheletti, Middendorf, Miramonti, Mitrica, Mockler, Mollerach,
  Montanet, Morello, Morlino, M{\"{u}}ller, M{\"{u}}ller, Muller, M{\"{u}}ller,
  Mussa, Naranjo, Nguyen, Niculescu-Oglinzanu, Niechciol, Niemietz, Niggemann,
  Nitz, Nosek, Novotny, No{\v{z}}ka, N{\'{u}}{\~{n}}ez, Oikonomou, Olinto,
  Palatka, Pallotta, Papenbreer, Parente, Parra, Paul, Pech, Pedreira, Pȩkala,
  Pe{\~{n}}a-Rodriguez, Pereira, Perlin, Perrone, Peters, Petrera, Phuntsok,
  Pierog, Pimenta, Pirronello, Platino, Plum, Poh, Porowski, Prado, Privitera,
  Prouza, Quel, Querchfeld, Quinn, Ramos-Pollan, Rautenberg, Ravignani, Ridky,
  Riehn, Risse, Ristori, Rizi, Roncoroni, Roth, Roulet, Rovero, Ruehl, Saffi,
  Saftoiu, Salamida, Salazar, Saleh, Salina, S{\'{a}}nchez, Sanchez-Lucas,
  Santos, Santos, Sarazin, Sarmento, Sarmiento-Cano, Sato, Schauer, Scherini,
  Schieler, Schimp, Schmidt, Scholten, Schov{\'{a}}nek, Schr{\"{o}}der,
  Schr{\"{o}}der, Schulz, Schumacher, Sciutto, Segreto, Shadkam, Shellard,
  Sigl, Silli, {\v{S}}m{\'{i}}da, Snow, Sommers, Sonntag, Soriano, Squartini,
  Stanca, Stasielak, Stassi, Stolpovskiy, Strafella, Streich, Suarez,
  Suarez-Dur{\'{a}}n, Sudholz, Suomij{\"{a}}rvi, Supanitsky, {\v{S}}up{\'{i}}k,
  Swain, Szadkowski, Taboada, Taborda, Timmermans, Tomankova, Tom{\'{e}},
  Travnicek, Trini, Tueros, Ulrich, Unger, Urban, Vali{\~{n}}o, Valore, van
  Aar, van Bodegom, van~den Berg, van Vliet, Varela, C{\'{a}}rdenas,
  V{\'{a}}zquez, Ventura, Verzi, Vicha, Villase{\~{n}}or, Vorobiov, Wahlberg,
  Wainberg, Walz, Watson, Weber, Weindl, Wiede{\'{n}}ski, Wiencke,
  Wilczy{\'{n}}ski, Wirtz, Wittkowski, Wundheiler, Yang, Yushkov, Zas,
  Zavrtanik, Zavrtanik, Zepeda, Zimmermann, Ziolkowski, Zong, Zuccarello,
  Collaboration, Kim, Schulze, Bauer, Corral-Santana,
  Gonz{\'{a}}lez-L{\'{o}}pez, Hartmann, Ishwara-Chandra, Mart{\'{i}}n, Mehner,
  Misra, Micha{\l}owski, Resmi, Collaboration, Paragi, Agudo, An, Beswick,
  Casadio, Frey, Jonker, Kettenis, Marcote, Moldon, Szomoru, Langevelde, Yang,
  Team, Cwiek, Cwiok, Czyrkowski, Dabrowski, Kasprowicz, Mankiewicz, Nawrocki,
  Opiela, Piotrowski, Wrochna, Zaremba, {\.{Z}}arnecki, Collaboration, Haggard,
  Nynka, Ruan, University, Bland, Booler, Devillepoix, de~Gois, Hancock, Howie,
  Paxman, Sansom, Towner, Network, Tonry, Coughlin, Stubbs, Denneau, Heinze,
  Stalder, Weiland, ATLAS, Eatough, Kramer, Kraus, Survey, Troja, Piro,
  Gonz{\'{a}}lez, Butler, Fox, Khandrika, Kutyrev, Lee, Ricci,
  S'́anchez-Ram'́irez, Veilleux, Watson, Wieringa, Burgess, van Eerten,
  Fontes, Fryer, Korobkin, Wollaeger, RIMAS, RATIR, Camilo, Foley, Goedhart,
  Makhathini, Oozeer, Smirnov, Fender, Woudt, \& MeerKAT}]{GW170817_MMA}
Abbott, B.~P., Abbott, R., Abbott, T.~D., {et~al.} 2017{\natexlab{a}}, The
  Astrophysical Journal Letters, 848, L12

\bibitem[{Abbott {et~al.}(2017{\natexlab{b}})Abbott, Abbott, Abbott, Acernese,
  Ackley, Adams, Adams, Addesso, Adhikari, Adya, Affeldt, Afrough, Agarwal,
  Agathos, Agatsuma, Aggarwal, Aguiar, Aiello, Ain, Ajith, Allen, Allen,
  Allocca, Altin, Amato, Ananyeva, Anderson, Anderson, Angelova, Antier,
  Appert, Arai, Araya, Areeda, Arnaud, Arun, Ascenzi, Ashton, Ast, Aston,
  Astone, Atallah, Aufmuth, Aulbert, AultONeal, Austin, Avila-Alvarez, Babak,
  Bacon, Bader, Bae, Baker, Baldaccini, Ballardin, Banagiri, Barayoga, Barclay,
  Barish, Barker, Barkett, Barone, Barr, Barsotti, Barsuglia, Barta, Bartlett,
  Bartos, Bassiri, Basti, Batch, Bawaj, Bayley, Bazzan, B'ecsy, Beer, Bejger,
  Belahcene, Bell, Bergmann, Bernuzzi, Bero, Berry, Bersanetti, Bertolini,
  Betzwieser, Bhagwat, Bhandare, Bilenko, Billingsley, Billman, Birch, Birney,
  Birnholtz, Biscans, Biscoveanu, Bisht, Bitossi, Biwer, Bizouard, Blackburn,
  Blackman, Blair, Blair, Blair, Bloemen, Bock, Bode, Boer, Bogaert, Bohe,
  Bondu, Bonilla, Bonnand, Boom, Bork, Boschi, Bose, Bossie, Bouffanais, Bozzi,
  Bradaschia, Brady, Branchesi, Brau, Briant, Brillet, Brinkmann, Brisson,
  Brockill, Broida, Brooks, Brown, Brunett, Buchanan, Buikema, Bulik, Bulten,
  Buonanno, Buskulic, Buy, Byer, Cabero, Cadonati, Cagnoli, Cahillane,
  Bustillo, Callister, Calloni, Camp, Canepa, Canizares, Cannon, Cao, Cao,
  Capano, Capocasa, Carbognani, Caride, Carney, Diaz, Casentini, Caudill,
  Cavagli`a, Cavalier, Cavalieri, Cella, Cepeda, Cerd'a-Dur'an, Cerretani,
  Cesarini, Chamberlin, Chan, Chao, Charlton, Chase, Chassande-Mottin,
  Chatterjee, Chatziioannou, Cheeseboro, Chen, Chen, Chen, Cheng, Chia,
  Chincarini, Chiummo, Chmiel, Cho, Cho, Chow, Christensen, Chu, Chua, Chua,
  Chung, Chung, Ciani, Ciolfi, Cirelli, Cirone, Clara, Clark, Clearwater,
  Cleva, Cocchieri, Coccia, Cohadon, Cohen, Colla, Collette, Cominsky,
  Constancio, Conti, Cooper, Corban, Corbitt, Cordero-Carri'on, Corley,
  Cornish, Corsi, Cortese, Costa, Coughlin, Coughlin, Coulon, Countryman,
  Couvares, Covas, Cowan, Coward, Cowart, Coyne, Coyne, Creighton, Creighton,
  Cripe, Crowder, Cullen, Cumming, Cunningham, Cuoco, Canton, D'alya,
  Danilishin, D'Antonio, Danzmann, Dasgupta, Costa, Dattilo, Dave, Davier,
  Davis, Daw, Day, De, DeBra, Degallaix, {De Laurentis}, Del'eglise, {Del
  Pozzo}, Demos, Denker, Dent, {De Pietri}, Dergachev, {De Rosa}, DeRosa, {De
  Rossi}, DeSalvo, de~Varona, Devenson, Dhurandhar, D'iaz, Dietrich, {Di
  Fiore}, {Di Giovanni}, {Di Girolamo}, {Di Lieto}, {Di Pace}, {Di Palma}, {Di
  Renzo}, Doctor, Dolique, Donovan, Dooley, Doravari, Dorrington, Douglas,
  'Alvarez, Downes, Drago, Dreissigacker, Driggers, Du, Ducrot, Dupej, Dwyer,
  Edo, Edwards, Effler, Eggenstein, Ehrens, Eichholz, Eikenberry, Eisenstein,
  Essick, Estevez, Etienne, Etzel, Evans, Evans, Factourovich, Fafone, Fair,
  Fairhurst, Fan, Farinon, Farr, Farr, Fauchon-Jones, Favata, Fays, Fee,
  Fehrmann, Feicht, Fejer, Fernandez-Galiana, Ferrante, Ferreira, Ferrini,
  Fidecaro, Finstad, Fiori, Fiorucci, Fishbach, Fisher, Fitz-Axen, Flaminio,
  Fletcher, Fong, Font, Forsyth, Forsyth, Fournier, Frasca, Frasconi, Frei,
  Freise, Frey, Frey, Fries, Fritschel, Frolov, Fulda, Fyffe, Gabbard, Gadre,
  Gaebel, Gair, Gammaitoni, Ganija, Gaonkar, Garcia-Quiros, Garufi, Gateley,
  Gaudio, Gaur, Gayathri, Gehrels, Gemme, Genin, Gennai, George, George,
  Gergely, Germain, Ghonge, Ghosh, Ghosh, Ghosh, Giaime, Giardina, Giazotto,
  Gill, Glover, Goetz, Goetz, Gomes, Goncharov, Gonz'alez, Castro, Gopakumar,
  Gorodetsky, Gossan, Gosselin, Gouaty, Grado, Graef, Granata, Grant, Gras,
  Gray, Greco, Green, Gretarsson, Groot, Grote, Grunewald, Gruning, Guidi, Guo,
  Gupta, Gupta, Gushwa, Gustafson, Gustafson, Halim, Hall, Hall, Hamilton,
  Hammond, Haney, Hanke, Hanks, Hanna, Hannam, Hannuksela, Hanson, Hardwick,
  Harms, Harry, Harry, Hart, Haster, Haughian, Healy, Heidmann, Heintze,
  Heitmann, Hello, Hemming, Hendry, Heng, Hennig, Heptonstall, Heurs, Hild,
  Hinderer, Hoak, Hofman, Holt, Holz, Hopkins, Horst, Hough, Houston, Howell,
  Hreibi, Hu, Huerta, Huet, Hughey, Husa, Huttner, Huynh-Dinh, Indik, Inta,
  Intini, Isa, Isac, Isi, Iyer, Izumi, Jacqmin, Jani, Jaranowski, Jawahar,
  Jim'enez-Forteza, Johnson, Johnson-McDaniel, Jones, Jones, Jonker, Ju,
  Junker, Kalaghatgi, Kalogera, Kamai, Kandhasamy, Kang, Kanner, Kapadia,
  Karki, Karvinen, Kasprzack, Kastaun, Katolik, Katsavounidis, Katzman, Kaufer,
  Kawabe, Kawaguchi, K'ef'elian, Keitel, Kemball, Kennedy, Kent, Key, Khalili,
  Khan, Khan, Khan, Khazanov, Kijbunchoo, Kim, Kim, Kim, Kim, Kim, Kim,
  Kimbrell, King, King, Kinley-Hanlon, Kirchhoff, Kissel, Kleybolte, Klimenko,
  Knowles, Koch, Koehlenbeck, Koley, Kondrashov, Kontos, Korobko, Korth,
  Kowalska, Kozak, Kr"amer, Kringel, Kr'olak, Kuehn, Kumar, Kumar, Kumar, Kuo,
  Kutynia, Kwang, Lackey, Lai, Landry, Lang, Lange, Lantz, Lanza, Larson,
  Lartaux-Vollard, Lasky, Laxen, Lazzarini, Lazzaro, Leaci, Leavey, Lee, Lee,
  Lee, Lee, Lee, Lehmann, Lenon, Leonardi, Leroy, Letendre, Levin, Li, Linker,
  Littenberg, Liu, Liu, Lo, Lockerbie, London, Lord, Lorenzini, Loriette,
  Lormand, Losurdo, Lough, Lousto, Lovelace, L"uck, Lumaca, Lundgren, Lynch,
  Ma, Macas, Macfoy, Machenschalk, MacInnis, Macleod, Hernandez,
  Maga{\~{n}}a-Sandoval, Zertuche, Magee, Majorana, Maksimovic, Man, Mandic,
  Mangano, Mansell, Manske, Mantovani, Marchesoni, Marion, M'arka, M'arka,
  Markakis, Markosyan, Markowitz, Maros, Marquina, Martelli, Martellini,
  Martin, Martin, Martynov, Mason, Massera, Masserot, Massinger, Masso-Reid,
  Mastrogiovanni, Matas, Matichard, Matone, Mavalvala, Mazumder, McCarthy,
  McClelland, McCormick, McCuller, McGuire, McIntyre, McIver, McManus, McNeill,
  McRae, McWilliams, Meacher, Meadors, Mehmet, Meidam, Mejuto-Villa, Melatos,
  Mendell, Mercer, Merilh, Merzougui, Meshkov, Messenger, Messick, Metzdorff,
  Meyers, Miao, Michel, Middleton, Mikhailov, Milano, Miller, Miller, Miller,
  Millhouse, Milovich-Goff, Minazzoli, Minenkov, Ming, Mishra, Mitra,
  Mitrofanov, Mitselmakher, Mittleman, Moffa, Moggi, Mogushi, Mohan, Mohapatra,
  Montani, Moore, Moraru, Moreno, Morriss, Mours, Mow-Lowry, Mueller, Muir,
  Mukherjee, Mukherjee, Mukherjee, Mukund, Mullavey, Munch, Mu{\~{n}}iz,
  Muratore, Murray, Napier, Nardecchia, Naticchioni, Nayak, Neilson, Nelemans,
  Nelson, Nery, Neunzert, Nevin, Newport, Newton, Ng, Nguyen, Nichols, Nielsen,
  Nissanke, Nitz, Noack, Nocera, Nolting, North, Nuttall, Oberling, O'Dea,
  Ogin, Oh, Oh, Ohme, Okada, Oliver, Oppermann, Oram, O'Reilly, Ormiston,
  Ortega, O'Shaughnessy, Ossokine, Ottaway, Overmier, Owen, Pace, Page, Page,
  Pai, Pai, Palamos, Palashov, Palomba, Pal-Singh, Pan, Pan, Pang, Pang,
  Pankow, Pannarale, Pant, Paoletti, Paoli, Papa, Parida, Parker, Pascucci,
  Pasqualetti, Passaquieti, Passuello, Patil, Patricelli, Pearlstone, Pedraza,
  Pedurand, Pekowsky, Pele, Penn, Perez, Perreca, Perri, Pfeiffer, Phelps,
  Piccinni, Pichot, Piergiovanni, Pierro, Pillant, Pinard, Pinto, Pirello,
  Pitkin, Poe, Poggiani, Popolizio, Porter, Post, Powell, Prasad, Pratt,
  Pratten, Predoi, Prestegard, Prijatelj, Principe, Privitera, Prodi,
  Prokhorov, Puncken, Punturo, Puppo, P"urrer, Qi, Quetschke, Quintero,
  Quitzow-James, Rabeling, Radkins, Raffai, Raja, Rajan, Rajbhandari,
  Rakhmanov, Ramirez, Ramos-Buades, Rapagnani, Raymond, Razzano, Read,
  Regimbau, Rei, Reid, Reitze, Ren, Reyes, Ricci, Ricker, Rieger, Riles, Rizzo,
  Robertson, Robie, Robinet, Rocchi, Rolland, Rollins, Roma, Romano, Romel,
  Romie, Rosi'nska, Ross, Rowan, R"udiger, Ruggi, Rutins, Ryan, Sachdev,
  Sadecki, Sadeghian, Sakellariadou, Salconi, Saleem, Salemi, Samajdar, Sammut,
  Sampson, Sanchez, Sanchez, Sanchis-Gual, Sandberg, Sanders, Sassolas, Sauter,
  Savage, Sawadsky, Schale, Scheel, Scheuer, Schmidt, Schmidt, Schnabel,
  Schofield, Sch"onbeck, Schreiber, Schuette, Schulte, Schutz, Schwalbe, Scott,
  Scott, Seidel, Sellers, Sengupta, Sentenac, Sequino, Sergeev, Shaddock,
  Shaffer, Shah, Shahriar, Shaner, Shao, Shapiro, Shawhan, Sheperd, Shoemaker,
  Shoemaker, Siellez, Siemens, Sieniawska, Sigg, Silva, Singer, Singh, Singhal,
  Sintes, Slagmolen, Smith, Smith, Smith, Somala, Son, Sonnenberg, Sorazu,
  Sorrentino, Souradeep, Spencer, Srivastava, Staats, Staley, Steinke,
  Steinlechner, Steinlechner, Steinmeyer, Stevenson, Stone, Stops, Strain,
  Stratta, Strigin, Strunk, Sturani, Stuver, Summerscales, Sun, Sunil, Suresh,
  Sutton, Swinkels, Szczepa'nczyk, Tacca, Tait, Talbot, Talukder, Tanner,
  T'apai, Taracchini, Tasson, Taylor, Taylor, Tewari, Theeg, Thies, Thomas,
  Thomas, Thomas, Thorne, Thrane, Tiwari, Tiwari, Tokmakov, Toland, Tonelli,
  Tornasi, Torres-Forn'e, Torrie, T"oyr"a, Travasso, Traylor, Trinastic,
  Tringali, Trozzo, Tsang, Tse, Tso, Tsukada, Tsuna, Tuyenbayev, Ueno, Ugolini,
  Unnikrishnan, Urban, Usman, Vahlbruch, Vajente, Valdes, van Bakel, van
  Beuzekom, van~den Brand, Broeck, Vander-Hyde, van~der Schaaf, van Heijningen,
  van Veggel, Vardaro, Varma, Vass, Vas'uth, Vecchio, Vedovato, Veitch, Veitch,
  Venkateswara, Venugopalan, Verkindt, Vetrano, Vicer'e, Viets, Vinciguerra,
  Vine, Vinet, Vitale, Vo, Vocca, Vorvick, Vyatchanin, Wade, Wade, Wade, Walet,
  Walker, Wallace, Walsh, Wang, Wang, Wang, Wang, Wang, Ward, Warner, Was,
  Watchi, Weaver, Wei, Weinert, Weinstein, Weiss, Wen, Wessel, Wessels,
  Westerweck, Westphal, Wette, Whelan, Whiting, Whittle, Wilken, Williams,
  Williams, Williamson, Willis, Willke, Wimmer, Winkler, Wipf, Wittel, Woan,
  Woehler, Wofford, Wong, Worden, Wright, Wu, Wysocki, Xiao, Yamamoto, Yancey,
  Yang, Yap, Yazback, Yu, Yu, Yvert, Zny, Zanolin, Zelenova, Zendri, Zevin,
  Zhang, Zhang, Zhang, Zhang, Zhao, Zhou, Zhou, Zhu, Zhu, Zimmerman, Zucker, \&
  Zweizig}]{GW170817_kilonova}
---. 2017{\natexlab{b}}, The Astrophysical Journal Letters, 850, L39

\bibitem[{Abbott {et~al.}(2019)Abbott, Abbott, Abbott, Abernathy, Acernese,
  Ackley, Adams, Adams, Addesso, Adhikari, Adya, Affeldt, Agathos, Agatsuma,
  Aggarwal, Aguiar, Aiello, Ain, Ajith, Akutsu, Allen, Allocca, Altin,
  Ananyeva, Anderson, Anderson, Ando, Appert, Arai, Araya, Araya, Areeda,
  Arnaud, Arun, Asada, Ascenzi, Ashton, Aso, Ast, Aston, Astone, Atsuta,
  Aufmuth, Aulbert, Avila-Alvarez, Awai, Babak, Bacon, Bader, Baiotti, Baker,
  Baldaccini, Ballardin, Ballmer, Barayoga, Barclay, Barish, Barker, Barone,
  Barr, Barsotti, Barsuglia, Barta, Bartlett, Barton, Bartos, Bassiri, Basti,
  Batch, Baune, Bavigadda, Bazzan, B{\'{e}}csy, Beer, Bejger, Belahcene,
  Belgin, Bell, Berger, Bergmann, Berry, Bersanetti, Bertolini, Betzwieser,
  Bhagwat, Bhandare, Bilenko, Billingsley, Billman, Birch, Birney, Birnholtz,
  Biscans, Bisht, Bitossi, Biwer, Bizouard, Blackburn, Blackman, Blair, Blair,
  Blair, Bloemen, Bock, Boer, Bogaert, Bohe, Bondu, Bonnand, Boom, Bork,
  Boschi, Bose, Bouffanais, Bozzi, Bradaschia, Brady, Braginsky, Branchesi,
  Brau, Briant, Brillet, Brinkmann, Brisson, Brockill, Broida, Brooks, Brown,
  Brown, Brown, Brunett, Buchanan, Buikema, Bulik, Bulten, Buonanno, Buskulic,
  Buy, Byer, Cabero, Cadonati, Cagnoli, Cahillane, {Calder{\'{o}}n Bustillo},
  Callister, Calloni, Camp, Cannon, Cao, Cao, Capano, Capocasa, Carbognani,
  Caride, {Casanueva Diaz}, Casentini, Caudill, Cavagli{\`{a}}, Cavalier,
  Cavalieri, Cella, Cepeda, {Cerboni Baiardi}, Cerretani, Cesarini, Chamberlin,
  Chan, Chao, Charlton, Chassande-Mottin, Cheeseboro, Chen, Chen, Cheng,
  Chincarini, Chiummo, Chmiel, Cho, Cho, Chow, Christensen, Chu, Chua, Chua,
  Chung, Ciani, Clara, Clark, Cleva, Cocchieri, Coccia, Cohadon, Colla,
  Collette, Cominsky, Constancio, Conti, Cooper, Corbitt, Cornish, Corsi,
  Cortese, Costa, Coughlin, Coughlin, Coulon, Countryman, Couvares, Covas,
  Cowan, Coward, Cowart, Coyne, Coyne, Creighton, Creighton, Cripe, Crowder,
  Cullen, Cumming, Cunningham, Cuoco, Canton, Danilishin, D'Antonio, Danzmann,
  Dasgupta, {Da Silva Costa}, Dattilo, Dave, Davier, Davies, Davis, Daw, Day,
  Day, De, DeBra, Debreczeni, Degallaix, {De Laurentis}, Del{\'{e}}glise, {Del
  Pozzo}, Denker, Dent, Dergachev, {De Rosa}, DeRosa, DeSalvo, Devine,
  Dhurandhar, D{\'{i}}az, Fiore, Giovanni, Girolamo, Lieto, Pace, Palma,
  Virgilio, Doctor, Doi, Dolique, Donovan, Dooley, Doravari, Dorrington,
  Douglas, {Dovale {\'{A}}lvarez}, Downes, Drago, Drever, Driggers, Du, Ducrot,
  Dwyer, Eda, Edo, Edwards, Effler, Eggenstein, Ehrens, Eichholz, Eikenberry,
  Eisenstein, Essick, Etienne, Etzel, Evans, Evans, Everett, Factourovich,
  Fafone, Fair, Fairhurst, Fan, Farinon, Farr, Farr, Fauchon-Jones, Favata,
  Fays, Fehrmann, Fejer, {Fern{\'{a}}ndez Galiana}, Ferrante, Ferreira,
  Ferrini, Fidecaro, Fiori, Fiorucci, Fisher, Flaminio, Fletcher, Fong,
  Forsyth, Fournier, Frasca, Frasconi, Frei, Freise, Frey, Frey, Fries,
  Fritschel, Frolov, Fujii, Fujimoto, Fulda, Fyffe, Gabbard, Gadre, Gaebel,
  Gair, Gammaitoni, Gaonkar, Garufi, Gaur, Gayathri, Gehrels, Gemme, Genin,
  Gennai, George, Gergely, Germain, Ghonge, Ghosh, Ghosh, Ghosh, Giaime,
  Giardina, Giazotto, Gill, Glaefke, Goetz, Goetz, Gondan, Gonz{\'{a}}lez,
  {Gonzalez Castro}, Gopakumar, Gorodetsky, Gossan, Gosselin, Gouaty, Grado,
  Graef, Granata, Grant, Gras, Gray, Greco, Green, Groot, Grote, Grunewald,
  Guidi, Guo, Gupta, Gupta, Gushwa, Gustafson, Gustafson, Hacker, Hagiwara,
  Hall, Hall, Hammond, Haney, Hanke, Hanks, Hanna, Hannam, Hanson, Hardwick,
  Harms, Harry, Harry, Hart, Hartman, Haster, Haughian, Hayama, Healy,
  Heidmann, Heintze, Heitmann, Hello, Hemming, Hendry, Heng, Hennig, Henry,
  Heptonstall, Heurs, Hild, Hirose, Hoak, Hofman, Holt, Holz, Hopkins, Hough,
  Houston, Howell, Hu, Huerta, Huet, Hughey, Husa, Huttner, Huynh-Dinh, Indik,
  Ingram, Inta, Ioka, Isa, Isac, Isi, Isogai, Itoh, Iyer, Izumi, Jacqmin, Jani,
  Jaranowski, Jawahar, Jim{\'{e}}nez-Forteza, Johnson, Jones, Jones, Jonker,
  Ju, Junker, Kagawa, Kajita, Kakizaki, Kalaghatgi, Kalogera, Kamiizumi, Kanda,
  Kandhasamy, Kanemura, Kaneyama, Kang, Kanner, Karki, Karvinen, Kasprzack,
  Kataoka, Katsavounidis, Katzman, Kaufer, Kaur, Kawabe, Kawai, Kawamura,
  K{\'{e}}f{\'{e}}lian, Keitel, Kelley, Kennedy, Key, Khalili, Khan, Khan,
  Khan, Khazanov, Kijbunchoo, Kim, Kim, Kim, Kim, Kim, Kim, Kimbrell, Kimura,
  King, King, Kirchhoff, Kissel, Klein, Kleybolte, Klimenko, Koch, Koehlenbeck,
  Kojima, Kokeyama, Koley, Komori, Kondrashov, Kontos, Korobko, Korth, Kotake,
  Kowalska, Kozak, Kr{\"{a}}mer, Kringel, Krishnan, Kr{\'{o}}lak, Kuehn, Kumar,
  Kumar, Kumar, Kuo, Kuroda, Kutynia, Kuwahara, Lackey, Landry, Lang, Lange,
  Lantz, Lanza, Lartaux-Vollard, Lasky, Laxen, Lazzarini, Lazzaro, Leaci,
  Leavey, Lebigot, Lee, Lee, Lee, Lee, Lee, Lehmann, Lenon, Leonardi, Leong,
  Leroy, Letendre, Levin, Li, Libson, Littenberg, Liu, Lockerbie, Lombardi,
  London, Lord, Lorenzini, Loriette, Lormand, Losurdo, Lough, Lousto, Lovelace,
  L{\"{u}}ck, Lundgren, Lynch, Ma, Macfoy, Machenschalk, MacInnis, Macleod,
  Maga{\~{n}}a-Sandoval, Majorana, Maksimovic, Malvezzi, Man, Mandic, Mangano,
  Mano, Mansell, Manske, Mantovani, Marchesoni, Marchio, Marion, M{\'{a}}rka,
  M{\'{a}}rka, Markosyan, Maros, Martelli, Martellini, Martin, Martynov, Mason,
  Masserot, Massinger, Masso-Reid, Mastrogiovanni, Matichard, Matone,
  Matsumoto, Matsushima, Mavalvala, Mazumder, McCarthy, McClelland, McCormick,
  McGrath, McGuire, McIntyre, McIver, McManus, McRae, McWilliams, Meacher,
  Meadors, Meidam, Melatos, Mendell, Mendoza-Gandara, Mercer, Merilh,
  Merzougui, Meshkov, Messenger, Messick, Metzdorff, Meyers, Mezzani, Miao,
  Michel, Michimura, Middleton, Mikhailov, Milano, Miller, Miller, Miller,
  Miller, Millhouse, Minenkov, Ming, Mirshekari, Mishra, Mitrofanov,
  Mitselmakher, Mittleman, Miyakawa, Miyamoto, Miyamoto, Miyoki, Moggi, Mohan,
  Mohapatra, Montani, Moore, Moore, Moraru, Moreno, Morii, Morisaki, Moriwaki,
  Morriss, Mours, Mow-Lowry, Mueller, Muir, Mukherjee, Mukherjee, Mukherjee,
  Mukund, Mullavey, Munch, Muniz, Murray, Mytidis, Nagano, Nakamura, Nakamura,
  Nakano, Nakano, Nakano, Nakao, Napier, Nardecchia, Narikawa, Naticchioni,
  Nelemans, Nelson, Neri, Nery, Neunzert, Newport, Newton, Nguyen, Ni, Nielsen,
  Nissanke, Nitz, Noack, Nocera, Nolting, Normandin, Nuttall, Oberling,
  Ochsner, Oelker, Ogin, Oh, Oh, Ohashi, Ohishi, Ohkawa, Ohme, Okutomi, Oliver,
  Ono, Ono, Oohara, Oppermann, Oram, O'Reilly, O'Shaughnessy, Ottaway,
  Overmier, Owen, Pace, Page, Pai, Pai, Palamos, Palashov, Palomba, Pal-Singh,
  Pan, Pankow, Pannarale, Pant, Paoletti, Paoli, Papa, Paris, Parker, Pascucci,
  Pasqualetti, Passaquieti, Passuello, Patricelli, Pearlstone, Pedraza,
  Pedurand, Pekowsky, Pele, {Pe{\~{n}}a Arellano}, Penn, Perez, Perreca, Perri,
  Pfeiffer, Phelps, Piccinni, Pichot, Piergiovanni, Pierro, Pillant, Pinard,
  Pinto, Pitkin, Poe, Poggiani, Popolizio, Post, Powell, Prasad, Pratt, Predoi,
  Prestegard, Prijatelj, Principe, Privitera, Prodi, Prokhorov, Puncken,
  Punturo, Puppo, P{\"{u}}rrer, Qi, Qin, Qiu, Quetschke, Quintero,
  Quitzow-James, Raab, Rabeling, Radkins, Raffai, Raja, Rajan, Rakhmanov,
  Rapagnani, Raymond, Razzano, Re, Read, Regimbau, Rei, Reid, Reitze, Rew,
  Reyes, Rhoades, Ricci, Riles, Rizzo, Robertson, Robie, Robinet, Rocchi,
  Rolland, Rollins, Roma, Romano, Romie, Rosi{\'{n}}ska, Rowan, R{\"{u}}diger,
  Ruggi, Ryan, Sachdev, Sadecki, Sadeghian, Sago, Saijo, Saito, Sakai,
  Sakellariadou, Salconi, Saleem, Salemi, Samajdar, Sammut, Sampson, Sanchez,
  Sandberg, Sanders, Sasaki, Sassolas, Sathyaprakash, Sato, Sato, Saulson,
  Sauter, Savage, Sawadsky, Schale, Scheuer, Schmidt, Schmidt, Schmidt,
  Schnabel, Schofield, Sch{\"{o}}nbeck, Schreiber, Schuette, Schutz, Schwalbe,
  Scott, Scott, Sekiguchi, Sekiguchi, Sellers, Sengupta, Sentenac, Sequino,
  Sergeev, Setyawati, Shaddock, Shaffer, Shahriar, Shapiro, Shawhan, Sheperd,
  Shibata, Shikano, Shimoda, Shoda, Shoemaker, Shoemaker, Siellez, Siemens,
  Sieniawska, Sigg, Silva, Singer, Singer, Singh, Singh, Singhal, Sintes,
  Slagmolen, Smith, Smith, Smith, Somiya, Son, Sorazu, Sorrentino, Souradeep,
  Spencer, Srivastava, Staley, Steinke, Steinlechner, Steinlechner, Steinmeyer,
  Stephens, Stevenson, Stone, Strain, Straniero, Stratta, Strigin, Sturani,
  Stuver, Sugimoto, Summerscales, Sun, Sunil, Sutton, Suzuki, Swinkels,
  Szczepa{\'{n}}czyk, Tacca, Tagoshi, Takada, Takahashi, Takahashi, Takamori,
  Talukder, Tanaka, Tanaka, Tanaka, Tanner, T{\'{a}}pai, Taracchini, Tatsumi,
  Taylor, Telada, Theeg, Thomas, Thomas, Thomas, Thorne, Thrane, Tippens,
  Tiwari, Tiwari, Tokmakov, Toland, Tomaru, Tomlinson, Tonelli, Tornasi,
  Torrie, T{\"{o}}yr{\"{a}}, Travasso, Traylor, Trifir{\`{o}}, Trinastic,
  Tringali, Trozzo, Tse, Tso, Tsubono, Tsuzuki, Turconi, Tuyenbayev, Uchiyama,
  Uehara, Ueki, Ueno, Ugolini, Unnikrishnan, Urban, Ushiba, Usman, Vahlbruch,
  Vajente, Valdes, van Bakel, van Beuzekom, van~den Brand, {Van Den Broeck},
  Vander-Hyde, van~der Schaaf, van Heijningen, van Putten, van Veggel, Vardaro,
  Varma, Vass, Vas{\'{u}}th, Vecchio, Vedovato, Veitch, Veitch, Venkateswara,
  Venugopalan, Verkindt, Vetrano, Vicer{\'{e}}, Viets, Vinciguerra, Vine,
  Vinet, Vitale, Vo, Vocca, Vorvick, Voss, Vousden, Vyatchanin, Wade, Wade,
  Wade, Wakamatsu, Walker, Wallace, Walsh, Wang, Wang, Wang, Wang, Ward,
  Warner, Was, Watchi, Weaver, Wei, Weinert, Weinstein, Weiss, Wen, We{\ss}els,
  Westphal, Wette, Whelan, Whiting, Whittle, Williams, Williams, Williamson,
  Willis, Willke, Wimmer, Winkler, Wipf, Wittel, Woan, Woehler, Worden, Wright,
  Wu, Wu, Yam, Yamamoto, Yamamoto, Yamamoto, Yancey, Yano, Yap, Yokoyama,
  Yokozawa, Yoon, Yu, Yu, Yuzurihara, Yvert, Zadro{\.{z}}ny, Zangrando,
  Zanolin, Zeidler, Zendri, Zevin, Zhang, Zhang, Zhang, Zhang, Zhao, Zhou,
  Zhou, Zhu, Zhu, Zucker, \& Zweizig}]{TheLIGOScientificCollaboration2019}
---. 2019, Physical Review X, 9, 31040

\bibitem[{Alfaro \& Rom{\'{a}}n-Z{\'{u}}{\~{n}}iga(2018)}]{Alfaro2018}
Alfaro, E.~J., \& Rom{\'{a}}n-Z{\'{u}}{\~{n}}iga, C.~G. 2018, Monthly Notices
  of the Royal Astronomical Society: Letters, 478, L110

\bibitem[{Amaro-Seoane \& Chen(2016)}]{Amaro-Seoane2016}
Amaro-Seoane, P., \& Chen, X. 2016, Monthly Notices of the Royal Astronomical
  Society, 458, 3075

\bibitem[{Andrews {et~al.}(2015)Andrews, Farr, Kalogera, \&
  Willems}]{Andrews2015}
Andrews, J.~J., Farr, W.~M., Kalogera, V., \& Willems, B. 2015, Astrophysical
  Journal, 801, 32

\bibitem[{Andrews \& Zezas(2019)}]{Andrews2019}
Andrews, J.~J., \& Zezas, A. 2019, Monthly Notices of the Royal Astronomical
  Society, 486, 3213

\bibitem[{Antognini {et~al.}(2014)Antognini, Shappee, Thompson, \&
  Amaro-seoane}]{Antognini2014}
Antognini, J.~M., Shappee, B.~J., Thompson, T.~A., \& Amaro-seoane, P. 2014,
  Monthly Notices of the Royal Astronomical Society, 439, 1079

\bibitem[{Arnould {et~al.}(2007)Arnould, Goriely, \& Takahashi}]{Arnould2007}
Arnould, M., Goriely, S., \& Takahashi, K. 2007, Physics Reports, 450, 97

\bibitem[{Askar {et~al.}(2018)Askar, Sedda, \& Giersz}]{Askar2018}
Askar, A., Sedda, M.~A., \& Giersz, M. 2018, Monthly Notices of the Royal
  Astronomical Society, 478, 1844

\bibitem[{Askar {et~al.}(2017)Askar, Szkudlarek, Gondek-Rosi{\'{n}}ska, Giersz,
  \& Bulik}]{Askar2017}
Askar, A., Szkudlarek, M., Gondek-Rosi{\'{n}}ska, D., Giersz, M., \& Bulik, T.
  2017, Monthly Notices of the Royal Astronomical Society, 464, 36

\bibitem[{Banerjee(2017)}]{Banerjee2017}
Banerjee, S. 2017, Monthly Notices of the Royal Astronomical Society, 467, 524

\bibitem[{Bastian {et~al.}(2013)Bastian, Cabrera-Ziri, Davies, \&
  Larsen}]{Bastian2013}
Bastian, N., Cabrera-Ziri, I., Davies, B., \& Larsen, S.~S. 2013, Monthly
  Notices of the Royal Astronomical Society, 436, 2852

\bibitem[{Bastian {et~al.}(2014)Bastian, Hollyhead, \&
  Cabrera-Ziri}]{Bastian2014}
Bastian, N., Hollyhead, K., \& Cabrera-Ziri, I. 2014, Monthly Notices of the
  Royal Astronomical Society, 445, 378

\bibitem[{Bastian \& Lardo(2018)}]{Bastian2018}
Bastian, N., \& Lardo, C. 2018, Annual Review of Astronomy and Astrophysics,
  56, 83

\bibitem[{Bekki \& Tsujimoto(2017)}]{Bekki2017}
Bekki, K., \& Tsujimoto, T. 2017, The Astrophysical Journal, 844, 34

\bibitem[{Belczy{\'{n}}ski {et~al.}(2002)Belczy{\'{n}}ski, Kalogera, \&
  Bulik}]{Belczynski2002}
Belczy{\'{n}}ski, K., Kalogera, V., \& Bulik, T. 2002, The Astrophysical
  Journal, 527, 407

\bibitem[{Belczynski {et~al.}(2008)Belczynski, Kalogera, Rasio, Taam, Zezas,
  Bulik, Maccarone, \& Ivanova}]{Belczynski2008}
Belczynski, K., Kalogera, V., Rasio, F.~A., {et~al.} 2008, The Astrophysical
  Journal Supplement Series, 174, 223

\bibitem[{Beniamini \& Piran(2016)}]{Beniamini2016}
Beniamini, P., \& Piran, T. 2016, Monthly Notices of the Royal Astronomical
  Society, 456, 4089

\bibitem[{Blaauw(1961)}]{Blaauw1961}
Blaauw, A. 1961, Bulletin of the Astronomical Institutes of the Netherlands,
  15, 265

\bibitem[{Bonetti {et~al.}(2019)Bonetti, Perego, Dotti, \&
  Cescutti}]{Bonetti2019}
Bonetti, M., Perego, A., Dotti, M., \& Cescutti, G. 2019, Monthly Notices of
  the Royal Astronomical Society, 490, 296

\bibitem[{Boyles {et~al.}(2011)Boyles, Lorimer, Turk, Mnatsakanov, Lynch,
  Ransom, Freire, \& Belczynski}]{Boyles2011}
Boyles, J., Lorimer, D.~R., Turk, P.~J., {et~al.} 2011, Astrophysical Journal,
  742, 51

\bibitem[{Bray \& Eldridge(2018)}]{Bray2018}
Bray, J.~C., \& Eldridge, J.~J. 2018, Monthly Notices of the Royal Astronomical
  Society, 480, 5657

\bibitem[{Breivik {et~al.}(2019)Breivik, Coughlin, Zevin, Rodriguez, Kremer,
  Ye, Andrews, Kurkowski, Larson, \& Rasio}]{COSMIC}
Breivik, K., Coughlin, S., Zevin, M., {et~al.} 2019, arXiv:1911.00903

\bibitem[{Brisken {et~al.}(2002)Brisken, Benson, \& Goss}]{Brisken2002}
Brisken, W.~F., Benson, J.~M., \& Goss, W.~M. 2002, The Astrophysical Journal,
  571, 906

\bibitem[{Cabrera-Ziri {et~al.}(2014)Cabrera-Ziri, Bastian, Davies, Magris,
  Bruzual, \& Schweizer}]{Cabrera-Ziri2014}
Cabrera-Ziri, I., Bastian, N., Davies, B., {et~al.} 2014, Monthly Notices of
  the Royal Astronomical Society, 441, 2754

\bibitem[{Cabrera-Ziri {et~al.}(2016)Cabrera-Ziri, Bastian, Hilker, Davies,
  Schweizer, Kruijssen, Mej{\'{i}}a-Narv{\'{a}}ez, Niederhofer, Brandt,
  Rejkuba, Bruzual, \& Magris}]{Cabrera-Ziri2016}
Cabrera-Ziri, I., Bastian, N., Hilker, M., {et~al.} 2016, Monthly Notices of
  the Royal Astronomical Society, 457, 809

\bibitem[{Carretta {et~al.}(2010)Carretta, Gratton, Lucatello, Bragaglia,
  Catanzaro, Leone, Momany, D'Orazi, Cassisi, D'Antona, \&
  Ortolani}]{Carretta2010a}
Carretta, E., Gratton, R.~G., Lucatello, S., {et~al.} 2010, Astrophysical
  Journal Letters, 722, 1

\bibitem[{Cescutti {et~al.}(2015)Cescutti, Romano, Matteucci, Chiappini, \&
  Hirschi}]{Cescutti2015}
Cescutti, G., Romano, D., Matteucci, F., Chiappini, C., \& Hirschi, R. 2015,
  Astronomy {\&} Astrophysics, 577, A139

\bibitem[{Chatterjee {et~al.}(2010)Chatterjee, Fregeau, Umbreit, \&
  Rasio}]{Chatterjee2010}
Chatterjee, S., Fregeau, J.~M., Umbreit, S., \& Rasio, F.~A. 2010,
  Astrophysical Journal, 719, 915

\bibitem[{Chatterjee {et~al.}(2013)Chatterjee, Rasio, Sills, \&
  Glebbeek}]{Chatterjee2013}
Chatterjee, S., Rasio, F.~A., Sills, A., \& Glebbeek, E. 2013, Astrophysical
  Journal, 777, 106

\bibitem[{Chornock {et~al.}(2017)Chornock, Berger, Kasen, Cowperthwaite,
  Nicholl, Villar, Alexander, Blanchard, Eftekhari, Fong, Margutti, Williams,
  Annis, Brout, Brown, Chen, Drout, Foley, Frieman, Fryer, Holz, Matheson,
  Metzger, Quataert, Rest, Sako, Scolnic, Smith, \&
  Soares-Santos}]{Chornock2017}
Chornock, R., Berger, E., Kasen, D., {et~al.} 2017, The Astrophysical Journal
  Letters, 848, L19

\bibitem[{Chruslinska {et~al.}(2018)Chruslinska, Belczynski, Klencki, \&
  Benacquista}]{Chruslinska2018}
Chruslinska, M., Belczynski, K., Klencki, J., \& Benacquista, M. 2018, Monthly
  Notices of the Royal Astronomical Society, 474, 2937

\bibitem[{Chruslinska {et~al.}(2019)Chruslinska, Nelemans, \&
  Belczynski}]{Chruslinska2019}
Chruslinska, M., Nelemans, G., \& Belczynski, K. 2019, Monthly Notices of the
  Royal Astronomical Society, 482, 5012

\bibitem[{Ciolfi {et~al.}(2017)Ciolfi, Kastaun, Giacomazzo, Endrizzi, Siegel,
  \& Perna}]{Ciolfi2017}
Ciolfi, R., Kastaun, W., Giacomazzo, B., {et~al.} 2017, Physical Review D, 95,
  063016

\bibitem[{Claeys {et~al.}(2014)Claeys, Pols, Izzard, Vink, \&
  Verbunt}]{Claeys2014}
Claeys, J. S.~W., Pols, O.~R., Izzard, R.~G., Vink, J., \& Verbunt, F. W.~M.
  2014, Astronomy {\&} Astrophysics, 563, A83

\bibitem[{Cohen(2011)}]{Cohen2011}
Cohen, J.~G. 2011, Astrophysical Journal Letters, 740, L38

\bibitem[{C{\^{o}}t{\'{e}} {et~al.}(2017)C{\^{o}}t{\'{e}}, Fryer, Belczynski,
  Korobkin, Chru{\'{s}}li{\'{n}}ska, Vassh, Mumpower, Lippuner, Sprouse,
  Surman, \& Wollaeger}]{Cote2017}
C{\^{o}}t{\'{e}}, B., Fryer, C.~L., Belczynski, K., {et~al.} 2017, The
  Astrophysical Journal, 855, 99

\bibitem[{C{\^{o}}t{\'{e}} {et~al.}(2019)C{\^{o}}t{\'{e}}, Eichler, Arcones,
  Hansen, Simonetti, Frebel, Fryer, Pignatari, Reichert, Belczynski, \&
  Matteucci}]{Cote2019}
C{\^{o}}t{\'{e}}, B., Eichler, M., Arcones, A., {et~al.} 2019, The
  Astrophysical Journal, 875, 106

\bibitem[{Cowan {et~al.}(2019)Cowan, Sneden, Lawler, Aprahamian, Wiescher,
  Langanke, Mart{\'{i}}nez-Pinedo, \& Thielemann}]{Cowan2019}
Cowan, J.~J., Sneden, C., Lawler, J.~E., {et~al.} 2019, arXiv:1901.01410

\bibitem[{Cowperthwaite {et~al.}(2017)Cowperthwaite, Berger, Villar, Metzger,
  Nicholl, Chornock, Blanchard, Fong, Margutti, Soares-Santos, Alexander,
  Allam, Annis, Brout, Brown, Butler, Chen, Diehl, Doctor, Drout, Eftekhari,
  Farr, Finley, Foley, Frieman, Fryer, Garc{\'{i}}a-Bellido, Gill, Guillochon,
  Herner, Holz, Kasen, Kessler, Marriner, Matheson, Neilsen, Quataert, Palmese,
  Rest, Sako, Scolnic, Smith, Tucker, Williams, Balbinot, Carlin, Cook, Durret,
  Li, Lopes, Louren{\c{c}}o, Marshall, Medina, Muir, Mu{\~{n}}oz, Sauseda,
  Schlegel, Secco, Vivas, Wester, Zenteno, Zhang, Abbott, Banerji, Bechtol,
  Benoit-L{\'{e}}vy, Bertin, Buckley-Geer, Burke, Capozzi, Rosell, Kind,
  Castander, Crocce, Cunha, D'Andrea, da~Costa, Davis, DePoy, Desai, Dietrich,
  Drlica-Wagner, Eifler, Evrard, Fernandez, Flaugher, Fosalba, Gaztanaga,
  Gerdes, Giannantonio, Goldstein, Gruen, Gruendl, Gutierrez, Honscheid, Jain,
  James, Jeltema, Johnson, Johnson, Kent, Krause, Kron, Kuehn, Kuropatkin,
  Lahav, Lima, Lin, Maia, March, Martini, McMahon, Menanteau, Miller, Miquel,
  Mohr, Neilsen, Nichol, Ogando, Plazas, Roe, Romer, Roodman, Rykoff, Sanchez,
  Scarpine, Schindler, Schubnell, Sevilla-Noarbe, Smith, Smith, Sobreira,
  Suchyta, Swanson, Tarle, Thomas, Thomas, Troxel, Vikram, Walker, Wechsler,
  Weller, Yanny, \& Zuntz}]{Cowperthwaite2017}
Cowperthwaite, P.~S., Berger, E., Villar, V.~A., {et~al.} 2017, The
  Astrophysical Journal Letters, 848, L17

\bibitem[{Davies {et~al.}(1994)Davies, Benz, Piran, \& Thielemann}]{Davies1994}
Davies, M.~B., Benz, W., Piran, T., \& Thielemann, F.~K. 1994, The
  Astrophysical Journal, 431, 742

\bibitem[{Delgado \& Thomas(1981)}]{Delgado1981}
Delgado, A., \& Thomas, H.-C. 1981, Astronomy {\&} Astrophysics, 96, 142

\bibitem[{Dessart {et~al.}(2009)Dessart, Ott, Burrows, Rosswog, \&
  Livne}]{Dessart2009}
Dessart, L., Ott, C.~D., Burrows, A., Rosswog, S., \& Livne, E. 2009,
  Astrophysical Journal, 690, 1681

\bibitem[{Dewi {et~al.}(2002)Dewi, Pols, Savonije, \& {Van Den
  Heuvel}}]{Dewi2002}
Dewi, J.~D., Pols, O.~R., Savonije, G.~J., \& {Van Den Heuvel}, E.~P. 2002,
  Monthly Notices of the Royal Astronomical Society, 331, 1027

\bibitem[{Dominik {et~al.}(2012)Dominik, Belczynski, Fryer, Holz, Berti, Bulik,
  Mandel, \& O'Shaughnessy}]{Dominik2012}
Dominik, M., Belczynski, K., Fryer, C., {et~al.} 2012, Astrophysical Journal,
  759, 52

\bibitem[{Eichler {et~al.}(1989)Eichler, Liviot, Piran, \&
  Schramm}]{Eichler1989}
Eichler, D., Liviot, M., Piran, T., \& Schramm, D.~N. 1989, Nature, 340, 126

\bibitem[{Fern{\'{a}}ndez \& Metzger(2013)}]{Fernandez2013}
Fern{\'{a}}ndez, R., \& Metzger, B.~D. 2013, Monthly Notices of the Royal
  Astronomical Society, 435, 502

\bibitem[{Fragione {et~al.}(2019)Fragione, Leigh, \& Perna}]{Fragione2018b}
Fragione, G., Leigh, N. W.~C., \& Perna, R. 2019, Monthly Notices of the Royal
  Astronomical Society, 488, 2825

\bibitem[{Fregeau {et~al.}(2004)Fregeau, Cheung, Zwart, \& Rasio}]{Fregeau2004}
Fregeau, J.~M., Cheung, P., Zwart, S. F.~P., \& Rasio, F.~A. 2004, Monthly
  Notices of the Royal Astronomical Society, 352, 1

\bibitem[{Fregeau {et~al.}(2003)Fregeau, Gurkan, Joshi, \& Rasio}]{Fregeau2003}
Fregeau, J.~M., Gurkan, M.~A., Joshi, K.~J., \& Rasio, F.~A. 2003, The
  Astronomical Journal, 593, 772

\bibitem[{Fregeau \& Rasio(2007)}]{Fregeau2007}
Fregeau, J.~M., \& Rasio, F.~A. 2007, The Astrophysical Journal, 658, 1047

\bibitem[{Freiburghaus {et~al.}(1999)Freiburghaus, Rosswog, \&
  Thielemann}]{Freiburghaus1999}
Freiburghaus, C., Rosswog, S., \& Thielemann, F.-K. 1999, The Astrophysical
  Journal, 525, L121

\bibitem[{Fryer \& Kalogera(1997)}]{Fryer1997}
Fryer, C., \& Kalogera, V. 1997, The Astrophysical Journal, 489, 244

\bibitem[{Fryer {et~al.}(2012)Fryer, Belczynski, Wiktorowicz, Dominik,
  Kalogera, \& Holz}]{Fryer2012}
Fryer, C.~L., Belczynski, K., Wiktorowicz, G., {et~al.} 2012, The Astrophysical
  Journal, 749, 14

\bibitem[{Giacobbo {et~al.}(2018)Giacobbo, Mapelli, \& Spera}]{Giacobbo2018a}
Giacobbo, N., Mapelli, M., \& Spera, M. 2018, Monthly Notices of the Royal
  Astronomical Society, 474, 2959

\bibitem[{Giesler {et~al.}(2018)Giesler, Clausen, \& Ott}]{Giesler2018}
Giesler, M., Clausen, D., \& Ott, C.~D. 2018, Monthly Notices of the Royal
  Astronomical Society, 477, 1853

\bibitem[{Goldstein {et~al.}(2015)Goldstein, Connaughton, Briggs, \&
  Burns}]{Goldstein2015}
Goldstein, A., Connaughton, V., Briggs, M.~S., \& Burns, E. 2015, The
  Astrophysical Journal, 818, 18

\bibitem[{Gratton {et~al.}(2012)Gratton, Carretta, \& Bragaglia}]{Gratton2012}
Gratton, R.~G., Carretta, E., \& Bragaglia, A. 2012, Astronomy and Astrophysics
  Review, 20, 49

\bibitem[{Grindlay {et~al.}(2006)Grindlay, Zwart, \& McMillan}]{Grindlay2006}
Grindlay, J., Zwart, S. F.~P., \& McMillan, S. 2006, Nature Physics, 2, 116

\bibitem[{Gurkan {et~al.}(2004)Gurkan, Freitag, \& Rasio}]{Gurkan2004}
Gurkan, M.~A., Freitag, M., \& Rasio, F.~A. 2004, The Astrophysical Journal,
  604, 632

\bibitem[{Halevi \& M{\"{o}}sta(2018)}]{Halevi2018}
Halevi, G., \& M{\"{o}}sta, P. 2018, Monthly Notices of the Royal Astronomical
  Society, 477, 2366

\bibitem[{Hansen {et~al.}(2017)Hansen, Simon, Marshall, Li, Carollo, DePoy,
  Nagasawa, Bernstein, Drlica-Wagner, Abdalla, Allam, Annis, Bechtol,
  Benoit-L{\'{e}}vy, Brooks, Buckley-Geer, Rosell, Kind, Carretero, Cunha,
  da~Costa, Desai, Eifler, Neto, Flaugher, Frieman, Garc{\'{i}}a-Bellido,
  Gaztanaga, Gerdes, Gruen, Gruendl, Gschwend, Gutierrez, James, Krause, Kuehn,
  Kuropatkin, Lahav, Miquel, Plazas, Romer, Sanchez, Santiago, Scarpine, Smith,
  Soares-Santos, Sobreira, Suchyta, Swanson, Tarle, \& Walker}]{Hansen2017}
Hansen, T.~T., Simon, J.~D., Marshall, J.~L., {et~al.} 2017, The Astronomical
  Journal, 838, 44

\bibitem[{Hansen {et~al.}(2018)Hansen, Holmbeck, Beers, Placco, Roederer,
  Frebel, Sakari, Simon, \& Thompson}]{Hansen2018}
Hansen, T.~T., Holmbeck, E.~M., Beers, T.~C., {et~al.} 2018, The Astrophysical
  Journal, 858, 92

\bibitem[{H{\'{e}}non(1971{\natexlab{a}})}]{Henon1971a}
H{\'{e}}non, M. 1971{\natexlab{a}}, Astrophysics and Space Science, 13, 284

\bibitem[{H{\'{e}}non(1971{\natexlab{b}})}]{Henon1971b}
---. 1971{\natexlab{b}}, Astrophysics and Space Science, 14, 151

\bibitem[{Hobbs {et~al.}(2005)Hobbs, Lorimer, Lyne, \& Kramer}]{Hobbs2005}
Hobbs, G., Lorimer, D.~R., Lyne, A.~G., \& Kramer, M. 2005, Monthly Notices of
  the Royal Astronomical Society, 360, 974

\bibitem[{Hollyhead {et~al.}(2015)Hollyhead, Bastian, Adamo, Silva-Villa, Dale,
  Ryon, \& Gazak}]{Hollyhead2015}
Hollyhead, K., Bastian, N., Adamo, A., {et~al.} 2015, Monthly Notices of the
  Royal Astronomical Society, 449, 1106

\bibitem[{Hong {et~al.}(2018)Hong, Vesperini, Askar, Giersz, Szkudlarek, \&
  Bulik}]{Hong2018}
Hong, J., Vesperini, E., Askar, A., {et~al.} 2018, Monthly Notices of the Royal
  Astronomical Society, 480, 5645

\bibitem[{Hotokezaka {et~al.}(2018)Hotokezaka, Beniamini, \&
  Piran}]{Hotokezaka2018}
Hotokezaka, K., Beniamini, P., \& Piran, T. 2018, International Journal of
  Modern Physics D, 27, 1842005.
\newblock
  \url{https://www.worldscientific.com/doi/abs/10.1142/S0218271818420051}

\bibitem[{Hotokezaka {et~al.}(2013)Hotokezaka, Kyutoku, Tanaka, Kiuchi,
  Sekiguchi, Shibata, \& Wanajo}]{Hotokezaka2013}
Hotokezaka, K., Kyutoku, K., Tanaka, M., {et~al.} 2013, Astrophysical Journal
  Letters, 778, L16

\bibitem[{Hotokezaka {et~al.}(2015)Hotokezaka, Piran, \& Paul}]{Hotokezaka2015}
Hotokezaka, K., Piran, T., \& Paul, M. 2015, Nature Physics, 11, 1042

\bibitem[{Hunter(2007)}]{matplotlib}
Hunter, J.~D. 2007, Computing in Science and Engineering, 9, 99

\bibitem[{Hurley {et~al.}(2000)Hurley, Pols, \& Tout}]{Hurley2000}
Hurley, J.~R., Pols, O.~R., \& Tout, C.~A. 2000, Monthly Notices of the Royal
  Astronomical Society, 315, 543

\bibitem[{Hurley {et~al.}(2002)Hurley, Tout, \& Pols}]{Hurley2002}
Hurley, J.~R., Tout, C.~A., \& Pols, O.~R. 2002, Monthly Notices of the Royal
  Astronomical Society, 329, 897

\bibitem[{Ito {et~al.}(2009)Ito, Aoki, Honda, \& Beers}]{Ito2009}
Ito, H., Aoki, W., Honda, S., \& Beers, T.~C. 2009, Astrophysical Journal, 698,
  37

\bibitem[{Ivanova {et~al.}(2005)Ivanova, Belczynski, Fregeau, \&
  Rasio}]{Ivanova2005}
Ivanova, N., Belczynski, K., Fregeau, J.~M., \& Rasio, F.~A. 2005, Monthly
  Notices of the Royal Astronomical Society, 358, 572

\bibitem[{Ivanova {et~al.}(2003)Ivanova, Belczynski, Kalogera, Rasio, \&
  Taam}]{Ivanova2003}
Ivanova, N., Belczynski, K., Kalogera, V., Rasio, F.~A., \& Taam, R.~E. 2003,
  The Astrophysical Journal, 592, 475

\bibitem[{Ivanova {et~al.}(2008)Ivanova, Heinke, Rasio, Belczynski, \&
  Fregeau}]{Ivanova2008}
Ivanova, N., Heinke, C.~O., Rasio, F.~A., Belczynski, K., \& Fregeau, J.~M.
  2008, Monthly Notices of the Royal Astronomical Society, 386, 553

\bibitem[{Ivanova \& Taam(2004)}]{Ivanova2004}
Ivanova, N., \& Taam, R.~E. 2004, The Astronomical Journal, 601, 1058

\bibitem[{Ji {et~al.}(2016)Ji, Frebel, Chiti, \& Simon}]{Ji2016}
Ji, A.~P., Frebel, A., Chiti, A., \& Simon, J.~D. 2016, Nature, 531, 610

\bibitem[{Ji {et~al.}(2019)Ji, Simon, Frebel, Venn, \& Hansen}]{Ji2019a}
Ji, A.~P., Simon, J.~D., Frebel, A., Venn, K.~A., \& Hansen, T.~T. 2019, The
  Astrophysical Journal, 870, 83

\bibitem[{Jones {et~al.}(2001)Jones, Oliphant, Peterson, \& Others}]{scipy}
Jones, E., Oliphant, T.~E., Peterson, P., \& Others. 2001, {SciPy: Open source
  scientific tools for Python}, , .
\newblock \url{http://www.scipy.org/}

\bibitem[{Joshi {et~al.}(2001)Joshi, Nave, \& Rasio}]{Joshi2001}
Joshi, K.~J., Nave, C.~P., \& Rasio, F.~A. 2001, The Astrophysical Journal,
  550, 691

\bibitem[{Joshi {et~al.}(2000)Joshi, Rasio, \& {Portegies Zwart}}]{Joshi2000}
Joshi, K.~J., Rasio, F.~A., \& {Portegies Zwart}, S.~P. 2000, ApJ, 540, 969

\bibitem[{Just {et~al.}(2015)Just, Bauswein, {Ardevol Pulpillo}, Goriely, \&
  Janka}]{Just2015}
Just, O., Bauswein, A., {Ardevol Pulpillo}, R., Goriely, S., \& Janka, H.~T.
  2015, Monthly Notices of the Royal Astronomical Society, 448, 541

\bibitem[{Kajino {et~al.}(2019)Kajino, Aoki, Balantekin, Diehl, Famiano, \&
  Mathews}]{Kajino2019}
Kajino, T., Aoki, W., Balantekin, A.~B., {et~al.} 2019, Progress in Particle
  and Nuclear Physics, 107, 109.
\newblock \url{https://doi.org/10.1016/j.ppnp.2019.02.008}

\bibitem[{Kalogera(1996)}]{Kalogera1996}
Kalogera, V. 1996, Astrophysical Journal, 471, 352

\bibitem[{Kasen {et~al.}(2017)Kasen, Metzger, Barnes, Quataert, \&
  Ramirez-Ruiz}]{Kasen2017}
Kasen, D., Metzger, B., Barnes, J., Quataert, E., \& Ramirez-Ruiz, E. 2017,
  Nature, 551, 80

\bibitem[{Kimmig {et~al.}(2015)Kimmig, Seth, Ivans, Strader, Caldwell,
  Anderton, \& Gregersen}]{Kimmig2015}
Kimmig, B., Seth, A., Ivans, I.~I., {et~al.} 2015, Astronomical Journal, 149,
  53

\bibitem[{Komiya \& Shigeyama(2016)}]{Komiya2016}
Komiya, Y., \& Shigeyama, T. 2016, The Astrophysical Journal, 830, 1

\bibitem[{Kremer {et~al.}(2018)Kremer, Ye, Chatterjee, Rodriguez, \&
  Rasio}]{Kremer2018}
Kremer, K., Ye, C.~S., Chatterjee, S., Rodriguez, C.~L., \& Rasio, F.~A. 2018,
  The Astrophysical Journal Letters, 855, L15

\bibitem[{Kremer {et~al.}(2019)Kremer, Rodriguez, Amaro-Seoane, Breivik,
  Chatterjee, Katz, Larson, Rasio, Samsing, Ye, \& Zevin}]{Kremer2019}
Kremer, K., Rodriguez, C.~L., Amaro-Seoane, P., {et~al.} 2019, Physical Review
  D, 99, 63003

\bibitem[{Kroupa(2001)}]{Kroupa2001}
Kroupa, P. 2001, Monthly Notices of the Royal Astronomical Society, 322, 231

\bibitem[{Kulkarni {et~al.}(1993)Kulkarni, Hut, \& McMillan}]{Kulkarni1993}
Kulkarni, S.~R., Hut, P., \& McMillan, S.~J. 1993, Nature, 364, 421

\bibitem[{Lattimer \& Schramm(1974)}]{Lattimer1974}
Lattimer, J.~M., \& Schramm, D.~N. 1974, The Astronomical Journal, 192, L145

\bibitem[{Lattimer \& Schramm(1976)}]{Lattimer1976}
---. 1976, The Astrophysical Journal, 210, 549

\bibitem[{Lattimer \& Yahil(1989)}]{Lattimer1989}
Lattimer, J.~M., \& Yahil, A. 1989, The Astronomical Journal, 340, 426

\bibitem[{Lee {et~al.}(2010)Lee, Ramirez-Ruiz, \& {De Van Ven}}]{Lee2010}
Lee, W.~H., Ramirez-Ruiz, E., \& {De Van Ven}, G. 2010, Astrophysical Journal,
  720, 953

\bibitem[{Leonard {et~al.}(1992)Leonard, Richer, \& Fahlman}]{Leonard1992}
Leonard, P. J.~T., Richer, H.~B., \& Fahlman, G.~G. 1992, The Astrophysical
  Journal, 104, 2104

\bibitem[{Leroy {et~al.}(2018)Leroy, Bolatto, Ostriker, Walter, Gorski,
  Ginsburg, Krieger, Meier, Mills, Ott, Rosolowsky, Thompson, Veilleux, \&
  Zschaechner}]{Leroy2018}
Leroy, A.~K., Bolatto, A.~D., Ostriker, E.~C., {et~al.} 2018, The Astrophysical
  Journal, 869, 126

\bibitem[{Li {et~al.}(2011)Li, Chornock, Leaman, Filippenko, Poznanski, Wang,
  Ganeshalingam, \& Mannucci}]{Li2011}
Li, W., Chornock, R., Leaman, J., {et~al.} 2011, Monthly Notices of the Royal
  Astronomical Society, 412, 1473

\bibitem[{Liang {et~al.}(2007)Liang, Zhang, Virgili, \& Dai}]{Liang2007}
Liang, E., Zhang, B., Virgili, F., \& Dai, Z.~G. 2007, The Astrophysical
  Journal, 662, 1111

\bibitem[{Lovegrove \& Woosley(2013)}]{Lovegrove2013}
Lovegrove, E., \& Woosley, S.~E. 2013, Astrophysical Journal, 769, 109

\bibitem[{MacFadyen \& Woosley(1999)}]{MacFadyen1999}
MacFadyen, A.~I., \& Woosley, S.~E. 1999, The Astrophysical Journal, 524, 262

\bibitem[{Macias \& Ramirez-Ruiz(2019)}]{Macias2019}
Macias, P., \& Ramirez-Ruiz, E. 2019, The Astrophysical Journal, 877, L24.
\newblock \url{http://dx.doi.org/10.3847/2041-8213/ab2049}

\bibitem[{Mackey {et~al.}(2007)Mackey, Wilkinson, Davies, \&
  Gilmore}]{Mackey2007}
Mackey, A.~D., Wilkinson, M.~I., Davies, M.~B., \& Gilmore, G.~F. 2007, Monthly
  Notices of the Royal Astronomical Society: Letters, 379, 40

\bibitem[{Mackey {et~al.}(2008)Mackey, Wilkinson, Davies, \&
  Gilmore}]{Mackey2008}
---. 2008, Monthly Notices of the Royal Astronomical Society: Letters, 386, 65

\bibitem[{Madau(2017)}]{Madau2017}
Madau, P. 2017, The Astrophysical Journal, 851, 50

\bibitem[{Marino {et~al.}(2009)Marino, Milone, Piotto, Villanova, Bedin,
  Bellini, \& Renzini}]{Marino2009}
Marino, A.~F., Milone, A.~P., Piotto, G., {et~al.} 2009, Astronomy and
  Astrophysics, 505, 1099

\bibitem[{Mart{\'{i}}nez-Pinedo {et~al.}(2012)Mart{\'{i}}nez-Pinedo, Fischer,
  Lohs, \& Huther}]{Martinez-Pinedo2012}
Mart{\'{i}}nez-Pinedo, G., Fischer, T., Lohs, A., \& Huther, L. 2012, Physical
  Review Letters, 109, 251104

\bibitem[{Martocchia {et~al.}(2018)Martocchia, Niederhofer, Dalessandro,
  Bastian, Kacharov, Usher, Cabrera-Ziri, Lardo, Cassisi, Geisler, Hilker,
  Hollyhead, Kozhurina-Platais, Larsen, Mackey, Mucciarelli, Platais, \&
  Salaris}]{Martocchia2018}
Martocchia, S., Niederhofer, F., Dalessandro, E., {et~al.} 2018, Monthly
  Notices of the Royal Astronomical Society, 4705, 4696

\bibitem[{McKinney(2010)}]{pandas}
McKinney, W. 2010, in Proceedings of the 9th Python in Science Conference, ed.
  S.~van~der Walt \& J.~Millman, 51--56.
\newblock \url{http://pandas.sourceforge.net}

\bibitem[{Metzger {et~al.}(2008)Metzger, Thompson, \& Quataert}]{Metzger2008}
Metzger, B.~D., Thompson, T.~A., \& Quataert, E. 2008, The Astrophysical
  Journal, 676, 1130

\bibitem[{Meyer(1989)}]{Meyer1989}
Meyer, B.~S. 1989, The Astrophysical Journal, 343, 254

\bibitem[{Meylan \& Heggie(1997)}]{Meylan1997}
Meylan, G., \& Heggie, D.~C. 1997, Astronomy and Astrophysics Review, 8, 1

\bibitem[{Miyaji {et~al.}(1980)Miyaji, Nomoto, Yokoi, \& Sugimoto}]{Miyagi1980}
Miyaji, S., Nomoto, K., Yokoi, K., \& Sugimoto, D. 1980, Publications of the
  Astronomical Society of Japan, 32, 303

\bibitem[{Moe \& {Di Stefano}(2017)}]{Moe2017}
Moe, M., \& {Di Stefano}, R. 2017, The Astrophysical Journal Supplement Series,
  230, 55

\bibitem[{Morscher {et~al.}(2015)Morscher, Pattabiraman, Rodriguez, Rasio, \&
  Umbreit}]{Morscher2015}
Morscher, M., Pattabiraman, B., Rodriguez, C., Rasio, F.~A., \& Umbreit, S.
  2015, Astrophysical Journal Letters, 800, 9

\bibitem[{M{\"{o}}sta {et~al.}(2018)M{\"{o}}sta, Roberts, Halevi, Ott,
  Lippuner, Haas, \& Schnetter}]{Moesta2018}
M{\"{o}}sta, P., Roberts, L.~F., Halevi, G., {et~al.} 2018, The Astrophysical
  Journal, 864, 171

\bibitem[{M{\"{o}}sta {et~al.}(2014)M{\"{o}}sta, Richers, Ott, Haas, Piro,
  Boydstun, Abdikamalov, Reisswig, \& Schnetter}]{Moesta2014}
M{\"{o}}sta, P., Richers, S., Ott, C.~D., {et~al.} 2014, Astrophysical Journal
  Letters, 785, L29

\bibitem[{Neijssel {et~al.}(2019)Neijssel, Vigna-G{\'{o}}mez, Stevenson,
  Barrett, Gaebel, Broekgaarden, de~Mink, Sz{\'{e}}csi, Vinciguerra, \&
  Mandel}]{Neijssel2019}
Neijssel, C.~J., Vigna-G{\'{o}}mez, A., Stevenson, S., {et~al.} 2019,
  arXiv:1906.08136

\bibitem[{Nomoto(1984)}]{Nomoto1984}
Nomoto, K. 1984, The Astrophysical Journal, 277, 791

\bibitem[{Nomoto(1987)}]{Nomoto1987}
---. 1987, The Astrophysical Journal, 322, 206

\bibitem[{Nomoto \& Kondo(1991)}]{Nomoto1991}
Nomoto, K., \& Kondo, Y. 1991, The Astrophysical Journal, 367, L19

\bibitem[{Oechslin {et~al.}(2007)Oechslin, Janka, \& Marek}]{Oechslin2007}
Oechslin, R., Janka, H.-T., \& Marek, A. 2007, Astronomy {\&} Astrophysics,
  467, 395

\bibitem[{Oliphant(2006)}]{numpy}
Oliphant, T.~E. 2006, {A guide to NumPy} (USA: Trelgol Publishing)

\bibitem[{Parker(2018)}]{Parker2018}
Parker, R.~J. 2018, Monthly Notices of the Royal Astronomical Society, 476, 617

\bibitem[{Pattabiraman {et~al.}(2013)Pattabiraman, Umbreit, Liao, Choudhary,
  Kalogera, Memik, \& Rasio}]{Pattabiraman2013}
Pattabiraman, B., Umbreit, S., Liao, W.~K., {et~al.} 2013, Astrophysical
  Journal, Supplement Series, 204, 16

\bibitem[{P{\'{e}}rez \& Granger(2007)}]{ipython}
P{\'{e}}rez, F., \& Granger, B.~E. 2007, IEEE Journals {\&} Magazines, 9, 21

\bibitem[{Perley {et~al.}(2016)Perley, Kr{\"{u}}hler, Schulze, {de Ugarte
  Postigo}, Hjorth, Berger, Cenko, Chary, Cucchiara, Ellis, Fong, Fynbo,
  Gorosabel, Greiner, Jakobsson, Kim, Laskar, Levan, Micha{\l}owski,
  Milvang-Jensen, Tanvir, Th{\"{o}}ne, \& Wiersema}]{Perley2016}
Perley, D.~A., Kr{\"{u}}hler, T., Schulze, S., {et~al.} 2016, The Astrophysical
  Journal, 817, 7

\bibitem[{Peters(1964)}]{Peters1964}
Peters, P.~C. 1964, Physical Review, 136, 1224

\bibitem[{Pfahl {et~al.}(2002)Pfahl, Rappaport, \& Podsiadlowski}]{Pfahl2002}
Pfahl, E., Rappaport, S., \& Podsiadlowski, P. 2002, The Astrophysical Journal,
  573, 283

\bibitem[{Placco {et~al.}(2015)Placco, Frebel, Lee, Jacobson, Beers, Pena,
  Chan, \& Heger}]{Placco2015}
Placco, V.~M., Frebel, A., Lee, Y.~S., {et~al.} 2015, Astrophysical Journal,
  809, 136

\bibitem[{Plummer(1911)}]{Plummer1911}
Plummer, H.~C. 1911, Monthly Notices of the Royal Astronomical Society, 71, 460

\bibitem[{Podsiadlowski {et~al.}(2004)Podsiadlowski, Langer, Poelarends,
  Rappaport, Heger, \& Pfahl}]{Podsiadlowski2004}
Podsiadlowski, P., Langer, N., Poelarends, a. J.~T., {et~al.} 2004, The
  Astrophysical Journal, 612, 1044

\bibitem[{Price-Whelan {et~al.}(2018)Price-Whelan, Sipőcz, G{\"{u}}nther, Lim,
  Crawford, Conseil, Shupe, Craig, Dencheva, Ginsburg, VanderPlas, Bradley,
  P{\'{e}}rez-Su{\'{a}}rez, de~Val-Borro, Aldcroft, Cruz, Robitaille, Tollerud,
  Ardelean, Babej, Bachetti, Bakanov, Bamford, Barentsen, Barmby, Baumbach,
  Berry, Biscani, Boquien, Bostroem, Bouma, Brammer, Bray, Breytenbach,
  Buddelmeijer, Burke, Calderone, Rodr{\'{i}}guez, Cara, Cardoso, Cheedella,
  Copin, Crichton, D{\'{A}}vella, Deil, Depagne, Dietrich, Donath, Droettboom,
  Earl, Erben, Fabbro, Ferreira, Finethy, Fox, Garrison, Gibbons, Goldstein,
  Gommers, Greco, Greenfield, Groener, Grollier, Hagen, Hirst, Homeier, Horton,
  Hosseinzadeh, Hu, Hunkeler, Ivezi{\'{c}}, Jain, Jenness, Kanarek, Kendrew,
  Kern, Kerzendorf, Khvalko, King, Kirkby, Kulkarni, Kumar, Lee, Lenz,
  Littlefair, Ma, Macleod, Mastropietro, McCully, Montagnac, Morris, Mueller,
  Mumford, Muna, Murphy, Nelson, Nguyen, Ninan, N{\"{o}}the, Ogaz, Oh, Parejko,
  Parley, Pascual, Patil, Patil, Plunkett, Prochaska, Rastogi, Janga, Sabater,
  Sakurikar, Seifert, Sherbert, Sherwood-Taylor, Shih, Sick, Silbiger,
  Singanamalla, Singer, Sladen, Sooley, Sornarajah, Streicher, Teuben, Thomas,
  Tremblay, Turner, Terr{\'{o}}n, van Kerkwijk, de~la Vega, Watkins, Weaver,
  Whitmore, Woillez, \& Zabalza}]{TheAstropyCollaboration2018}
Price-Whelan, A.~M., Sipőcz, B.~M., G{\"{u}}nther, H.~M., {et~al.} 2018, The
  Astronomical Journal, 156, 123

\bibitem[{Qian \& Woosley(1996)}]{Qian1996}
Qian, Y.~Z., \& Woosley, S.~E. 1996, The Astrophysical Journal, 471, 331

\bibitem[{Ramirez-Ruiz {et~al.}(2015)Ramirez-Ruiz, Trenti, MacLeod, Roberts,
  Lee, \& Saladino-Rosas}]{Ramirez-Ruiz2015}
Ramirez-Ruiz, E., Trenti, M., MacLeod, M., {et~al.} 2015, Astrophysical Journal
  Letters, 802, L22

\bibitem[{Raskutti {et~al.}(2016)Raskutti, Ostriker, \& Skinner}]{Raskutti2016}
Raskutti, S., Ostriker, E.~C., \& Skinner, M.~A. 2016, The Astrophysical
  Journal, 829, 130

\bibitem[{Roberts {et~al.}(2012)Roberts, Reddy, \& Shen}]{Roberts2012}
Roberts, L.~F., Reddy, S., \& Shen, G. 2012, Physical Review C - Nuclear
  Physics, 86, 065803

\bibitem[{Robitaille {et~al.}(2013)Robitaille, Tollerud, Greenfield,
  Droettboom, Bray, Aldcroft, Davis, Ginsburg, Price-Whelan, Kerzendorf,
  Conley, Crighton, Barbary, Muna, Ferguson, Grollier, Parikh, Nair,
  G{\"{u}}nther, Deil, Woillez, Conseil, Kramer, Turner, Singer, Fox, Weaver,
  Zabalza, Edwards, Bostroem, Burke, Casey, Crawford, Dencheva, Ely, Jenness,
  Labrie, Lim, Pierfederici, Pontzen, Ptak, Refsdal, Servillat, \&
  Streicher}]{TheAstropyCollaboration2013}
Robitaille, T.~P., Tollerud, E.~J., Greenfield, P., {et~al.} 2013, Astronomy
  {\&} Astrophysics, 558, A33

\bibitem[{Rodriguez {et~al.}(2018)Rodriguez, Amaro-seoane, Chatterjee, Kremer,
  Rasio, Samsing, Ye, \& Zevin}]{Rodriguez2018c}
Rodriguez, C.~L., Amaro-seoane, P., Chatterjee, S., {et~al.} 2018, Physical
  Review D, 98, 123005

\bibitem[{Rodriguez {et~al.}(2016)Rodriguez, Chatterjee, \&
  Rasio}]{Rodriguez2016a}
Rodriguez, C.~L., Chatterjee, S., \& Rasio, F.~A. 2016, Physical Review D, 93,
  084029

\bibitem[{Rodriguez {et~al.}(2015)Rodriguez, Morscher, Pattabiraman,
  Chatterjee, Haster, \& Rasio}]{Rodriguez2015a}
Rodriguez, C.~L., Morscher, M., Pattabiraman, B., {et~al.} 2015, Physical
  Review Letters, 115, 051101

\bibitem[{Roederer(2011)}]{Roederer2011a}
Roederer, I.~U. 2011, Astrophysical Journal Letters, 732, L17

\bibitem[{Roederer {et~al.}(2016)Roederer, Mateo, Bailey, Spencer, Crane, \&
  Shectman}]{Roederer2016}
Roederer, I.~U., Mateo, M., Bailey, J.~I., {et~al.} 2016, Monthly Notices of
  the Royal Astronomical Society, 455, 2417

\bibitem[{Roederer \& Sneden(2011)}]{Roederer2011}
Roederer, I.~U., \& Sneden, C. 2011, Astronomical Journal, 142, 22

\bibitem[{Roederer \& Thompson(2015)}]{Roederer2015}
Roederer, I.~U., \& Thompson, I.~B. 2015, Monthly Notices of the Royal
  Astronomical Society, 449, 3889

\bibitem[{Rosswog \& Liebend(1999)}]{Rosswog1999}
Rosswog, S., \& Liebend, M. 1999, Astronomy {\&} Astrophysics, 341, 499

\bibitem[{Ruffert {et~al.}(1997)Ruffert, Janka, Takahashi, \&
  Schaefer}]{Ruffert1997}
Ruffert, M., Janka, H.~T., Takahashi, K., \& Schaefer, G. 1997, Astronomy {\&}
  Astrophysics, 319, 122

\bibitem[{Safarzadeh {et~al.}(2018)Safarzadeh, Ramirez-Ruiz, Andrews, Fragos,
  Macias, \& Scannapieco}]{Safarzadeh2018}
Safarzadeh, M., Ramirez-Ruiz, E., Andrews, J.~J., {et~al.} 2018, The
  Astrophysical Journal, 872, 105

\bibitem[{Safarzadeh {et~al.}(2019)Safarzadeh, Sarmento, \&
  Scannapieco}]{Safarzadeh2019b}
Safarzadeh, M., Sarmento, R., \& Scannapieco, E. 2019, The Astrophysical
  Journal, 876, 10

\bibitem[{Safarzadeh \& Scannapieco(2017)}]{Safarzadeh2017}
Safarzadeh, M., \& Scannapieco, E. 2017, Monthly Notices of the Royal
  Astronomical Society, 471, 2088

\bibitem[{Saio(2004)}]{Saio2004}
Saio, H. 2004, The Astrophysical Journal, 615, 444

\bibitem[{Saio \& Nomoto(1985)}]{Saio1985}
Saio, H., \& Nomoto, K. 1985, Astronomy and Astrophysics, 150, L21

\bibitem[{Sakari {et~al.}(2018)Sakari, Placco, Farrell, Roederer, Wallerstein,
  Beers, Ezzeddine, Frebel, Hansen, Holmbeck, Sneden, Cowan, Venn, Davis,
  Matijevic, Wyse, Bland-Hawthorn, Chiappini, Freeman, Gibson, Grebel, Helmi,
  Kordopatis, Kunder, Navarro, Reid, Seabroke, Steinmetz, \&
  Watson}]{Sakari2018}
Sakari, C.~M., Placco, V.~M., Farrell, E.~M., {et~al.} 2018, The Astrophysical
  Journal, 868, 110

\bibitem[{Samsing {et~al.}(2014)Samsing, MacLeod, \&
  Ramirez-Ruiz}]{Samsing2014}
Samsing, J., MacLeod, M., \& Ramirez-Ruiz, E. 2014, The Astrophysical Journal,
  784, 71

\bibitem[{Sana {et~al.}(2011)Sana, James, \& Gosset}]{Sana2011}
Sana, H., James, G., \& Gosset, E. 2011, Monthly Notices of the Royal
  Astronomical Society, 416, 817

\bibitem[{Schwab {et~al.}(2010)Schwab, Podsiadlowski, \&
  Rappaport}]{Schwab2010}
Schwab, J., Podsiadlowski, P., \& Rappaport, S. 2010, Astrophysical Journal,
  719, 722

\bibitem[{Shen {et~al.}(2015)Shen, Cooke, Ramirez-Ruiz, Madau, Mayer, \&
  Guedes}]{Shen2015}
Shen, S., Cooke, R.~J., Ramirez-Ruiz, E., {et~al.} 2015, Astrophysical Journal,
  807, 115

\bibitem[{Siegel(2019)}]{Siegel2019a}
Siegel, D.~M. 2019, arXiv:1901.09044

\bibitem[{Siegel {et~al.}(2019)Siegel, Barnes, \& Metzger}]{Siegel2019}
Siegel, D.~M., Barnes, J., \& Metzger, B.~D. 2019, Nature, 569, 241

\bibitem[{Siegel {et~al.}(2014)Siegel, Ciolfi, \& Rezzolla}]{Siegel2014}
Siegel, D.~M., Ciolfi, R., \& Rezzolla, L. 2014, Astrophysical Journal Letters,
  785, 6

\bibitem[{Siegel \& Metzger(2017)}]{Siegel2017}
Siegel, D.~M., \& Metzger, B.~D. 2017, Physical Review Letters, 119, 231102

\bibitem[{Sigurdsson \& Phinney(1993)}]{Sigurdsson1993}
Sigurdsson, S., \& Phinney, E.~S. 1993, The Astrophysical Journal, 39, 631

\bibitem[{Skinner \& Ostriker(2015)}]{Skinner2015}
Skinner, M.~A., \& Ostriker, E.~C. 2015, Astrophysical Journal, 809, 187

\bibitem[{Sneden {et~al.}(1997)Sneden, Shetrone, Kraft, Smith, Langer, \&
  Prosser}]{Sneden1997}
Sneden, C., Shetrone, M.~D., Kraft, R.~P., {et~al.} 1997, The Astronomical
  Journal, 114, 1964

\bibitem[{Sobeck {et~al.}(2011)Sobeck, Kraft, Sneden, Preston, Cowan, Smith,
  Thompson, Shectman, \& Burley}]{Sobeck2011}
Sobeck, J.~S., Kraft, R.~P., Sneden, C., {et~al.} 2011, Astronomical Journal,
  141, 18

\bibitem[{Spitzer(1987)}]{Spitzer1987}
Spitzer, L. 1987, {Dynamical evolution of globular clusters} (Princeton, N.J. :
  Princeton University Press, c1987)

\bibitem[{Stanek {et~al.}(2006)Stanek, Gnedin, Beacom, Gould, Johnson,
  Kollmeier, Modjaz, Pinsonneauly, Pogge, \& Weinberg}]{Stanek2006}
Stanek, K., Gnedin, O., Beacom, J., {et~al.} 2006, ACTA Astronomica, 56, 333

\bibitem[{Tanvir {et~al.}(2017)Tanvir, Levan, Gonzalez-Fernandez, Korobkin,
  Mandel, Rosswog, Hjorth, D'Avanzo, Fruchter, Fryer, Kangas, Milvang-Jensen,
  Rosetti, Steeghs, Wollaeger, Cano, Copperwheat, Covino, D'Elia, Postigo,
  Evans, Even, Fairhurst, Jaimes, Fontes, Fujii, Fynbo, Gompertz, Greiner,
  Hodosan, Irwin, Jakobsson, Jorgensen, Kann, Lyman, Malesani, McMahon,
  Melandri, O'Brien, Osborne, Palazzi, Perley, Pian, Piranomonte, Rabus, Rol,
  Rowlinson, Schulze, Sutton, Thoene, Ulaczyk, Watson, Wiersema, \&
  Wijers}]{Tanvir2017}
Tanvir, N.~R., Levan, A.~J., Gonzalez-Fernandez, C., {et~al.} 2017, The
  Astrophysical Journal Letters, 848, L27

\bibitem[{Tauris {et~al.}(2013)Tauris, Langer, Moriya, Podsiadlowski, Yoon, \&
  Blinnikov}]{Tauris2013}
Tauris, T.~M., Langer, N., Moriya, T.~J., {et~al.} 2013, Astrophysical Journal
  Letters, 778, L23

\bibitem[{Tauris {et~al.}(2015)Tauris, Langer, \& Podsiadlowski}]{Tauris2015}
Tauris, T.~M., Langer, N., \& Podsiadlowski, P. 2015, Monthly Notices of the
  Royal Astronomical Society, 451, 2123

\bibitem[{Tauris {et~al.}(2017)Tauris, Kramer, Freire, Wex, Janka, Langer,
  Podsiadlowski, Bozzo, Chaty, Kruckow, van~den Heuvel, Antoniadis, Breton, \&
  Champion}]{Tauris2017}
Tauris, T.~M., Kramer, M., Freire, P. C.~C., {et~al.} 2017, The Astrophysical
  Journal, 846, 170

\bibitem[{Thompson {et~al.}(2001)Thompson, Burrows, \& Meyer}]{Thompson2001}
Thompson, T.~A., Burrows, A., \& Meyer, B.~M. 2001, The Astrophysical Journal,
  562, 887

\bibitem[{Thompson {et~al.}(2004)Thompson, Chang, \& Quataert}]{Thompson2004}
Thompson, T.~A., Chang, P., \& Quataert, E. 2004, The Astrophysical Journal,
  611, 380

\bibitem[{Tsang \& Milosavljevi{\'{c}}(2018)}]{Tsang2018}
Tsang, B.~T., \& Milosavljevi{\'{c}}, M. 2018, Monthly Notices of the Royal
  Astronomical Society, 478, 4142

\bibitem[{van~de Voort {et~al.}(2015)van~de Voort, Quataert, Hopkins,
  Kere{\v{s}}, \& Faucher-Gigu{\'{e}}re}]{VandeVoort2015}
van~de Voort, F., Quataert, E., Hopkins, P.~F., Kere{\v{s}}, D., \&
  Faucher-Gigu{\'{e}}re, C.~A. 2015, Monthly Notices of the Royal Astronomical
  Society, 447, 140

\bibitem[{Vanzella {et~al.}(2018)Vanzella, Calura, Meneghetti, Castellano,
  Caminha, Mercurio, Cupani, Rosati, Grillo, Gilli, Mignoli, Fiorentino,
  Arcidiacono, Lombini, \& Cortecchia}]{Vanzella2018}
Vanzella, E., Calura, F., Meneghetti, M., {et~al.} 2018, Monthly Notices of the
  Royal Astronomical Society, 483, 3618

\bibitem[{Verbunt {et~al.}(2017)Verbunt, Igoshev, \& Cator}]{Verbunt2017}
Verbunt, F., Igoshev, A., \& Cator, E. 2017, Astronomy and Astrophysics, 608,
  15

\bibitem[{Vigna-G{\'{o}}mez {et~al.}(2018)Vigna-G{\'{o}}mez, Neijssel,
  Stevenson, Barrett, Belczynski, Justham, de~Mink, M{\"{u}}ller,
  Podsiadlowski, Renzo, Sz{\'{e}}csi, \& Mandel}]{Vigna-Gomez2018}
Vigna-G{\'{o}}mez, A., Neijssel, C.~J., Stevenson, S., {et~al.} 2018, Monthly
  Notices of the Royal Astronomical Society, 481, 4009

\bibitem[{Villar {et~al.}(2017)Villar, Guillochon, Berger, Metzger,
  Cowperthwaite, Nicholl, Alexander, Blanchard, Chornock, Eftekhari, Fong,
  Margutti, \& Williams}]{Villar2017}
Villar, V.~A., Guillochon, J., Berger, E., {et~al.} 2017, The Astrophysical
  Journal Letters, 851, L21

\bibitem[{Vink \& de~Koter(2005)}]{Vink2005}
Vink, J.~S., \& de~Koter, A. 2005, Astronomy {\&} Astrophysics, 442, 587

\bibitem[{Vink {et~al.}(2001)Vink, de~Koter, \& Lamers}]{Vink2001}
Vink, J.~S., de~Koter, A., \& Lamers, H. J. G. L.~M. 2001, Astronomy {\&}
  Astrophysics, 369, 574

\bibitem[{Wallner {et~al.}(2015)Wallner, Faestermann, Feige, Feldstein, Knie,
  Korschinek, Kutschera, Ofan, Paul, Quinto, Rugel, \& Steier}]{Wallner2015}
Wallner, A., Faestermann, T., Feige, J., {et~al.} 2015, Nature Communications,
  6, 1

\bibitem[{Wanderman \& Piran(2010)}]{Wanderman2010}
Wanderman, D., \& Piran, T. 2010, Monthly Notices of the Royal Astronomical
  Society, 406, 1944

\bibitem[{Webb \& Leigh(2015)}]{Webb2015}
Webb, J.~J., \& Leigh, N.~W. 2015, Monthly Notices of the Royal Astronomical
  Society, 453, 3278

\bibitem[{Webbink(1984)}]{Webbink1984}
Webbink, R. 1984, Astrophysical Journal, 277, 355

\bibitem[{Wehmeyer {et~al.}(2015)Wehmeyer, Pignatari, \&
  Thielemann}]{Wehmeyer2015}
Wehmeyer, B., Pignatari, M., \& Thielemann, F.~K. 2015, Monthly Notices of the
  Royal Astronomical Society, 452, 1970

\bibitem[{Winteler {et~al.}(2012)Winteler, K{\"{a}}ppeli, Perego, Arcones,
  Vasset, Nishimura, Liebend{\"{o}}rfer, \& Thielemann}]{Winteler2012}
Winteler, C., K{\"{a}}ppeli, R., Perego, A., {et~al.} 2012, Astrophysical
  Journal Letters, 750, 5

\bibitem[{Wong {et~al.}(2010)Wong, Willems, \& Kalogera}]{Wong2010}
Wong, T.-W., Willems, B., \& Kalogera, V. 2010, The Astrophysical Journal, 721,
  1689

\bibitem[{Woosley(1993)}]{Woosley1993}
Woosley, S.~E. 1993, The Astronomical Journal, 405, 273

\bibitem[{Woosley(2016)}]{Woosley2016}
---. 2016, The Astrophysical Journal, 836, 244

\bibitem[{Worley {et~al.}(2013)Worley, Hill, Sobeck, \& Carretta}]{Worley2013}
Worley, C.~C., Hill, V., Sobeck, J., \& Carretta, E. 2013, Astronomy {\&}
  Astrophysics, 553, A47

\bibitem[{Ye {et~al.}(2019)Ye, Kremer, Chatterjee, Rodriguez, \&
  Rasio}]{Ye2019}
Ye, C.~S., Kremer, K., Chatterjee, S., Rodriguez, C.~L., \& Rasio, F.~A. 2019,
  The Astrophysical Journal, 877, 122

\bibitem[{Yong {et~al.}(2006)Yong, Aoki, Lambert, \& Paulson}]{Yong2006}
Yong, D., Aoki, W., Lambert, D.~L., \& Paulson, D.~B. 2006, The Astrophysical
  Journal, 639, 918

\bibitem[{Yong \& Grundahl(2008)}]{Yong2008}
Yong, D., \& Grundahl, F. 2008, The Astrophysical Journal, 672, L29

\bibitem[{Zevin {et~al.}(2019)Zevin, Samsing, Rodriguez, Haster, \&
  Ramirez-Ruiz}]{Zevin2019a}
Zevin, M., Samsing, J., Rodriguez, C., Haster, C.-J., \& Ramirez-Ruiz, E. 2019,
  The Astrophysical Journal, 871, 15

\bibitem[{Zwart {et~al.}(2010)Zwart, McMillan, \& Gieles}]{PortegiesZwart2010}
Zwart, S.~P., McMillan, S., \& Gieles, M. 2010, Annual Review of Astronomy and
  Astrophysics, 48, 431

\end{thebibliography}

\appendix
\section{Treatment of SN Kicks Following a CE Phase}\label{Appendix}

\begin{figure*}[b!]\label{fig:defaultBSE}
    \centering
    \includegraphics[width=0.98\textwidth]{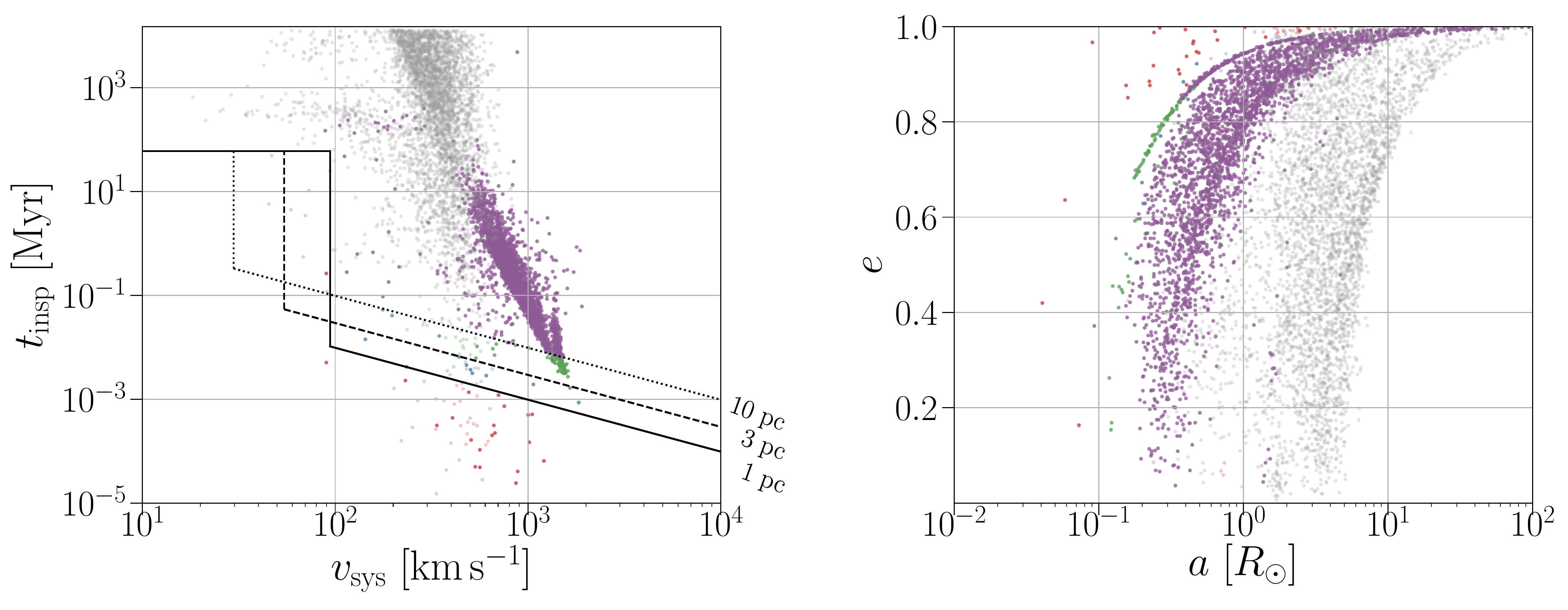}
    \caption{Analogous to Model A in Figure \ref{fig:primordial}, but for a population model where the pre-CE donor mass is used when determining the effect of the second SN. }
\end{figure*}

Besides the standard updates previously implemented in the \texttt{COSMIC} version of \texttt{BSE}, one notable change in our population models is how SN kicks are implemented during CE evolution. 
By default, \texttt{BSE} treats this in a self-inconsistent way that leads to artificially inflated post-SN systemic velocities. 

Following a CE phase, the final separation between the core of the donor star and its compact companion is calculated using standard energetics arguments as in  \cite{Webbink1984}. 
The orbital hardening during a CE phase is calculated by the energy necessary to eject the envelope: 
\begin{equation}
    E_{\rm bind,\,i}  = \alpha_{\rm CE} (E_{\rm orb,\,f} - E_{\rm orb,\,i}),
\end{equation}
\begin{equation}
    -\frac{G}{\lambda} \frac{M_{\rm c}(M_{\star}-M_{\rm c})}{R} = - \alpha_{\rm CE} \frac{G M_{\rm c} m_{\rm NS}}{2} \left( \frac{1}{a_{\rm f}} - \frac{1}{a_{\rm i}} \right).
\end{equation}
This can be solved for the final orbital separation as 
\begin{equation}
    a_{\rm f} = \frac{\alpha_{\rm CE}\,\lambda\,R\,m_{\rm NS}\,a_{\rm i}}
    {2 (M_\star-M_{\rm c})\,a_{\rm i} + \alpha_{\rm CE}\,\lambda\,R\,m_{\rm NS}},
\end{equation}
where $\alpha_{\rm CE}$ is the CE efficiency parameter, $\lambda$ is the envelope binding energy parameter that adjusts the amount of energy needed from the orbit to eject the envelope, $R$ is the envelope radius, and $M_{\star}$ is the initial mass of the donor star, and $M_{\rm c}$ is the core mass of the donor star. 

If a star evolves into an NS or BH immediately after the CE, the function for implementing the SN kick is called. 
However, in default \texttt{BSE}, at this point the pre-SN mass of the exploding star is reset to its value prior to the CE (it does not account for the envelope mass that is ejected during the CE spiral-in to bring the pre-SN binary to its tightened orbital configuration), leading to a large amount of mass loss during the SN at a tight pre-SN orbital separation. 
This issue is especially drastic when dealing with CEs involving an evolved naked He star, as the cores of the stars can reach extremely hardened orbits with separations of $\approx\,0.1\,R_{\odot}$. 
The contribution of the mass-loss Blaauw kick on the post-SN systemic velocity is 
\begin{equation}
    v_{\rm sys,\ Blaauw} \approx 220 
    \left(\frac{\Delta M_{\rm SN}}{M_{\odot}}\right)
    \left(\frac{a_{\rm pre}}{R_{\odot}}\right)^{-1/2}
    \left[2.8 + \left(\frac{\Delta M_{\rm SN}}{M_{\odot}}\right)\right]^{-1/2}
    \rm{km\,s}^{-1},
\end{equation}
where $\Delta M_{\rm SN}$ is the mass lost in the SN, $a_{\rm pre}$ is the pre-SN semi-major axis, and NS masses are assumed to be $1.4\,M_{\odot}$. 

The increased mass loss therefore leads to systematically higher post-SN systemic velocities. 
For example, a DNS progenitor at an orbital separation of $0.1\,R_{\odot}$ that loses $1.5\,M_{\odot}$ in the SN will achieve a post-SN systemic velocity solely from the Blaauw kick of $\approx$\,500 km\,s$^{-1}$. 
Therefore, even without SN natal kicks, newly born DNSs can be boosted to post-SN systemic velocities of $\gtrsim 1000$ km\,s$^{-1}$ solely due to mass loss in the SN, as seen in Figure \ref{fig:defaultBSE}. 
Though the successful Case BB CE can lead to hardened binaries that merge in $\lesssim\,1\,\mathrm{Myr}$, the amplified systemic velocities for these DNS systems cause most to escape the cluster environment before gravitational radiation leads to inspiral and merger. 
Thus, even with a successful Case BB CE, enrichment probabilities are only a few percent. 
In addition to systemic velocities, this inconsistency can strongly affect post-SN orbital properties, SN survival, and merger rates from populations that are modeled using a modified \texttt{BSE} framework.

\section{Effect of SN Mass Loss on Enrichment Candidates}\label{AppendixB}

\begin{figure*}[b!]\label{fig:blaauw}
\centering
\includegraphics[width=0.7\textwidth]{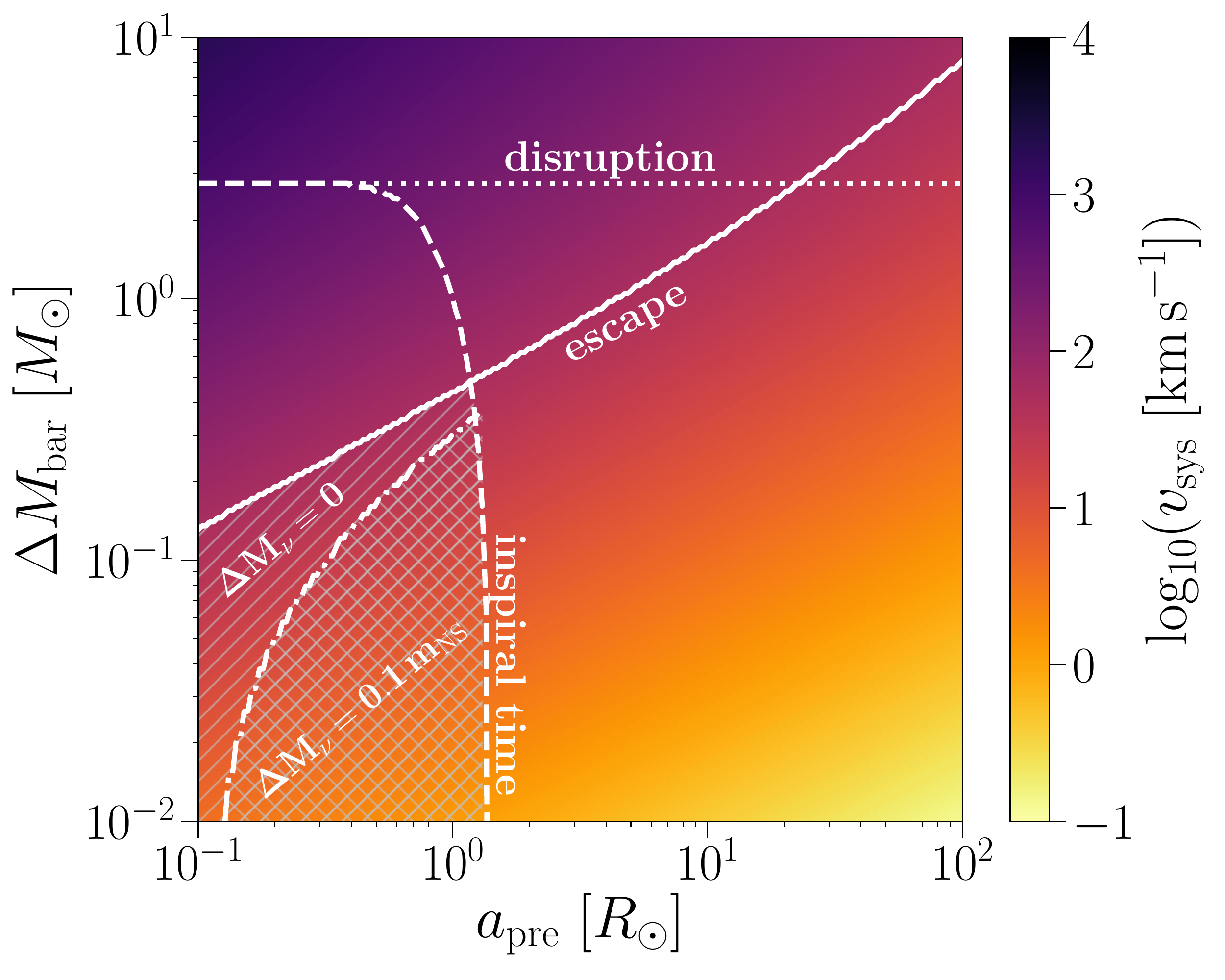}
\caption{Viable enrichment candidates from DNS systems that remain bound to a fiducial GC. 
We solely examine the effect of the Blaauw kick from mass loss in the SN; $v_{\rm sys}$ is the post-SN systemic velocity assuming zero natal kicks. 
$\Delta M_{\rm bar}$ is the baryonic mass loss, such that the total mass lost in the SN is $\Delta M_{\rm SN} = \Delta M_{\rm bar} + \Delta M_{\nu}$ with $\Delta M_{\nu}$ being the mass loss from neutrinos due to the collapse of the stellar core into an NS. 
Circular pre-SN orbits and NS masses of $m_{\rm NS} = 1.4\,M_{\odot}$ are assumed. 
The solid white line marks the escape velocity from the half-mass-radius of a GC progenitor with $M = 10^6\,M_{\odot}$ and $R_{\rm vir} = 3$ pc. 
The dashed white line marks an inspiral time of 100 Myr. 
Systems that satisfy these two criteria fall below and to the left of these two lines (in the white hatched region) and are viable enrichment candidates that remain bound to the GC. 
The diagonal hatched region assumes that $\Delta M_{\nu} = 0$, i.e.\ there is no neutrino mass loss. 
The dotted--dashed line and cross-hatched region assume that $\Delta M_{\nu} = 0.1\,m_{\rm NS}$, further constraining the possible combinations of $\Delta M_{\rm bar}$ and $a_{\rm pre}$ that will lead to fast-merging systems bound to the GC. 
Systems above the dotted white line are disrupted due to losing over half their total mass. }
\end{figure*}

Small natal kicks for some NSs at birth are necessary to explain the retention fraction of NSs in GCs \citep{Pfahl2002}. 
These retained systems are predicted to form from ECSNe or the accretion-induced collapse of a white dwarf \citep{Ye2019}. 
However, retention of NSs formed in binary systems also depends on the Blaauw kick, which is anti-correlated with the square root of the binary orbital separation at the time of the SN. 

As shown in Figure \ref{fig:primordial}, most DNSs born as a binary pair have post-SN systemic velocities that would exceed the escape velocity of young GCs. 
This is particularly apparent for systems that undergo Case BB MT, as even small amounts of mass loss in their extremely hardened state prior to SN will result in post-SN systemic velocities of $\gtrsim 100$ km\,s$^{-1}$. 
This does not conflict with the number of pulsars observed in GCs today since isolated NSs do not receive a Blaauw kick at formation --- the natal kick alone controls the post-SN velocity and therefore whether or not it remains bound to the cluster. 

Hardened DNS systems that underwent Case BB MT make up the bulk of the systems that contribute to GC enrichment fractions. 
In their tight orbital configurations, the natal kick is typically small compared to the pre-SN orbital velocity, and the post-SN systemic velocity is mostly controlled by the mass lost in the SN rather than the natal kick that the exploding star received \citep{Kalogera1996}. 
Lower amounts of mass loss in the SN could therefore lead to more of these systems being retained and amplifying the number of enrichment candidates (i.e.\ moving the purple points in the left column of Figure~\ref{fig:primordial} to the left). 

In Figure \ref{fig:blaauw} we show the impact of mass loss in the SN and pre-SN separation on the post-SN systemic velocity. 
For simplicity, we assume no natal kicks, a circular pre-SN orbit, and NSs masses of $m_{\rm NS}=1.4\,M_{\odot}$, such that the post-SN systemic velocity reduces to an exact expression that is solely dependent on the orbital separation and mass loss: 
\begin{equation}\label{eq:blaauw}
    v_{\rm sys}|_{v_{\rm k}=0} = \frac{\Delta M_{\rm SN}}{2}
    \sqrt{\frac{G}{a_{\rm pre} (2m_{\rm NS}+\Delta M_{\rm SN})}}.
\end{equation}
\noindent The entire hatched region marks the pre-SN separations and mass loss that would form a DNS that is bound to a fiducial GC and merges within 100 Myr. 
Stable Case BB MT can deplete the envelopes of the donor stars such that the mass lost in the subsequent SNe can be as low $\sim$\,0.1\,$M_{\odot}$ and harden pre-SN orbits to $a_{\rm pre}\,\lesssim\,1\,R_{\odot}$ \citep{Tauris2015}, safely within the hatched region. 
However, regardless of the level at which the envelope of the donor star is stripped, mass lost from neutrinos during the collapse of the stellar core results in a non-negligible contribution to the systemic velocity of the post-SN binary \citep{Lattimer1989,Lovegrove2013}, particularly in binaries that have been significantly hardened. 
The dotted--dashed line in Figure \ref{fig:blaauw} further constrains the parameter space by assuming 10\% of the NS mass is lost in neutrino emission during the SN; under this assumption bound systems that are viable enrichment candidates only reside in the narrow cross-hatched region. 

Fortuitous orientations of the natal kick can increase the parameter space that will lead to DNSs with low enough systemic velocities such that they remain bound to the cluster. 
However, assuming isotropic natal kicks, this will only happen for a small number of DNS systems. 
Given how fine-tuned these conditions must be, we expect the GC enrichment contribution from bound, fast-merging DNSs to be small regardless of the physical prescriptions underlying Case BB stripping, with the bulk of the enrichment candidates coming from DNSs that are unbound from the GC and merge before evacuating the cluster environment. 

\end{document}